\documentclass[structabstract]{aa}    
\usepackage{amssymb,amsmath} 
\usepackage{graphicx}  
\usepackage{txfonts}
\usepackage{natbib}
\usepackage{color}   
\usepackage{threeparttable}   
\usepackage{booktabs}
\usepackage{wasysym}
\newcommand{\degree}{\ensuremath{^\circ}}
\bibpunct{(}{)}{;}{a}{}{,} % to follow the A&A style
\def\mathbi#1{\textbf{\em #1}}
\def\ProDiMo{{\sc ProDiMo\ }}

\def\H2O{H$_2$O}
\def\pd{protoplanetary disk}

\begin{document}  
\title{Radiation thermo-chemical models of protoplanetary disks}
\subtitle{Grain and polycyclic aromatic hydrocarbon charging}
\author{W. F. Thi\inst{1}, G. Lesur\inst{2}, P. Woitke\inst{3,4},
  I. Kamp\inst{5}, Ch. Rab\inst{5}, A. Carmona\inst{6}}  
\titlerunning{Grain charging in protoplanetary disks} 
\institute{
Max Planck Institute for Extraterrestrial Physics, Giessenbachstrasse, 85741 Garching, Germany
 \and
Universit\'{e} Grenoble-Alpes, CNRS, Institut de Plan\'{e}tologie et
  d'Astrophysique (IPAG) UMR 5274, Grenoble, F-38041, France
   \and
    SUPA, School of Physics \&
  Astronomy, University of St. Andrews, North Haugh, St. Andrews KY16
  9SS, UK 
   \and
Centre for Exoplanet Science, University of St Andrews, St Andrews, UK 
  \and 
  Kapteyn Astronomical Institute, P.O. Box 800, 9700 AV
  Groningen, The Netherlands  \and
 IRAP, Universit\'{e} de Toulouse, CNRS, UPS, CNES, 14 Avenue Edouard Belin, Toulouse, F-31400, France
} 
\authorrunning{W.-F. Thi}
%\date{Received 2013; accepted 2013}
  \abstract
% Context  
  {Disks around pre-main-sequence stars evolve over time by turbulent viscous spreading. The main contender to explain the strength of the turbulence is the Magneto-Rotational-Instability (MRI) model, whose efficiency depends on the disk ionization fraction.}
  {We aim at computing self-consistently the chemistry including PAH charge chemistry, the grain charging and an estimate of an effective value of the turbulence $\alpha$ parameter in order to find observational signatures of disk turbulence.}
  {We introduced PAH and grain charging physics and their interplay with other gas-phase  reactions in the physico-chemical code {\sc ProDiMo}. Non-ideal magnetohydrodynamics effects such as Ohmic and ambipolar diffusion are parametrized to derive an effective value for the turbulent parameter $\alpha_{\rm eff}$. We explored the effects of turbulence heating and line broadening on CO isotopologue sub-millimeter lines.}
  { The spatial distribution of $\alpha_{\rm eff}$ depends on various unconstrained disk parameters such as the magnetic parameter $\beta_{\mathrm{mag}}$ or the cosmic ray density distribution inside the \pd s. The inner disk midplane shows the presence of the so-called "dead-zone" where the turbulence is quasi-inexistent. The disk is heated mostly by thermal accommodation on dust grains in the dead-zone, by viscous heating outside the dead-zone up to a few hundred astronomical units, and by chemical heating in the outer disk. The CO rotational lines probe the warm molecular disk layers where the turbulence is at its maximum. However, the effect of turbulence on the CO line profiles is minimal and difficult to distinguish from the thermal broadening.}  
 {Viscous heating of the gas in the disk midplane outside the dead-zone is efficient. The determination of $\alpha$ from CO rotational line observations alone is challenging.}


\maketitle
%________________________________________________________________
\section{Introduction}\label{introduction}

Pre-main-sequence stars (TTauri and HerbigAe stars) are surrounded by
planet-forming disks in a Keplerian rotation \citep{Williams2011,Espaillat2014prpl.conf..497E}. The disks are massive at
the early stage reaching gas masses of 10$^{-2}$ M$_\odot$ or even higher
and disappear after a few million years \citep{Alexander2014prpl.conf..475A}. The disk material can evaporate, form giant planets, 
or fall onto the star. Gas can accrete from the inner rim of disks to the star at a rate $\dot{M}$ of
10$^{-9}$-10$^{-7}$ M$_\odot$ yr$^{-1}$
\citep{Calvet2000prpl.conf..377C,Muzerolle2004ApJ...617..406M,Johns-Krull2000ApJ...539..815J,Hartmann1998ApJ...495..385H}, although episodic accretion rate of up to 10$^{-5}$ M$_\odot$ yr$^{-1}$ can occur for young disks \citep{Audard2014prpl.conf..387A}. Accretion shocks onto the stellar surfaces result in ultraviolet excess
emission. Gas accretion is also traced by optical emission lines
\citep{Mendigutia2012A&A...543A..59M,Mendigutia2011A&A...535A..99M,GarciaLopez2006A&A...459..837G}. 
 
The most promising explanation to drive mass accretion in disks is
turbulence driven by magneto-rotational instability because molecular
viscosity is too weak (MRI; see \citealt[]{Balbus1991ApJ...376..214B,Balbus2011ppcd.book..237B,Fromang2006A&A...457..343F,Bai2015ApJ...798...84B,Simon2013ApJ...775...73S,Simon2015ApJ...808..180S,Bhat2017MNRAS.472.2569B,Bethune2016A&A...589A..87B,OKeeffee2014MNRAS.441..571O,Fleming2003ApJ...585..908F,Davis2010ApJ...713...52D}).  Alternatively, a weak turbulence may be generated purely hydrodynamic instabilities, such as the vertical shear instability \citep{Nelson2004MNRAS.350..849N,Lin2015ApJ...811...17L}, gravitational instability for self-gravitating disks \citep{Gammie2001ApJ...553..174G,Forgan2012MNRAS.426.2419F,Hirose2017MNRAS.469..561H}, vertical shear instability \citep{Flock2017ApJ...835..230F}, zombie vortex instability, and baroclinic instabilities \citep{Klahr2003ApJ...582..869K,Lyra2011A&A...527A.138L}.

Magneto-rotational instability is an ideal magnetohydrodynamics (MHD) phenomenon that is effective only if the gas is
sufficiently ionized to couple dynamically to the magnetic fields
\citep{Gammie1996ApJ...457..355G,Bai2015ApJ...798...84B}. Gas turbulence efficiency is
characterized by the factor $\alpha$ so that the large-scale 
turbulence is $\nu=\alpha c_\mathrm{s}h$, where $c_\mathrm{s}$ is the gas sound speed and $h$ is disk pressure scale-height representing the fluid typical scale. For subsonic turbulence, the value of $\alpha$ should be lower than unity, with an usually assumed value of 0.01.

Many non-ideal Magnetohydrodynamics (MHD) dissipation effects can restrict the development of MRI
turbulence or even suppress it \citep{Jin1996ApJ...457..798J}. The Ohmic and ambipolar diffusion depend on the abundances of the charge carriers among other factors. Many MHD
simulations were carried out to study the extent of ''dead zones'' defined as regions where dissipation overcomes the MRI turbulence \citep{Wardle1999MNRAS.303..239W,Sano2002ApJ...570..314S,Bai2011ApJ...736..144B,Gressel2015ApJ...801...84G}. On the other
hand, Hall diffusion  revives the MRI under certain conditions (MRI-Hall; \citealt{Lesur2014A&A...566A..56L,Wardle2012MNRAS.422.2737W,Sano2002ApJ...570..314S,Balbus2001ApJ...552..235B}).
  
The charge carriers in disks are either electrons, atomic and
molecular ions, polycyclic aromatic hydrocarbons (PAHs) or dust grains \citep{Bai2011ApJ...736..144B}.  The abundances of the carriers vary throughout the disk due
to ionization processes (photoionization by UV and X-ray photons,
cosmic rays), recombination with an electron, and charge exchanges. The
rates of these processes are functions of the gas composition, the gas
and dust temperatures, the UV, X-ray, and density fluxes inside the
disks \citep{Rab2017A&A...603A..96R}. Therefore, the computations of the abundances require a detailed
gas and dust physical and chemical modelling together with the continuum
and line radiative-transfer. \cite{Bai2011ApJ...739...51B} and \cite{Perez-Becker2011ApJ...735....8P} explored the role of grains in the efficiency of MRI but did not consider the effects of radiation on the grain charging and considered PAHs as small grains and not as macro-molecules.

Ionization fraction in realistic \pd s was modelled
with different levels of sophistication. The aim of those studies was
to determine the extent of the dead-zone, the area in disks where the
ionization fraction is too low for the magnetic field to couple to the neutral gas, and hence
for MRI to sustain turbulence. No observational evidence exists either for or against the presence of a
dead-zone yet. The dead-zone encompasses the region of planet formation in
disks. Interestingly one of the questions in planet-formation is the influence of
low turbulence on the growth of planetesimals.

\citet{Ilgner2006A&A...445..205I} considered viscous heating and
radiative cooling and tested the effects of different chemical networks
on the ionization fraction in disks. The source of ionization is
stellar X-rays. Subsequently, \citet{Ilgner2006A&A...455..731I} studied
the effects caused by X-ray flares. The effect of turbulent mixing on
gas chemistry was explored in
\citet{Ilgner2006A&A...445..223I, Heinzeller2011ApJ...731..115H}. \citet{Fromang2002MNRAS.329...18F}
modelled the ionization fraction and the size of the dead-zone in
disks. \citet{Perez-Becker2011ApJ...735....8P} focused on the UV
ionized layers of disks and argued that this layer cannot sustain the
accretion rate generated at larger
radii. \citet{Dzyurkevich2013ApJ...765..114D} studied the dependence
of disk parameters such as temperature, surface density profile, and
gas-to-dust mass ratio on the MRI efficiency. The sources of ionisations in the  inner disks are discussed by \citet{Desch2015ApJ...811..156D}.
They found that the ambipolar diffusion controls the location of the dead zone. \citet{Ivlev2016ApJ...833...92I} considered the ionization and the effect of dust charging in \pd s. The role played by dust grains in the ionization fraction in disks has also been discussed in \cite{Salmeron2008MNRAS.388.1223S}. \cite{Simon2011ApJ...743...17S} performed non-ideal MHD simulations, predicting turbulence velocities in the disk midplane ranging from 0.01 times the sound speed in the dead-zone to 0.1 outside.

Observational constraints on the turbulence factor $\alpha$ in \pd s have been obtained for the \object{TW Hya} \citep{Teague2016A&A...592A..49T,Flaherty2018ApJ...856..117F} and \object{HD 163296} \citep{Flaherty2015ApJ...813...99F,Flaherty2017ApJ...843..150F} disks by determining the contribution of the turbulence broadening to the total line width using ALMA data.  \cite{Teague2016A&A...592A..49T}  found a turbulence value of $(0.2-0.4)\  c_ s$ in the \object{TW Hya}  disk, while \cite{Flaherty2018ApJ...856..117F} constrained $\alpha$ to be between  0.04  $c_ s$ and  0.13 $c_ s$. \cite{Flaherty2015ApJ...813...99F} provided a low upper limit of  $\mathrm{v}_{\rm turb} <  0.03\ c_{\rm s}$ for the turbulence in the 
\object{HD 163296} disk. Earlier studies with the IRAM Plateau de Bure interferometer provided limits of
$\mathrm{v}_{\rm turb} \leq (0.3-0.5)\ c_{\rm s}$ \citep{Dartois2003A&A...399..773D,Pietu2007A&A...467..163P}. \cite{Hughes2011ApJ...727...85H}
constrained the value $\mathrm{v}_{\rm turb} \leq 0.1\ c_{\rm s}$ for the \object{TW Hya} disk and $\mathrm{v}_{\rm turb} \leq 0.4\ c_{\rm s}$ for the \object{HD 163296} disk
using data obtained by the Smithsonian Millimeter Array. The accuracy of the estimated values is limited by the uncertainties in inverting the \pd\ thermal structure from the observations. \citet{Hartmann2018MNRAS.474...88H} argue that a low viscosity disk model driven by hydrodynamic turbulence is compatible with the observed disk mass accretion rates and the measured low turbulence width.

In this paper, we explore further the effects of a detailed treatment of the physics and chemistry of PAHs and grain charging on the disk ionization. For this purpose, we implemented MRI-turbulence  heating and cooling in the photo-chemical disk code {\sc ProDiMo}. We considered far-ultraviolet from the star and the accretion excess, X-ray, and cosmic rays as sources of ionization. We used CO isotopologue rotational lines as potential tracers of turbulence in disks because CO is the most abundant molecule in disks after H$_2$, is widespread through the disks, and because its chemistry is well understood. 
The numerical study of MHD processes in triggering disk turbulence is beyond the scope of this paper. Instead, we parametrize the onset and effects of turbulence on the gas by simple empirical formula. This study focuses on the role of the gas and grain chemistry in the disk's ionization, which in turns can affect the non-ideal MHD coefficients. The scope of the {\sc ProDiMo} code is to be able to derive disk parameters such as $\alpha$ by matching high signal-to-noise observations carefully considering the relevant disk physico-chemical processes.

The paper is organized as follows: the \ProDiMo code is introduced in Sect.~\ref{prodimo}; a brief discussion on the gas-phase and PAH chemistry are given in Sect.~\ref{Gas_phase_chemistry}  and ~\ref{PAH_chemistry}; the dust charging physics implemented in the code is described in Sect.~\ref{dust_charging}; the prescription of our MRI-driven turbulence model is provided in Sect.~\ref{mri_turbulence}; analytical results on the influence of non-ideal MHD are shown in Sect.~\ref{analytical_results}; the \pd\ model and the model results are presented in Sect.~\ref{disk_model} and Sect.~\ref{disk_results}; our findings are discussed in Sect.~\ref{discussion} and we conclude in Sect.~\ref{conclusions}.

\section{Gas and dust charge exchange reactions}
\label{mri_implementation}

\subsection{ProDiMo}\label{prodimo}

{\sc ProDiMo} is a code built to model the gas and dust grain physics
and chemistry
\citep{Woitke2009A&A...501..383W,Kamp2010A&A...510A..18K,Woitke2016A&A...586A.103W}. It has been
used to model disk Spectral Energy Distributions
(SEDs, \citealt{Thi2011MNRAS.412..711T}), water deuteration chemistry
\citep{Thi2010MNRAS.407..232T} CO rovibrational emissions including
UV-fluorescence \citep{thi2013}, and many {\it
  Herschel} observations from the {\it GASPS} large programme.  X-ray
physics are implemented
\citep{Aresu2012A&A...547A..69A,Meijerink2012A&A...547A..68M,Aresu2011A&A...526A.163A}. {\sc ProDIMo} has been designed to run within a few CPU-hours per model such that automatic fittings of the observed continuum emission and of thousands of gas lines are feasible. Such fitting procedure requires running thousands of models, which would be too time-consuming with a full 3D non-ideal MHD radiation chemical code.
As such, {\sc ProDiMo} does not solve the gas hydrodynamic equations. The disk density structures are parametrized. Only the vertical hydrostatic structure can be self-consistently re-adjusted with the gas temperature. 
In this study we chose to  model disks with fixed vertical hydrostatic structures.
A detailed discussion of the different physics and their implementations are given in the articles listed above. Here we summarize the main features relevant to the modelling of the
MRI-turbulence.

In our chemical modelling, we included gas and ice species as well as
PAHs. The grain charge is computed for grains of mean radius
$\langle a \rangle$. The grains can be up to four times positively or negatively
charged. The photoionization and photodissociation rates are computed
from the cross-sections and UV field calculated from 2D continuum
radiative transfer
\citep{vanDishoeck2011arXiv1106.3917V}. Self-shielding is taken into
account. The gas temperature at each location in the disk is computed
by balancing the heating rate with the cooling rate contrary to
the previous study of MRI in disks where the gas temperature follows a
power-law distribution in radius $T \propto r^{-p}$ and the disk is
isothermal in the vertical direction or considered viscous heating
only \citep{Hughes2011ApJ...727...85H,Flaherty2015ApJ...813...99F,Teague2016A&A...592A..49T}. 

In {\sc ProDiMo} heating agents include photoelectrons ejected
from PAHs \citep{Bakes1994ApJ...427..822B} and dust grains (photoelectric effects), chemical heating,
photoionization heating, and in this study release of viscous
energy. Atomic and molecular lines can heat or cool the gas by
absorption or emission. At high densities, the gas and the dust grains
exchange energy by thermal contact. The thermal accommodation can heat
or cool the gas depending on the sign of $\Delta
T$=$T_{\mathrm{gas}}$-$T_{\mathrm{dust}}$
\citep{Burke1983ApJ...265..223B}. If $\Delta T$ is positive, the
gas-dust accommodation will cool the gas and vice-versa. 
The codes generates detailed line fluxes and profiles from non-LTE radiative 
transfer, that can be compared directly to observations.

Knowledge of the precise ionization fraction at each location in the disks is
paramount for the onset of MRI-driven turbulence. A few additional
features have been implemented to improve the physics and chemistry
that regulate the charge distribution in \pd s. Most of the features concern a better
treatment of the charging of the PAHs and dust grains.

%---------------------------------------------------------------------------------------
\subsection{Gas phase chemistry}\label{Gas_phase_chemistry}

The gas-phase chemistry includes photodissociation, ion-neutral, neutral-neutral, as well as a few colliders and three-body reactions. The chemical network is discussed in details in \cite{Kamp2017arXiv170707281K}.\\
At high gas temperatures, ionization by collisions with hydrogen atoms
or molecules and with electrons is possible.
\begin{equation}
\mathrm{H} + \mathrm{H}_2 \rightarrow \mathrm{H}^- + \mathrm{H}_2^+,
\end{equation}
with the rates provided by \cite{Hollenbach1980ApJ...241L..47H}.  For
the thermal ionization by hydrogen,
\cite{Hollenbach1980ApJ...241L..47H} suggest using the rate for
electrons scaled by a factor 1.7$\times$10$^{-4}$.
Collisional ionization of metals can also occur
\begin{equation}
\mathrm{M} + \mathrm{N} \rightarrow \mathrm{M}^- + \mathrm{N}+\mathrm{e}^-,
\end{equation}
where $M$ is a neutral metal (Fe, Mg, S, Na, Si, ...) and $N$ is either H, H$_2$, or He.
\subsection{PAH charge exchange chemistry}\label{PAH_chemistry}

PAHs can become the main charge carriers in disks because of their
abundances and electron affinity.  Details on the PAH chemistry can be found in \cite{Kamp2017arXiv170707281K} and Thi et al., submitted.

In protoplanetary disks, their abundances are lower by a factor
$f_{\mathrm{PAH}}$=10$^{-1}$-10$^{-3}$ compared to their interstellar
abundance of 3$\times$10$^{-7}$ \citep{Tielens2008ARA&A..46..289T}.
We chose to use the peri-condensed circumcoronene (C$_{54}$H$_{18}$) as typical PAHs
that are large enough to remain unaffected by photodissociation in disks around
HerbigAe stars \citep{Visser2007A&A...466..229V}. The circumcoronene can be once negatively-charged
(PAH$^-$) and three times positively charged by absorbing a UV photon with
energy below 13.6 eV or by charge exchange reactions (PAH$^+$,
PAH$^{2+}$, PAH$^{3+}$, see Table~\ref{tab_PAH}). The effective radius of
a PAH is computed by \citep{Weingartner2001ApJS..134..263W}
\begin{equation}
a_{\mathrm{PAH}}=10^{-7}\left(\frac{N_{\mathrm{C}}}{468}\right)^{1/3}\ \mathrm{cm}, 
\end{equation}  
where $N_{\mathrm{C}}$ is the number of carbon atoms in the PAH. The
radius for the circumcoronene is
$a_{\mathrm{PAH}}$(C$_{54}$H$_{18}$)=4.686 $\times$ 10$^{-8}$ cm. The
PAH ionization potential can either be taken from the literature when
they are measured or estimated \citep{Weingartner2001ApJS..134..263W}
\begin{equation}
IP_{\mathrm{PAH}} = W_0 + (Z_{\mathrm{PAH}}+0.5)\frac{e^2}{a_{\mathrm{PAH}}}+(Z_{\mathrm{PAH}}+2)\frac{e^2}{a_{\mathrm{PAH}}}\frac{0.3\times 10^{-8}}{a_{\mathrm{PAH}}}\ {\mathrm{erg}}, 
\end{equation}
where $W_0$ is the work function assumed to be 4.4 eV
($7.05\times10^{-12}$ erg), and $Z_{\mathrm{PAH}}$ is the charge of
the PAH. The ionization potentials (I.P.) for circumcoronene are listed in
Table~\ref{tab_PAH}.

PAHs are not formed or destroyed in our chemical network and only
exchange charges with other positively-charged species (for examples
H$^+$, He$^+$, Mg$^+$, Fe$^+$, Si$^+$, S$^+$, HCO$^+$, ...). Chemical
reaction rates involving PAHs are highly uncertain. Most of the rates
are extrapolations from a few existing laboratory or theoretical rates. PAH freeze-out is
presented in \cite{Kamp2017arXiv170707281K}.

% ----------
\begin{table}
\begin{center}
  \caption{Circumcoronene electron affinity and ionization
    potential. The measured (lit.) and computed (WD2001) values are for the Electron Affinity (E.A.) and Ionization Potential (I.P.)
    shown.\label{tab_PAH}}
\begin{tabular}{llll}
                  \toprule
                  & E.A. (eV) & I.P. (eV) lit. & I.P WD2001\\
\noalign{\smallskip}   
\hline

\smallskip
C$_{54}$H$_{18}$       & 1.3 & 5.9 & 6.2 \\
\smallskip
C$_{54}$H$_{18}^+$       & ... & 8.8 & 9.4\\
C$_{54}$H$_{18}^{2+}$       & ... & 12.9 & 12.5\\
\noalign{\smallskip}     
\bottomrule
\end{tabular}   
\end{center}
\end{table}
% ----------
On disk surfaces, PAH are mostly positively charged because of the photoelectric effect. We compute the PAH photoejection rate using detailed PAH cross-sections and disk UV fields computed by radiative transfer \citep{Woitke2016A&A...586A.103W}. A free electron can also attach on PAHs. A detailed discussion on PAH charging is provided in Thi et al., submitted.

\subsection{Dust grain charging}\label{dust_charging}  

Dust grains with radius can be major charge carriers in the interstellar medium in
general and in \pd s in particular. Dust grain charging is studied in the field of plasma physics
\citep{Mishra2015doi:10.1063/1.4907664}.

We implemented the silicate dust grain charging physics of \citet{Draine1987ApJ...320..803D}, 
\citet{Weingartner2001ApJS..134..263W}, and \citet{Umebayashi1980PASJ...32..405U}  with a couple of
differences. We considered an average grain of radius $\langle a \rangle \equiv a$ and
a geometric cross-section $\sigma_{\mathrm{dust}}=\pi a^2$. The charge
of the grain is $Z_{\mathrm{d}}$ with a discrete charge distribution function
$f(Z_{\mathrm{d}})$ with a minimum charge $Z_{\mathrm{min}}$ and
maximum charge $Z_{\mathrm{max}}$, such that
$\sum^{Z_{\mathrm{max}}}_{Z_{\mathrm{min}}} f(Z_{\mathrm{d}})=1$.

\subsubsection{Silicate dust grain ionization potential}\label{IP}

Detailed quantum mechanical calculations of the work function of oxide and silicate clusters show variations for different silicate compositions \citep{Rapp2012_angeo-30-1661-2012}. They noticed that silicates have an increase in the work function by 2 eV compared to the oxides (MgO, FeO) and attributed this effect to the presence of the silicon. When a grain is covered by a thick water ice mantle, the work function will be modified. Water ice has a work function of 8.7 eV \citep{Rapp2009_angeo-27-2417-2009,Baron1978}, comparable to that of the magnesium-rich silicate. Therefore, the work function $W_0$ for bulk silicate is assumed to be typically 8 eV whether the grain is coated by an icy mantle or not.\\
When a grain is positively charged, the
Coulomb attraction between the grain and an electron effectively
increases the effective work function by an extra work function
that depends on the grain charge $Z_{\mathrm{d}}$. The effective work
function $W_{\mathrm{eff}}$, equivalent to the ionization potential for an atom or
molecule, becomes \citep{Weingartner2001ApJS..134..263W}
\begin{equation}
IP = W_{\mathrm{eff}} = W_0 + W_{\mathrm{c}},
\end{equation}
where $W_{\mathrm{c}}$ is the extra work function defined by
\begin{equation} 
W_{\mathrm{c}} = \left(Z_{\mathrm{d}}+\frac{1}{2}\right) \frac{e^2}{a},\label{eqn_Wc}
\end{equation}
where $e$ is an elementary charge.When the grain is positively-charged ($Z_\mathrm{d}>1$).
The value of the constant added to the grain charge is the correction
due to the finite size of a perfectly spherical grain and is
controversial.  A value of 3/8 has also been proposed instead of 1/2
\citep{Wong2003PhRvB..67c5406W}. 

\subsubsection{Photoejection and photodetachment}\label{dust_photoelectric}

An electron can be ejected upon absorption of a photon with energy
$\Theta$ higher than the ionization potential of the grain. The photoejection process concerns
positively-charged and neutral grains, while the photodetachment process 
concerns negatively-charged grains.

The photodetachment cross-sections for negatively-charged grains and for
neutral and positively-charged grains respectively are calculated
following the prescription of \citet{Weingartner2001ApJS..134..263W}. 

\begin{figure}[!ht]
\centering
 \resizebox{\hsize}{!}{\includegraphics[angle=0,width=10cm,height=7cm,trim=55 55 60 50, clip]{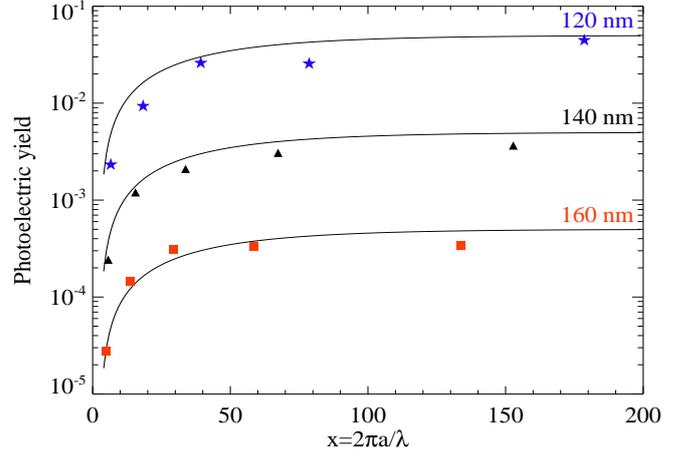}}
\caption{Analytical fits to the \citet{Abbas2006ApJ...645..324A}
  experimental data, shown as stars, triangles, and squares, at wavelengths 120 nm, 140 nm, and 160 nm.}  
  \label{fig_abbas}          
\end{figure}

The photoejection yield $\eta_{\mathrm{b}}$ for the silicate bulk varies as function of
the photon energy $\Theta$ following equation 17 of
\citet{Weingartner2001ApJS..134..263W}
\begin{equation}
\eta_{\mathrm{b}}(\Theta) = \frac{0.5(\Theta/W_0)}{1+5(\Theta/W_0)},
\end{equation}
where $W_0$ is the silicate bulk photoejection yield assumed to be 8
eV, see also \citep{Kimura2016MNRAS.459.2751K}.
The bulk yield is enhanced for very small grains by a factor
$\eta_{\mathrm{small}}$ according to equation 13 of
\citet{Weingartner2001ApJS..134..263W}. The effective yield becomes
$\eta_{\mathrm{eff}}=\eta_{\mathrm{b}}\eta_{\mathrm{small}}$. For
large grains ($x=2\pi a/\lambda>$5) we used an analytical fit to the
experimental data of \citet{Abbas2006ApJ...645..324A} for the effective
yield
\begin{equation}
\eta_{\mathrm{eff}}(x,\lambda)= (1-e^{-0.025(x-x_0)})\times10^{-0.05(\lambda-\lambda_0)-1.3},
\end{equation}
where $x_0=2.5$, $\lambda_0$=120 nm. The formula reproduces the
experimental data within factor of a few. The photoelectric yield at
three wavelengths for different grain sizes is shown in
Fig.~\ref{fig_abbas}.
% ----------------
% ---------
The energy threshold for photoejection of a grain of charge $Z_{\mathrm{d}}$ reads
\begin{equation}
h\nu_{\mathrm{pe}}=W_{\mathrm{eff}}. 
\end{equation}
When a grain is negatively charged the photodetachment threshold
energy is \citep{Weingartner2001ApJS..134..263W}
\begin{equation}
h\nu_{\mathrm{pd}}(Z_{\mathrm{d}}<0)=EA(Z_{\mathrm{d}}+1,a)+E_{\mathrm{min}}(Z_{\mathrm{d}},a),
\end{equation}
where the electron affinity is
\begin{equation}
EA(Z_{\mathrm{d}},a)=W_0-E_{\mathrm{bg}}+\left(Z_{\mathrm{d}}+\frac{1}{2}\right)\frac{e^2}{a}. 
\end{equation}
The band gap $E_{\mathrm{bg}}=0$ for metals and semimetals while it can be several eVs for other materials.
\citet{Weingartner2001ApJS..134..263W} assumed $(W_0-E_{\mathrm{bg}})=3$ eV for a silicate grain (a band gap of 5~eV). 
The value for water ice is much lower at 0.8 eV \citep{doCouto2006129}.\\
The value for $E_{\mathrm{min}}$ follows the definition of \citet{Weingartner2001ApJS..134..263W}
\begin{equation}
E_{\mathrm{min}}(Z_{\mathrm{d}}<0,a) = -(Z_{\mathrm{d}}+1)\frac{e^2}{a}\left[1+\left(\frac{27\ \AA}{a} \right)^{0.75}\right]^{-1}.
\end{equation}
For a singly negatively-charged grain, $Z_{\mathrm{d}}$=-1 and $a$ = 1 $\mu$m,
$h\nu_{\mathrm{pd}}(Z_{\mathrm{d}},a)\simeq$ 2.9 eV or
$\lambda_{\mathrm{pd}}\simeq$~0.4 $\mu$m for a pure silicate grains and $h\nu_{\mathrm{pd}}(Z_{\mathrm{d}},a)\simeq$ 0.7 eV or
$\lambda_{\mathrm{pd}}\simeq$~1.66 $\mu$m for pure water ice grains in the near infrared. The threshold energy for photodetachment is
much lower than for photoejection. For disks, the dust extinction is
lower at longer wavelengths and at the same time the stellar
luminosity is higher in the blue. 
Both effects combine to make photodetachment very efficient. The combined rate for photoejection and photodetachment is
\begin{equation}
k_{\mathrm{pe}}=\pi a^2 \int_{\nu_{\mathrm{pe}}}^{\nu_{\mathrm{max}}} \eta_{\mathrm{eff}}Q_{\mathrm{abs}}J_\nu d\nu +\pi a^2 \int_{\mathrm{\nu_{\mathrm{pd}}}}^{\nu_{\mathrm{max}}} \eta_{\mathrm{pd}} Q_{\mathrm{abs}} J_\nu d\nu,
\end{equation}
where $Q_{\mathrm{abs}}$ is the frequency-dependent absorption efficiency, $J_\nu$ is the
specific mean intensity at the frequency $\nu$ computed in by the continuum
radiative transfer. 

\subsubsection{Electron attachment}\label{dust_electron_attachment}

The electron attachment rate coefficient onto neutral grains is
\begin{equation}
 k_{\mathrm{e,n}}=
n_{\mathrm{e}}S_{\mathrm{e}}\sqrt{\frac{8kT_{\mathrm{e}}}{\pi m_{\mathrm{e}}}}\sigma_{\mathrm{dust}}\ [\text{electrons s}^{-1}],\label{eqn_elec_attachment}
\end{equation}
where $T_{\mathrm{e}}$ is the electron temperature assumed equal to
the gas kinetic temperature, $S_{\mathrm{e}}$ is the sticking
coefficient for electron attachment, assumed to be 0.5 (50\% of the
encounters are assumed elastic), $\sigma_{\mathrm{dust}}$ is the
geometrical cross-section of a grain of radius $a$, and
$m_{\mathrm{e}}$ is the mass of an electron. \cite{Umebayashi1980PASJ...32..405U} discussed theoretically the sticking coefficient of electrons on grains. 
They found that the sticking probability depends on the surface composition and is in most cases larger than 0.3.\\
For positively-charged grains ($Z_{\mathrm{d}}>0$) the electron
recombination rate is enhanced by Coulomb attraction
\begin{equation}
 k_{\mathrm{e,+}}=k_{\mathrm{e,n}}\left(1+\frac{W_{\mathrm{c}}}{kT}\right),
\end{equation}
where $W_{\mathrm{c}}$ is the extra work function (see eq.~\ref{eqn_Wc}). On the contrary,
the recombination rate is lower if the grain is negatively charged
($Z_{\mathrm{d}}<0$)
\begin{equation}
k_{\mathrm{e,-}}=k_{\mathrm{e,n}}\exp{\left(\frac{W_{\mathrm{c}}}{kT}\right)}.
\end{equation}

\subsubsection{Thermionic emission}\label{dust_thermionic_emission}

We considered the thermionic emission of electrons by hot dust grains
following the Richardson-Dushman theory \citep{Ashcroft,Sodha2014}
\begin{equation}
  k_{\mathrm{RD}} = \frac{4\pi m_e k^2}{h^3}T_d^2(1-r)\exp{\left(-\frac{W_{\mathrm{therm}}}{kT_d}\right)}\sigma_{\mathrm{dust}}\ \text{[electrons s}^{-1}],\label{eqn_thermionic}
\end{equation}
where $T_d$ is the dust temperature, $r$ is the so-called reflection parameter. It corresponds to the fraction of electrons that have enough energy to escape at grain surfaces but do not do so. It is assumed to have a value of 0.5. The thermionic emission work $W_{\mathrm{therm}}$ is equal to 
$W_{\mathrm{eff}}$ for positively-charged grains and to $EA+E_{\mathrm{min}}$ for negatively-charged grains.

\subsubsection{Collisional electron detachment}\label{coll_elec_detachment}

An atomic hydrogen or molecular hydrogen impinging onto the grain can
either transfer momentum or energy to the grain surface such that an
electron is ejected.
\begin{equation}
k_{\mathrm{cd}}^{\mathrm{d}}= n_{\mathrm{n}}\sigma_{\mathrm{dust}}\sqrt{\frac{8kT}{\pi m_{\mathrm{n}}}}\exp{\left(-\frac{W_{\mathrm{cd}}}{kT}\right)},
\end{equation}
where the work $W_{\mathrm{cd}}$ is equal to $W_{\mathrm{eff}}$ for positively-charged grains and to $EA(Z_\mathrm{d}+1)+E_{\mathrm{min}}$ for negatively-charged grains.

\subsubsection{Charge exchange between ions and dust grains}

Gas-phase species and dust grains can exchange charges. \cite{Weingartner2001ApJ...563..842W} considered the effect of ion-charged grains on the ionization of the interstellar gas.
Cations can recombine with an electron from the grains. Likewise, a positively-charged grain can capture an electron from atoms and 
molecules with ionization potential lower than that of the grain. \\
For negatively-charged grains, the non-dissociative exchange reaction 
can be written as a suite of reactions ($n \geq 0$)
\begin{equation}
\mathrm{gr}^{n-} + \mathrm{X_g}^+ \rightarrow \mathrm{gr}^{(n-1)-} + \mathrm{X_s} + \Delta E_\mathrm{ads}(X_g) + \Delta E_\mathrm{Coulomb}
\end{equation}
\begin{equation}
\mathrm{gr}^{n-} + \mathrm{X_s}^+ \rightarrow \mathrm{gr}^{(n-1)-} + \mathrm{X_s} + \Delta E_\mathrm{rec} + \Delta E_\mathrm{relax}
\end{equation}
where $X_g^+$ and $X_s^+$ are gas-phase and surface ions respectively, or 
\begin{equation}
\mathrm{gr}^{n-} + \mathrm{X_s}^+ \rightarrow \mathrm{gr}^{(n-1)-} + \mathrm{X_g} + \Delta E_\mathrm{rec} - \Delta E_\mathrm{ads}(X_g).
\end{equation}
The gas-phase species X$_g^+$ acquires the approach energy that is the sum of the adsorption energy ($\Delta E_\mathrm{ads}(X_g)$) and the Coulomb energy $\Delta E_\mathrm{Coulomb}$. $\Delta E_\mathrm{rec}$ is the energy released/required by the recombination reaction and corresponds to the ionization potential minus the electron affinity for negatively-charged grains or the work function of the solid for neutral grains. 
The excess energy can be radiated away or transfer to the surface ($\Delta E_\mathrm{relax}<0$). It can also be used to break the weak bond ($\Delta E_\mathrm{ads}(X_s)\sim$~0.1-0.2 eV) between the species $X_s$ and the surface. Therefore, we assumed that the neutral atom immediately leaves the grain surface after the recombination.\\
For molecular ion recombination, extra outcomes are possible. The recombination results first into an excited species
\begin{equation}
\mathrm{gr}^{n-} + \mathrm{AH_g}^+ \rightarrow \mathrm{gr}^{(n-1)-} + \mathrm{AH_s^*} + \Delta E_{\rm rec}
\end{equation}
When the excess energy of the electronically excited neutral species AH$^*$ cannot transfer efficiently to the grain surface or radiated away (because the species has a low dipole moment), the recombination is dissociative, 
\begin{equation}
\mathrm{AH_s^*}  \rightarrow \mathrm{A_g} + \mathrm{H_g} + \Delta E_\mathrm{gdiss,rec},
\end{equation}
where the molecule dissociates into species that leave immediately the surface carrying the excess energy as kinetic energy; or one of the products remain on the grain surface
\begin{equation}
\mathrm{AH_s^*}  \rightarrow \mathrm{A_s} + \mathrm{H_g} + \Delta E_\mathrm{sdiss,rec}.
\end{equation}
Here the heavier product A tends to easily transfer the excess energy to the surface or possesses many degrees of freedom and, therefore, can radiate efficiently the excess energy away.
On the other hand, the inefficient transfer of energy from the light hydrogen atom to a heavy surface element results in the atom leaving the surface.\\
Cations can also exchange their positive charge with a neutral or
positively-charged grain with charge $+n$ with ($n \ge 0$)
\begin{equation}
\mathrm{gr}^{n+} + \mathrm{X_g}^+ \rightarrow \mathrm{gr}^{(n+1)+} + \mathrm{X}_g.
\end{equation}
For large silicate grains, 
\cite{Aikawa1999ApJ...527..262A} performed classical computation for the recombination of HCO$^+$ with negatively-charged grains and concluded that the recombination results in the dissociation of the molecules. Another example is given by
\begin{equation}
\mathrm{gr} + \mathrm{NH}^+ \rightarrow \mathrm{gr}^+ + \mathrm{N} + \mathrm{H}.
\end{equation}
Both recombination and charge exchange reactions proceed with the rate
\begin{equation}
 k_{\mathrm{gr,ion}}=
n_{\mathrm{ion}}S_{\mathrm{ion}}\sqrt{\frac{8kT_{\mathrm{ion}}}{\pi m_{\mathrm{ion}}}}\sigma_{\mathrm{dust}} \max\left(0,1-\frac{Z_{\mathrm{d}}e^{2}}{akT_{\mathrm{ion}}}\right)e^{-E_{\mathrm{therm}}/kT_{\mathrm{ion}}},
\end{equation}
where $E_{\mathrm{therm}}$ is an energy equal to the
endothermicity of the reaction. For an exothermic charge exchange, the
energy is null ($E_{\mathrm{therm}}$=0). The ion temperature $T_{\mathrm{ion}}$ is equal to the
gas thermal temperature $T_{\mathrm{gas}}$ and we assumed
$S_{\mathrm{ion}}$=1, similar to
\citet{Weingartner2001ApJS..134..263W}. The term in parentheses is positive for negatively-charged grains, enhancing cation recombinations. At the same time, it ensures that
cation exchanges with highly-positive grains are prevented due to a repulsive potential. \\
The energetic barrier is defined by
\begin{equation}
\begin{array}{rcl}
E_{\mathrm{therm}} & = & \Delta E_\mathrm{rec} - \Delta E_\mathrm{ads}(X_g) - \Delta E_\mathrm{Coulomb} \\
                                         & \simeq  & \max\left[IP_{\mathrm{gr}}-IP_{\mathrm{neu}} , 0 \right]\\\
\end{array}
\end{equation}
for neutral and positively-charged grains. We assume that the adsorption and Coulomb energy are negligible. For negatively-charged grains
\begin{equation}
E_{\mathrm{therm}} = \max\left[\left(EA(Z_{\mathrm{d}}+1,a)+E_{\mathrm{min}}(Z_{\mathrm{d}},a)\right)-IP_{\mathrm{neu}}, 0 \right]\\
\end{equation}
where $IP_{\mathrm{neu}}$ is the ionization energy of the neutral species.
Again for $Z_{\mathrm{d}}$=-1 and $a$ = 1 $\mu$m, $EA(Z_{\mathrm{d}}+1,a)\simeq$2.99 eV and $E_{\mathrm{min}}(Z_{\mathrm{d}}<0,a)\simeq$ 2.9 eV, thus
 \begin{equation}
E_{\mathrm{therm}} = \max\left[5.89 - IP_{\mathrm{neu}}, 0 \right]\ \mathrm{in\ eV}.
\end{equation}
The reality is probably much more complex. The Coulomb interaction acquired by the approaching ion can lead to the tunnelling of a surface electron through the grain work function, resulting in a recombination in the gas-phase \citep{Tielens1987ASSL..134..397T}.\\
Positively-charged grains ($n>$1) can transfer their charge to a neutral gas-phase species,
\begin{equation}
\mathrm{gr}^{n+} + \mathrm{X_g} \rightarrow \mathrm{gr}^{(n-1)+} + \mathrm{X_g}^+.
\end{equation}
The positive grains can induce a field that polarizes the neutral species. Since the grain has a large geometrical cross-section, the rate is assumed to be the largest value between a Langevin-type rate and that of a gas impinging on a neutral grain
\begin{equation}
\begin{split}
k_{\mathrm{gr^+,n}}= n_{\mathrm{n}} \times \max \left[2 \pi |Z_\mathrm{d}| e \sqrt{\frac{\alpha_{\mathrm{pol}}}{mu}},S_{\mathrm{n}}\sqrt{\frac{8kT_{\mathrm{n}}}{\pi m_{\mathrm{n}}}}\sigma_{\mathrm{dust}}\right] \\
\times e^{-E_{\mathrm{therm}}/kT_{\mathrm{n}}},
\end{split}
\end{equation}
where the reduced mass is basically that of the gas-phase species $\mu = m_{\mathrm{n}}$ and the polarizability is $\alpha_{\mathrm{pol}}=10^{-24}$
where the barrier term becomes
\begin{equation}
E_{\mathrm{therm}} = max \left[IP_{\mathrm{neu}}-IP_{\mathrm{gr}}, 0 \right]
\end{equation}
in practically all cases, the geometrical rate dominates,
\begin{equation}
\begin{split}
k_{\mathrm{gr^+,n}}= n_{\mathrm{n}} S_{\mathrm{n}}\sqrt{\frac{8kT_{\mathrm{n}}}{\pi m_{\mathrm{n}}}}\sigma_{\mathrm{dust}} e^{-E_{\mathrm{therm}}/kT_{\mathrm{n}}},
\end{split}
\end{equation}
Anions can also react with positively-charged grains
\begin{equation} 
\mathrm{gr}^{n+} + \mathrm{X_g^-} \rightarrow \mathrm{gr}^{(n-1)+} + \mathrm{X_g}.
\end{equation}
However, we did not have anions in the chemical network at this current stage.
% --------------------------
\begin{table}
\begin{center}
  \caption{Examples of work function and ionization potentials ($\le$13.6 eV).\label{tab_IP}}          
\begin{tabular}{lc}     % columns   
\toprule
\noalign{\smallskip}        
\multicolumn{1}{c}{Solid/Species} & \multicolumn{1}{c}{Work function/I.P. (eV)}\\
\hline
\noalign{\smallskip}
(MgSiO$_3$)$_3$        & 9.2  \\
(FeSiO$_3$)$_3$         & 8.6\\
(Mg$_2$SiO$_4$)$_4$ & 7.6\\ 
(FeOH)$_4$                   & 5.5\\
(MgOH)$_4$                  & 5.0\\
water ice                        & 8.7\\
\noalign{\smallskip}
\hline
\noalign{\smallskip} 
H/H$^+$                        & 13.6\\
Cl/Cl$^+$                      & 11.48\\
C/C$^+$                       & 11.26\\
S/S$^+$                       & 10.36\\  
Si/Si$^+$                        & 8.1517\\
Fe/Fe$^+$                    & 7.90\\
Mg/Mg$^+$                  & 7.646\\
Na/Na$^+$                   & 5.139\\
K/K$^+$                        & 4.341\\
\noalign{\smallskip}
\hline
\noalign{\smallskip}
NH/NH$^+$                  & 13.47 \\
H$_2$O/H$_2$O$^+$  & 12.6\\
NH$_2$/NH$_2^+$       &11.09\\
NH$_3$/NH$_3^+$      & 10.2\\
HCO/HCO$^+$             & 8.14 \\
H$_3$O/H$_3$O$^+$  & 4.95 \\
NH$_4$/NH$_4^+$       & 4.73\\
\noalign{\smallskip}
\hline
\noalign{\smallskip}
neutral silicate grain  & 8.0 (assumed)\\
charged grain ($Z_{\mathrm{d}}=-1$, 1 $\mu$m) & 2.7\\
charged grain ($Z_{\mathrm{d}}=1$, 1 $\mu$m) & 8.002\\
\noalign{\smallskip}
\bottomrule
\end{tabular}
\end{center}
Ref. {\sc NIST} database \citep{Nist2005}.
\end{table}
Table~\ref{tab_IP} gives a few examples of species ionization potentials.
Interestingly, charge exchange between neutral dust grains and the chemically important molecular ions H$_3$O$^+$,
NH$_4^+$ are endothermic, while reaction with HCO$^+$ is possibly exothermic.\\
Charge exchange reactions will result in the positive charges being carried by the species with the lowest ionization potential, 
whose value is lower than the energy required to detach an electron from a neutral grain. The cations with low
ionization potential will preferably recombine with free electrons and not with negatively-charged grains because of the velocity and
also on the extra work required to remove an electron from a grain ($\sim$2.7 eV).

\subsubsection{Minimum and maximum grain charges}\label{dust_max_charges}

The maximum positive charge that a grain can acquire is determined by
the highest energy of the photons if the dominant ionization process is photoejection.

Assuming a grain radius $a$ in micron
and $W_0$=8 eV and a maximum photon energy of
$h\nu_{\mathrm{max}}$=13.6 eV, an elementary charge $e$ in c.g.s. of 4.803206815 $\times$ 10$^{-10}$, 
and the conversion 1 eV = $1.60217657 \times 10^{-12}$ erg,  the maximum charge, assuming that photo-ejection is the dominant 
electron ejection process, reads
\begin{equation}\label{eq_Zmax}
Z_{\mathrm{max}}=\left(h\nu_{\mathrm{max}}-W_0\right)a/e^2-0.5\simeq 3884\left(\frac{a}{\mu\mathrm{m}}\right).
\end{equation}

For the minimum grain charge, we adopted again the formalism of \citet{Weingartner2001ApJS..134..263W}
\begin{equation}
Z_{\mathrm{min}} = \mathrm{int}\left( \frac{U_{\mathrm{ait}}}{14.4}\frac{a}{\AA}\right) +1,
\end{equation}
where
\begin{equation}
\frac{U_{\mathrm{ait}}}{\mathrm{V}} \simeq \left\{ 
\begin{array}{l l}
  3.9+1200(a/\mu m)+0.0002(\mu m/a) & \quad \text{for carbonaceous},\\
  2.5+700(a/\mu m)+0.0008(\mu m/a) & \quad \text{for silicate}\\ \end{array} \right.
\end{equation}
For a 1 micron radius silicate grain, $U_{\mathrm{ait}}\simeq 702 $ V and
$Z_\mathrm{min}\simeq -487499.$ This large value corresponds to the maximum negative charges on
a silicate grain of radius $a$ without considering any electron ejection process including charge exchange processes. In practice, the maximum amount of negative charges on a grain is limited
by the balance between electron recombination and electron emission.
Assuming a balance between electron attachment and thermionic emission in the absence of UV photons ($k_{\mathrm{RD}}(Z_d<1)=k_{\mathrm{e,-}}$), we can derive an upper limit to the amount of negative charges on a grain from formula \ref{eqn_elec_attachment} and \ref{eqn_thermionic}.
\begin{equation}
Z_{\mathrm{d}} \simeq \frac{1}{2}\frac{a}{e^2}(f-2)(W_0-E_\mathrm{bg}+kT\ln{B})-\frac{1}{2},\label{eq_Zmin2}
\end{equation}
where
\begin{equation}
B = \frac{n_{\mathrm{e}}S_{\mathrm{e}}\sqrt{8kT_{\mathrm{g}}/\pi m_{\mathrm{e}}}}{(4\pi m_{\mathrm{e}} k^2/h^3) T_{\mathrm{d}}^2(1-r)}.\\
\end{equation}
Assuming that $S_{\mathrm{e}}=1$, $r$=0.5, and $T_{\mathrm{d}}=T_{\mathrm{g}}=T$,
\begin{equation}
B \simeq 5.27 \times 10^{-8} T^{-3/2} n_{\mathrm{e}},
\end{equation}
and 
\begin{equation}
f=\left(1+\left(\frac{2.7\times 10^{-7}}{a}\right)^{0.75}\right)^{-1}
\end{equation}
For dense regions with n$_{\mathrm{e}}=10^{12}$ cm$^{-3}$ and $T=100$~K,  the term $B$ is greater than unity. The term
$kT\ln{B} \simeq 2.37 \times 10^{-5} T \ln{(5.27 \times 10^{-8} T^{-3/2} n_{\mathrm{e}})}$ eV is always negligible compared to $W_0$ (=8 eV),
thus a strong  lower limit to the grain charge can be derived
\begin{equation}
Z_{\mathrm{d}} > \frac{1}{2}(f-2)\frac{a}{e^2}(W_0-E_{\mathrm{bg}})-\frac{1}{2}.\label{eq_Zmin3}
\end{equation}
Assuming elementary charge in c.g.s of $e=4.80321 \times 10^{-10}$ statC and $W_0$=8 eV, $E_\mathrm{bg}$=5 eV, and $f\sim$1,
we obtain
\begin{equation}
Z_{\mathrm{d}} > -2107 \left(\frac{a}{\mu\mathrm{m}}\right)\label{eq_Zlimit}
\end{equation}
In this study, we did not include photoejection due to X-ray photons, and we adopted in our numerical models the limits
 \begin{equation}
 |Z_{\mathrm{d}}| < 4000\left(\frac{a}{\mu\mathrm{m}}\right).
 \end{equation}
 
\subsection{Dust charge estimates for dense UV-obscured regions}\label{charge_analytical_noUV} 

\begin{figure}[!ht]
\centering
 \resizebox{\hsize}{!}{\includegraphics[angle=0,width=10cm,height=7cm,trim=55 55 60 50, clip]{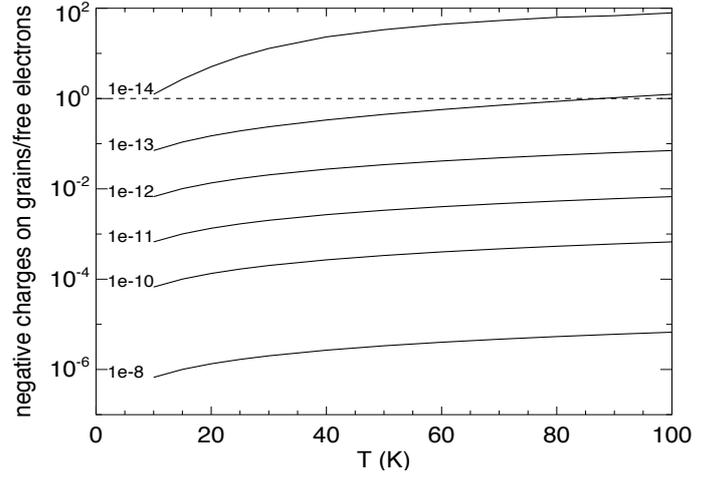}}
\caption{Ratio between the amount of negative charges on grains and free electrons for a grain of radius 1 micron and different total ionization fraction (from 10$^{-14}$ to 10$^{-8}$) as function of the gas kinetic temperature.}  
\label{fig_ratio_charge_grain_free_electron}          
\end{figure}

In disk regions where photoemission can be neglected (even photodetachment requires photons in the optical blue domain), one can estimate the grain charging at equilibrium by balancing the electron recombination with the charge exchange between the ions and the grains. Here we also neglect the effects of UV photons created by interaction of cosmic rays with the gas. The grain charge is the solution of the non-linear transcendental equation assuming equal sticking coefficients $S_{\mathrm{e}}=S_{\mathrm{i}}$ \citep{Evans1994duun.book.....E}
\begin{equation}
\left(1-\frac{Z_{\mathrm{d}}e^2}{akT} \right)\exp{-\left(\frac{(Z_{\mathrm{d}}+0.5)e^2}{akT}\right)}\frac{n_{\mathrm{ion}}}{n_{\mathrm{elec}}}=\sqrt{\frac{m_{\mathrm{ion}}}{m_{\mathrm{elec}}}},
\end{equation}
 where we can set $x=Z_{\mathrm{d}}e^2/akT$. Using the medium global neutrality $n_\mathrm{elec}=n_{\mathrm{ion}}+Z_{\mathrm{d}}n_{\mathrm{d}}$, the equation becomes
 \begin{equation}
 (1-x)e^{-x} \simeq \sqrt{\frac{m_{\mathrm{ion}}}{m_{\mathrm{elec}}}}\left(1+x \frac{n_\mathrm{d}}{n_\mathrm{ion}}\frac{akT}{e^2}\right)
 \end{equation}
 A representation of the ratio between charges on grain surfaces and free electrons in an obscured region is shown in Fig.~\ref{fig_ratio_charge_grain_free_electron} for a 1 micron radius grain.  Grains are negatively charged because of the difference in velocity between the electrons and the ions.
 % ----------------
% ---------
 Assuming $n_{\mathrm{ion}} \sim n_{\mathrm{elec}}$ an approximate solution to the equation is
 \begin{equation}
 Z_{\mathrm{d}}\sim-6(2.7+0.4\ln{(n_{\mathrm{nucl}})}) \left(\frac{a}{\mathrm{\mu m}}\right)\left(\frac{T}{100\mathrm{K}}\right),
\end{equation}
where $n_{\mathrm{nucl}}$ is the average number of nucleons of the cations. If the main ion is a molecular ion with 19 nucleons (HCO$^+$), the approximation becomes
\begin{equation}
 Z_{\mathrm{d}}\sim-23 \left(\frac{a}{\mathrm{\mu m}}\right)\left(\frac{T}{100\mathrm{K}}\right).\label{Zestimate_noUV}
\end{equation}
The estimated negative charge is much lower than the limit set by the balance between electron recombination and thermionic emission. The approximation is valid when the
charge carriers are dominated by the free electrons and by the ions. Interestingly, the dust negative charging increases with the gas temperature. 

Using the limit on the number of
negative charges on a dust grain (see formula \ref{eq_Zlimit}), we can derive the maximum fractional abundance of negative charges on grains
\begin{equation}
\chi (\mathrm{d})=\frac{| Z_{\mathrm{d}}| n_{\mathrm{d}} }{n_{\mathrm{<H>}}} \leq 6 \times 10^{-12} \left(\frac{\mathrm{\mu m^2}}{a^2}\right)\left(\frac{100}{gd}\right).\label{eq_max_elec_on_grains}
\end{equation}
One can also use the estimate of the dust charge in obscured disk regions (equation \ref{Zestimate_noUV}). The estimated fractional abundance of negative charges on grains becomes
\begin{equation}
\chi (\mathrm{d})=\frac{| Z_{\mathrm{d}}| n_{\mathrm{d}} }{n_{\mathrm{<H>}}}  \sim 6.7 \times 10^{-14} \left(\frac{\mathrm{\mu m^2}}{a^2}\right)\left(\frac{100}{gd}\right)\left(\frac{T}{100\mathrm{K}}\right).\label{max_dust_charge_noUV}
\end{equation}
This last equation together with Fig.~\ref{fig_ratio_charge_grain_free_electron}  shows that, unless the ionization fraction is below 10$^{-13}$, the negative charges on large silicate grains (with a radius greater than one micron) are negligible in UV-obscured regions. 

Dust charging is stochastic by nature and dust grains exhibit a distribution of charge states. In addition, the charge on a given grain can fluctuate over time \citep{Piel2010,Cui279018,Matthews2013ApJ...776..103M}. Extensive numerical studies suggest that the dust charge distribution about its equilibrium value $Z_{\mathrm{d}}$ when photons are absent has a standard deviation of $\delta Z_{\mathrm{d}} =0.5 | Z_{\mathrm{d}}|^{1/2}$. 
The simultaneous existence of positively and negatively charged grains is possible for  small grains of 0.1 $\mu$m in radius or smaller at 10~K.

The numerical implementation of the simultaneous computation of the grain charge distribution, the PAH restricted chemistry, and the gas-phase chemistry is explained in more details in Appendix~\ref{numerical_implementation}. When photon-induced electron ejection is present, the number of negative charges on grains will be even lower.

\section{MRI-driven turbulence prescription for hydrostatic \pd\  models.}\label{mri_turbulence}

In this section, we describe our model non-ideal MHD driven turbulence gas heating and cooling and line broadening.
% ----------- 
\subsection{Ideal MHD value of $\alpha$}\label{ideal_mhd}

The ideal-MHD MRI value for the turbulence parameter $\alpha$, which we call $\alpha_{\mathrm{ideal}}$, can be evaluated from the results of
detailed local 3D-MHD simulations. These simulations have shown that
$\alpha_{\mathrm{ideal}}$ can be related to $\beta_{\mathrm{mag}}$, which is the ratio of the thermal $P_{\mathrm{therm}}(r,z)$
to magnetic pressures $P_{\mathrm{mag}}(r,z)$, by the function
\begin{equation}
\alpha_{\mathrm{ideal}} = \left(\frac{2}{\beta_{\mathrm{mag}}}\right)^{\delta},
\label{alpha_MRI_delta}
\end{equation}
where $\delta$ is a parameter between 0.5 and 1. The $\beta_{\mathrm{mag}}$ (beta magnetic) parameter is related to other gas parameters
\begin{equation}
\beta_{\mathrm{mag}}(r,z) \equiv \frac{P_{\mathrm{therm}}(r,z)}{P_{\mathrm{mag}}(r,z)}=8 \pi \frac{\rho c_{\mathrm{s}}^2}{B_{\mathrm{z}}(r)^2}=2\left(\frac{c_{\mathrm{s}}}{\mathrm{v_A}}\right)^2 =\frac{2}{3}\frac{E_{\mathrm{th}}}{E_{\mathrm{mag}}},
\end{equation}
where $E_{\mathrm{th}}$ and $E_{\mathrm{mag}}$ are the thermal and magnetic energy respectively, $\rho$ is the gas mass density, $\mathrm{c}_{\mathrm{s}}=\sqrt{kT/\mu_{\mathrm{n}}}$ is the sound
speed with mean molecular mass $\mu_{\mathrm{n}}=2.2$ amu (atomic mass unit), $B_z(r)$ is the vertical component of the magnetic field, and
v$_{\mathrm{A}}$  is the Alfv\`{e}n speed in the disk vertical
direction defined by
\begin{equation}
\mathrm{v_A}=\frac{B_{\mathrm{z}}}{\sqrt{4\pi\rho}}.
\end{equation}
%\begin{equation} 
%\mathrm{v_A} = \left(\frac{2}{\beta}\right)^{1/2}c_{\mathrm{s}}.
%\end{equation}
The values for $\alpha_{\mathrm{ideal}}$ are bound between 2$\times$10$^{-3}$  and 0.1 for $\beta_{\mathrm{mag}}$ between 1 and 10$^6$.  A high value of $\beta_{\mathrm{mag}}$ means that thermal motions in the plasma is important while $\beta_{\mathrm{mag}}<1$ means that the dynamic is dictated by the magnetic field. 
% ----------------  
\begin{figure}[!ht]
\centering  
 \resizebox{\hsize}{!}    {\includegraphics[angle=0,width=10cm,height=8cm,trim=55 45 80 320, clip]
{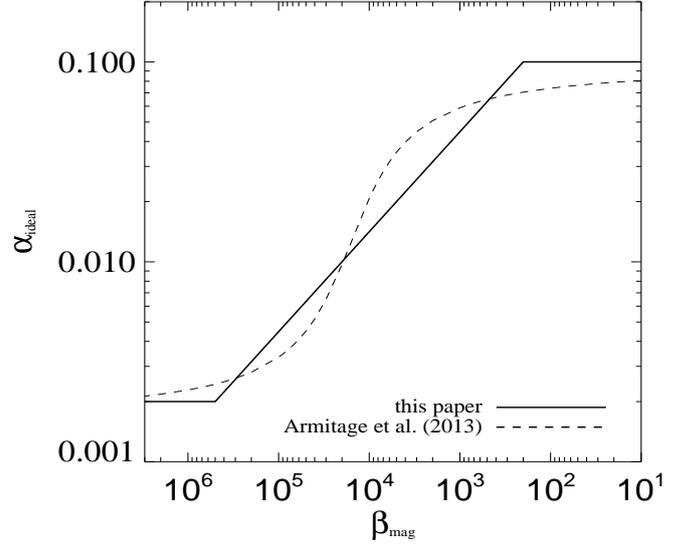}}
\caption{Variation of the value of $\alpha_{\mathrm{ideal}}$ a function of the value of $\beta_{\mathrm{mag}}$ for two prescriptions. For the Lesur model, we adopted $\delta$=0.5}  
  \label{fig_alpha_from_beta}          
\end{figure}   
% ---------    
Alternatively, we can use
the prescription  of \cite{Armitage2013ApJ...778L..14A}
\begin{equation}
 \log \alpha_{\mathrm{ideal}}  =  A + B \tan^{-1} \left[ \frac{C - \log \beta}{D} \right],
\end{equation}
where the value for the constants are: $A=-1.9$, $B=0.57$, $C=4.2$ and
$D=0.5$. Both functionals are plotted in Fig.~\ref{fig_alpha_from_beta} with $\delta=0.5$.
We assume that $\beta_{\mathrm{mid}}$ does not vary with radius in the midplane.  This is equivalent to assume
that the magnetic field is accreted with the gas \citep{Guilet2014MNRAS.441..852G}
\begin{equation}
\beta_{\mathrm{mag}}(r,z) = \beta_{\mathrm{mid}} \left(\frac{P_{\mathrm{therm}}(r,z)}{P_{\mathrm{therm}}(r,0)}\right).
\end{equation}
where the constant $\beta_{\mathrm{mid}}$ is a free parameter of the disk models. The vertical component of the magnetic field varies with radius but constant in the vertical direction 
(see Fig.~\ref{fig_Bfield}).
In both formulations, we set $\alpha_{\mathrm{ideal}}$, the ideal-MHD MRI value of $\alpha$ in the
absence of resistivities, to a very low value (10$^{-10}$), if $\beta_{\mathrm{mag}} \leq 1$, i.e. when the flow is directed
by the magnetic field. Since the gas thermal pressure decreases with height, $
\beta_{\mathrm{mag}}$ will also decrease. The ideal MRI value of
$\alpha_{\mathrm{ideal}}$ will thus increase together with the disk height until
$\beta_{\mathrm{mag}}$ reaches 1, whose location can be considered as the base of a MHD-driven wind. The different plasma resistivities in the case of 
non-ideal MHD will decrease the value of $\alpha_{\mathrm{ideal}}$ to
an effective value $\alpha_{\mathrm{eff}}$. 
In the subsequent sections, we will describe the different non-MHD resistivities before presenting the prescription for the value of $\alpha_{\mathrm{eff}}$ that is used in the chemical disk models.

% ------------------
\subsection{Non-ideal MHD resistivities}\label{non_ideal_mhd}

In the absence of strong X-ray radiation, the gas in protoplanetary
disks is mostly neutral with a maximum ionization fraction reaching a
few 10$^{-4}$ when all the carbon is ionized in the upper
atmospheres (weakly-charged plasma). Hydrogen is predominantly molecular. The ionization fraction at disk surfaces can be much high
when X-ray ionization is included, such that a significant fraction of hydrogens is in the form of protons. 

We considered a fluid that is globally neutral with velocity $\mathbf{v}$($\mathbf{r}$) and density $\rho$($\mathbf{r}$). A charged
species (electron, atom, molecule, PAH, or dust grain) $j$ has mass
$m_j$, charge $Z_j$, number density $n_j$, and drift velocity relative
to neutral $\mathbf{v}_j$. The current density $\mathbi{J}$ including all the charge-bearing species $j$
(electrons, ions, PAHs, dust grains) with charge $Z_j$ is
\begin{equation}
\mathbi{J}=e\sum_j n_j Z_j \mathbf{v}_j.
\end{equation}
For a weakly ionized gas, the Lorentz force and the drag with the
neutral gas force dominate over the inertia, gas pressure and
gravitational forces in the equation of motion for the charge species
\begin{equation}
n_jZ_je(\mathbi{E}+\frac{\mathbf{v}_j}{c}\times \mathbi{B})=n_j\gamma_j\rho m_j \mathbf{v}_j,\label{electron_motion}
\end{equation}
where 
\begin{equation}
  \gamma_j \equiv \frac{\langle \sigma\mathbf{v} \rangle_j}{m_j+m}
\end{equation}
is the drag coefficient (cm$^{3}$ g$^{-1}$ s$^{-1}$) with $\langle
\sigma\mathbf{v} \rangle_j$ being the average collision rate between
the charge species and the neutral gas of average mass $m$. The drag
forces measure the rate of momentum exchange via the collisions.
Using the gas global neutrality $\sum_j n_jZ_j=0$ we obtain
\begin{equation}
\frac{\mathbi{J}\times\mathbi{B}}{c} = \sum_j n_j\gamma_j\rho m_j \mathbf{v}_j
\end{equation}
The inversion of equation \ref{electron_motion} leads to the generalized Ohm's law, which characterizes the non-ideal MHD effects
\citep{Wardle1999MNRAS.303..239W,Norman1985A&A...147..247N}
\begin{equation}
\mathbi{J}=\sigma_\mathrm{O}\mathbi{E}'_\parallel+\sigma_{H}(\hat{\mathbi{B}}
\times\mathbi{E}'_\perp)+\sigma_{P}\mathbi{E}'_\perp,
\end{equation}
where $\mathbi{E}'_\parallel$ and $\mathbi{E}'_\perp$ denote the decomposition of $\mathbi{E}'$ into vectors parallel and perpendicular to the magnetic field $\mathbi{B}$ respectively, and the $\hat{}$ means unit
vector. $\sigma_\mathrm{O}$ is the Ohm conductivity, also referred as the conductivity parallel to the field ($\sigma_{\mathrm{O}}=\sigma_\parallel$), $\sigma_\mathrm{H}$
is the Hall conductivity, and $\sigma_\mathrm{P}$ is the Pedersen
conductivity. In cgs units, the conductivities have units of s$^{-1}$. The knowledge of the conductivities (or equivalently the
resistivities) is central to the efficiency of MRI-driven turbulence
in disks. The conductivities depend on the charge carriers and hence on the disk chemistry. Conversely, the decay of the turbulence provides energy to heat the gas and turbulence affects the line widths, hence the line cooling.

\subsubsection{Ohm resistivity}\label{ohm_resistivity}

We consider a neutral gas of number density $n_{\mathrm{<H>}}$
(cm$^{-3}$) at gas temperature $T_{\mathrm{gas}}$ (K) with a mean mass
$m_{\mathrm{n}}$ (grams), and mass density $\rho_{\mathrm{n}}$ (grams
cm$^{-3}$). We define for each charged species $j$ ($j$=e, atomic and molecular ions, PAH, gr,
where $gr$ stands for dust grains) the units-free plasma $\beta_j$ parameter (which should not be
confused with $\beta_{\mathrm{mag}}$) as
\begin{equation}
\beta_j \equiv \frac{|Z_j|eB}{m_jc}\frac{1}{\gamma_{j} \rho_{\mathrm{n}}},
\end{equation}
where $Z_j$ is the charge of species $j$ of mass $m_j$ (in grams), and $B$ is the
magnetic field strength, for which we only consider the vertical
component $B_{\mathrm{z}}$ (the units are the Gauss, denoted Gs, with 1 Gs = 1 cm$^{-1/2}$ g$^{1/2}$ s$^{-1}$) . 
$e$ is the elementary charge (in units of statC). Finally, $\gamma_{j}=\langle \sigma\mathbf{v}
\rangle_{j}/(m_{\mathrm{n}}+m_j)$ is the drag frequencies of the
  positively- or negatively-charged species $j$ of number density
  $n_j$ with the neutrals. %The elemental charge in cgs is $\simeq$4.8032$\times$10$^{-10}$ statC.
  If a species $j$ has $|\beta_j| \ll 1$, this
  means that it is well coupled with the neutral gas.   On the other hand, if $|\beta_j| \gg 1$, the charged species is well coupled with the magnetic field. Using the terms introduced above, the Ohm electrical conductivity is defined as 
\begin{equation}
\sigma_{\mathrm{O}} \equiv \frac{ec}{B}\sum_j n_j|Z_j|\beta_j = 
\sum_j \sigma_{\mathrm{O},j} = \frac{e^2}{\rho_{\mathrm{n}}}\sum_j \frac{Z_j^2 n_j}{m_j \gamma_{j}}.
\end{equation}
The collision rates $\langle \sigma\mathbf{v} \rangle_{jn}$, which appear in the drag frequencies term 
$\gamma_{j}$, are thermal velocity-averaged values of the cross-sections $\sigma$ (without subscript). The
electron-neutral collision rate is
\begin{equation}
\langle \sigma\mathbf{v} \rangle_{\mathrm{e,n}} \approx 8.28\times10^{-10}\sqrt{T}\ \mathrm{cm}^3\ \mathrm{s}^{-1}.
\end{equation}
The atomic and molecular ions have average collision rates with
neutral species following the Langevin rate 
\begin{equation}
\langle \sigma\mathbf{v} \rangle_{\mathrm{ion,n}} \approx 1.9\times10^{-9}\ \mathrm{cm}^3\ \mathrm{s}^{-1}.
\end{equation}
Specifically for the HCO$^+$-H$_2$ system, \citet{Flower2000MNRAS.313L..19F} provided the formula
\begin{equation}
\langle \sigma\mathbf{v} \rangle_{\mathrm{ion,n}} \approx 8.5\times10^{-10}\ T^{0.24} \mathrm{cm}^3\ \mathrm{s}^{-1},
\end{equation}
which shows a  weak temperature-dependency. At 100~K, the rate becomes 2.5 $\times$ 10$^{-9}$.
For PAH ions, we use the largest value between the temperature-independent
Langevin and the geometrical rate 
\begin{equation}
\langle \sigma\mathbf{v} \rangle_{\mathrm{PAH\ ion,n}} \approx max \left[1.9\times10^{-9},\pi a_{\mathrm{PAH}}^2 \left(\frac{8kT_{\mathrm{gas}}}{\pi m_{\mathrm{n}}}\right)^{1/2} \right]\ \mathrm{cm}^3\ \mathrm{s}^{-1},
\end{equation}
where $a_{\mathrm{PAH}}$ is the radius of the PAH. The average collision rate between the negatively-charged grains with
the neutral gas species is approximated by 
\begin{equation}
\langle \sigma\mathbf{v} \rangle_{\mathrm{dust,n}} \approx \pi  a^2 \left(\frac{8kT_{\mathrm{gas}}}{\pi m_{\mathrm{n}}}\right)^{1/2}\ \mathrm{cm}^3\ \mathrm{s}^{-1}.
\end{equation}
We have assumed that the grains are basically immobile w.r.t the gas. For a fully molecular gas $m_\mathrm{n}\approx$2.2 amu = 3.65317 $\times$10$^{-24}$ grams,
\begin{equation}
\langle \sigma\mathbf{v} \rangle_{\mathrm{dust,n}} \approx 3 \times 10^{-3} \left(\frac{ a^2 }{\mathrm{\mu m^2}}\right) \left(\frac{T_{\mathrm{gas}}}{100\ \mathrm{K}}\right)^{1/2}\ \mathrm{cm}^3\ \mathrm{s}^{-1}.
\end{equation}
%The Ohm conductivity due to charge-dust grains is
%\begin{equation}
%\sigma_{\mathrm{O,d}} \approx \frac{(\langle Z_{\mathrm{d}} e \rangle)^2
%\end{equation}
%The Ohm electrical conductivity due to grains is
%\begin{equation}
%\sigma_{\mathrm{dust,O}}= \frac{(Z_{\mathrm{d}}e)^2
%\end{equation}
%The Ohm resistivity is related to the conductivity by
%\begin{equation}
%\eta_\mathrm{O} =\frac{c^2}{4\pi\sigma_{\mathrm{O}}}
%\end{equation}
The Ohm diffusivity is characterized by the dimensionless Elsasser number, which is defined as the ratio between Lorentz and Coriolis force
\begin{equation}
\Lambda_{\rm Ohm} \equiv \frac{B\mathrm{_z}^2}{4 \pi \rho \eta_\mathrm{O}\Omega} \equiv \frac{\mathrm{v^2_A}}{\eta_\mathrm{O}\Omega} \equiv \left(\frac{4\pi \sigma_{\mathrm{O}}}{\Omega}\right)\left(\frac{\mathrm{v_A}}{c}\right)^2,\label{Elsasser_Ohm}
\end{equation}
where
\begin{equation}
\Omega =
\sqrt{\frac{GM_*}{r^{3}}}\simeq2\times10^{-7}\left(\frac{M_*}{M_\odot}\right)^{1/2}\left(\frac{r}{\mathrm{au}}\right)^{-3/2}\
\mathrm{rad}\ \mathrm{s}^{-1} 
\end{equation}
is the angular speed, which is in the case of a protoplanetary disk assumed to be in Keplerian rotation, where $G$ is the gravitational constant, $M_*$
is the mass of the central star ($M_\odot$ is the mass of the Sun), 
and $R$ and $r$ the distance in the disk from
the star in cm and in au respectively. One au (astronomical units) corresponds to  $1.4959787 \times 10^{13}$ cm. The Ohm resistivity is $\eta_\mathrm{O}=(c^2/4\pi)/\sigma_\mathrm{O}$. The Ohm Elsasser number can be rewritten as the sum of the contribution of each charged species to the Ohm conductivity:
\begin{equation}
\Lambda_{\rm Ohm} \equiv \frac{4\pi}{\Omega}\left(\frac{\mathrm{v_A}}{c}\right)^2 (\sigma_{\mathrm{e,O}}+\sigma_{\mathrm{ions,O}}+\sigma_{\mathrm{PAH,O}}+\sigma_{\mathrm{dust,O}}).\label{Elsasser_Ohm2}
\end{equation}

For MRI-driven turbulence to fully develop, the turbulence development timescale $\tau_{\mathrm{ideal}}$ has to be shorter than the damping by ohmic diffusion timescale $\tau_{\mathrm{damp}}$:
\begin{equation}
\tau_{\mathrm{ideal}}\sim \frac{h}{V_\mathrm{A}} < \tau_{\mathrm{damp}}\sim \frac{h^2}{\eta_\mathrm{O}}.
\end{equation}
This criterion translates to 
\begin{equation}
\frac{h \mathrm{v}_{\mathrm{A}}}{\eta_\mathrm{O}} > 1
\end{equation}
In the approximation of vertical isothermal disk ($h=c_{\mathrm{s}}/\Omega$) and in the condition of  $\mathrm{v}_{\mathrm{A}}=c_{\mathrm{s}}$ ($\beta_\mathrm{mag}$=2, $\alpha_\mathrm{ideal}=1$), the criterion becomes
\begin{equation}
\Lambda_{\rm Ohm} \equiv \frac{\mathrm{v^2_A}}{\eta_\mathrm{O}\Omega}>1. \label{Elsasser_Ohm_Criterium}
\end{equation}
This simple estimate is supported by non-ideal MHD simulations \citep{Sano2002ApJ...570..314S}.
Although the disks are not isothermal in the vertical direction, we will use this criterion in our models. The ohmic Elsasser number criterion can be seen as the rate of dissipation of the magnetic energy ($\sim B^2$) compared to the energy dissipated by ohmic resistivity.
% ----------
\subsubsection{Ambipolar diffusion}\label{ambipolar}

\begin{figure}[!ht]
\centering
 \resizebox{\hsize}{!}{\includegraphics[angle=0,width=10cm,height=7cm,trim=55 55 60 50, clip]
{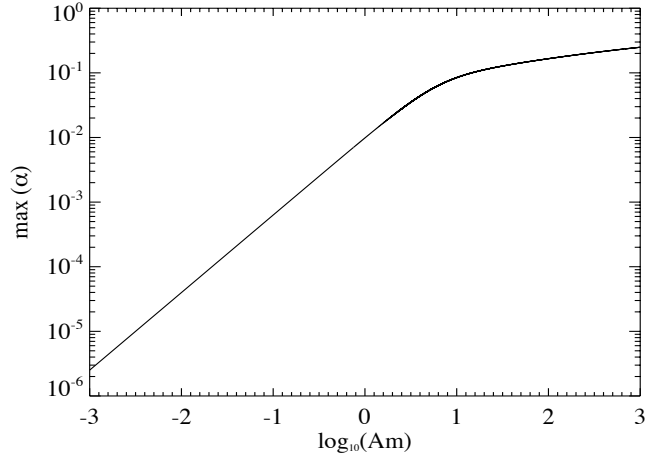}}
\caption{max($\alpha$) as function of the ambipolar diffusion term $Am$.}
  \label{fig_max_Am}          
\end{figure}

The difference between the ion and neutral (atoms and molecules, PAHs, and dust grains)
velocities is responsible for ambipolar diffusion \citep{Bittencourt2004}. The Elsasser
number for ambipolar diffusion is
\begin{equation}
Am \equiv \frac{\mathrm{v^2_A}}{\eta_\mathrm{A}\Omega},\label{eq_ambipolar1}
\end{equation}
where the ambipolar resistivity $\eta_\mathrm{A}$ is
\citep{Wardle1999MNRAS.303..239W}
%\begin{equation}
%\eta_\mathrm{A} =\frac{D^2c^2}{4\pi\sigma_{\mathrm{A}}}
%\end{equation}
\begin{equation}
\eta_\mathrm{A} =D^2\left(\frac{c^2}{4\pi}\frac{\sigma_{\mathrm{P}}}{\sigma_{\perp}^2}-\eta_{\mathrm{O}}\right),\label{eq_ambipolar2}
\end{equation}
where $\sigma_{\mathrm{P}}$ the Pedersen conductivity,
$\sigma_{\perp}$ is the total conductivity perpendicular to the field, and
$\eta_{\mathrm{O}}$ is the Ohm conductivity. The $D$ factor prevents
efficient ambipolar diffusion in strongly ionized gas,
$D=\rho_{\mathrm{neu}}/\rho_{\mathrm{gas}}$ where $\rho_{\mathrm{neu}}$
and $\rho_{\mathrm{gas}}$ are the mass density of the neutral gas and
the total gas respectively \citep{Pandey2008MNRAS.385.2269P}.
The total perpendicular conductivity is
\begin{equation}
\sigma_{\perp}  = \sqrt{\sigma_{\mathrm{H}}^2+\sigma_{\mathrm{P}}^2},
\end{equation} 
where the Pedersen conductivity is
\begin{equation}
\sigma_{\mathrm{P}}  \equiv \frac{ec}{B} \sum_j n_jZ_j\left(\frac{\beta_j}{1+\beta_j^2}\right). 
\end{equation}
%and the associated Pedersen resistivity is
%\begin{equation}
%\eta_\mathrm{P} =\frac{c^2}{4\pi\sigma_{\mathrm{P}}}.
%\end{equation}
The Hall conductivity $\sigma_{\mathrm{H}}$ is defined by
\begin{equation}
\sigma_{\mathrm{H}} \equiv \frac{ec}{B}\sum_j \frac{n_jZ_j}{1+\beta_j^2}.
\end{equation}
When all the charged species are well coupled to the neutral gas ($|\beta_j| \ll 1$ for all $j$), $\sigma_{\mathrm{O}} \approx \sigma_{\mathrm{P}} \gg | \sigma_{\mathrm{H}}|$ the conductivity is scalar (isotropic) and the regime is resistive. When all the charged species are well coupled to the field, then $\sigma_{\mathrm{O}} \gg \sigma_{\mathrm{P}} \gg | \sigma_{\mathrm{H}}|$ ambipolar diffusion dominates \citep{Wardle1999MNRAS.303..239W}. The ambipolar resistivity becomes
\begin{equation}
\eta_\mathrm{A} =D^2\left(\eta_\mathrm{P}-\eta_{\mathrm{O}}\right)\approx D^2 \eta_\mathrm{P}.\label{eq_ambipolar3}
\end{equation}
%\begin{equation}
%\eta_{\mathrm{A}}\equiv \frac{D^2B^2}{4\pi \gamma_i\rho_i \rho}\label{ambipolar2},
%\end{equation}
%where $j=e, i, PAH, g$. 
%Equivalent definitions of the ambipolar diffusion number are
%\citep{Sano2002ApJ...570..314S,Turner2010ApJ...708..188T}
%\begin{equation}
%Am \equiv D^{-2}\frac{\gamma_i \rho_i}{ \Omega}
%\end{equation}
%or
%\begin{equation}
%Am \equiv D^{-2}\frac{n_i\langle \sigma\mathbf{v} \rangle_i}{\Omega}.
%\end{equation}
%The momentum-transfer cross section of a singly charged ion due to
%electrical polarization force is given by the Langevin formula
%\citep{Sano2000ApJ...543..486S}.
%\begin{equation}
%\sigma_{i-n}=2.41\left(\frac{\alpha_p e^2}{\mu \mathrm{v}^2}\right)^{1/2},
%\end{equation}
%where $\alpha_p$ \, $e$ the electron charge,
%$\mu=m_{\mathrm{i}}\mu_{\mathrm{n}}/(m_{\mathrm{i}}+\mu_{\mathrm{n}})$
%is the reduced mass, and v the relative speed. Assuming mostly
%collisions with H$_2$, $\langle \sigma\mathbf{v}
%\rangle_{i-H_2}=2.0\times 10^{-9}\sqrt{m_{\mathrm{H}}/\mu}$ cm$^{3}$
%s$^{-1}$, which does not depend on the temperature as it is the case for Langevin-type rates. 
In addition to the neutral gas, we included ambipolar conductivity between neutral PAHs,
neutral grains and the ions.
%The Elsasser number for ambipolar diffusion does not depend on the
%magnetic field but on the number density of ions and the disk location
%only \citep{Perez-Becker2011ApJ...727....2P}, but without correction for highly ionized gas.
%\begin{equation}
%Am \simeq \left(\frac{\eta_{\mathrm{ion}}}{10^{-8}}\right)\left(\frac{n_{\mathrm{<H>}}}{10^{10}\mathrm{cm}^{-3}}\right)\left(\frac{r}{\mathrm{au}}\right)^{3/2},
%\end{equation}
%where $\eta_{\mathrm{ion}} \simeq \chi(\mathrm{e})$ if charges on grains are negligible.
%The ambipolar diffusion number $Am$ measures the degree of coupling between the plasma and the neutral gas. 
Based on intensive numerical simulations,
\citet{Bai2011ApJ...736..144B} parametrized the maximum value of
$\alpha$ for a given $Am$
\begin{equation}
max (\alpha) = \frac{1}{2}\left[\left(\frac{50}{Am^{1.2}}\right)^2+\left(\frac{8}{Am^{0.3}}+1\right)^2\right]^{-1/2}.\label{max_alpha_ambipolar}
\end{equation}
The maximum possible value for $\alpha$ is shown in
Fig.~\ref{fig_max_Am}. For $Am=1$, $max (\alpha)\simeq 10^{-2}$. It should be noted that the formula provides an upper limit on the value of $\alpha$ due to ambipolar diffusion.
% ----------------
% ---------  
%The induction number is
%\begin{equation}
%I \equiv \frac{\mathrm{v^2_A}}{\Omega}  
%\end{equation}
The ambipolar diffusion prevents the MRI to fully
developed at disk surfaces where the neutral-charges coupling is
inefficient because of the low density. In the midplane, the gas
density is high but the ionization fraction is low at least in the
inner disk region. Ambipolar diffusion is most likely the most
efficient in the intermediate-height disk layers.

\section{Non-ideal MHD functional form for $\alpha$ and analytical results}~\label{analytical_results} 
% -------------------
\begin{figure*}%[!ht]
\centering  
\resizebox{\hsize}{!}{
\includegraphics[angle=0,width=10cm,height=7cm,trim=55 55 60 50, clip]{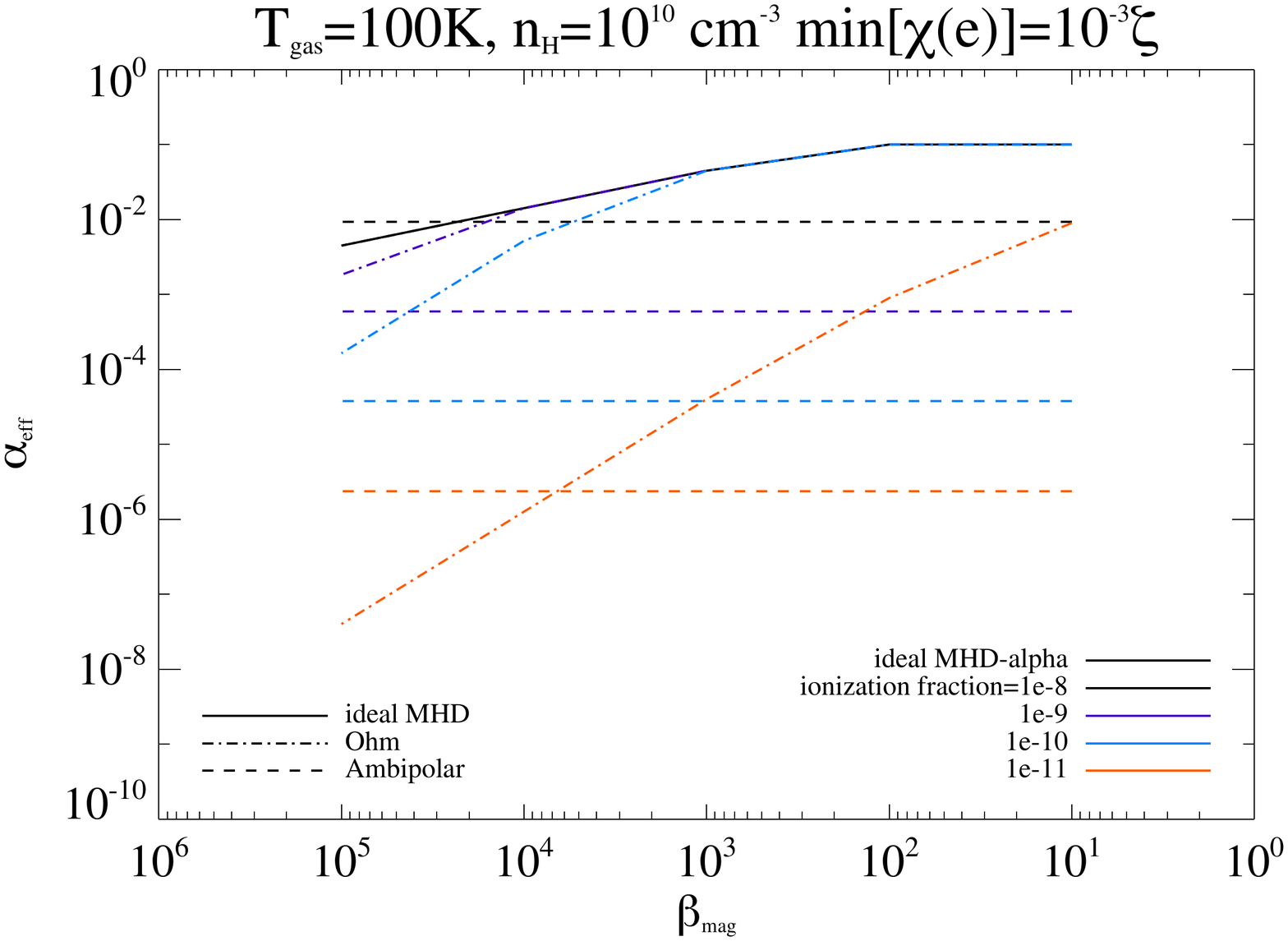}
\includegraphics[angle=0,width=10cm,height=7cm,trim=55 55 60 50, clip]{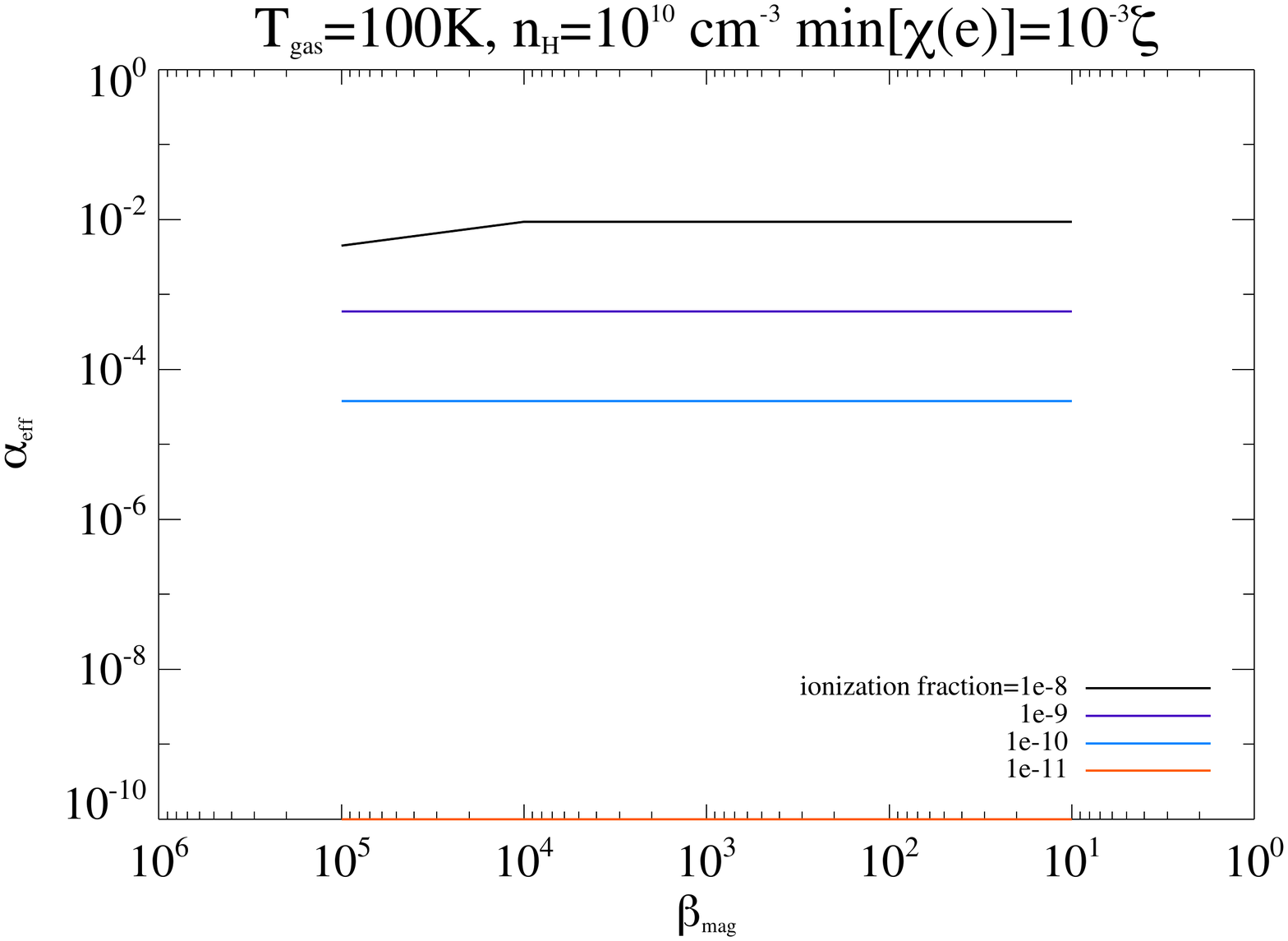}}
\resizebox{\hsize}{!}{
\includegraphics[angle=0,width=10cm,height=7cm,trim=55 55 60 50, clip]{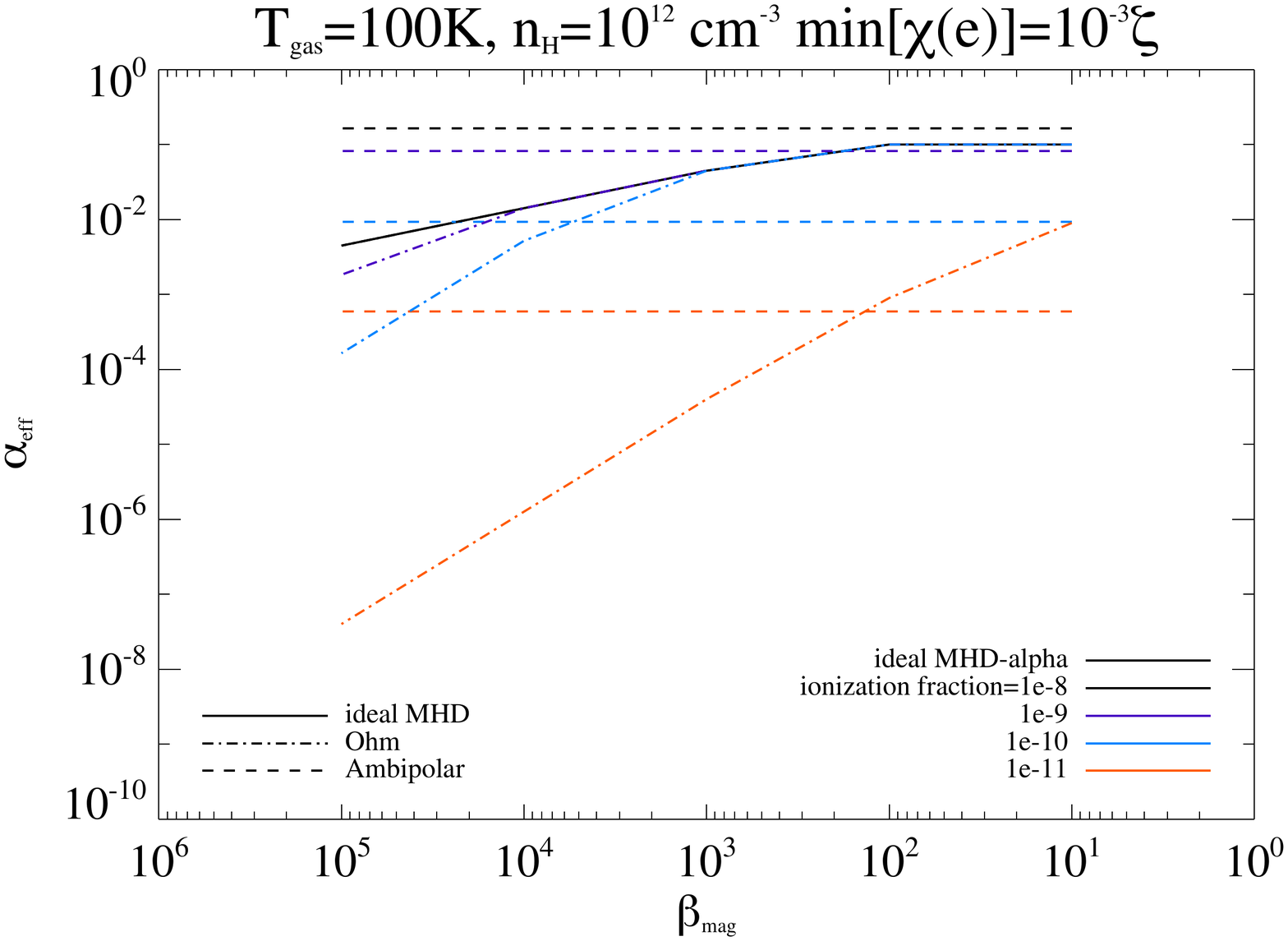}
\includegraphics[angle=0,width=10cm,height=7cm,trim=55 55 60 50, clip]{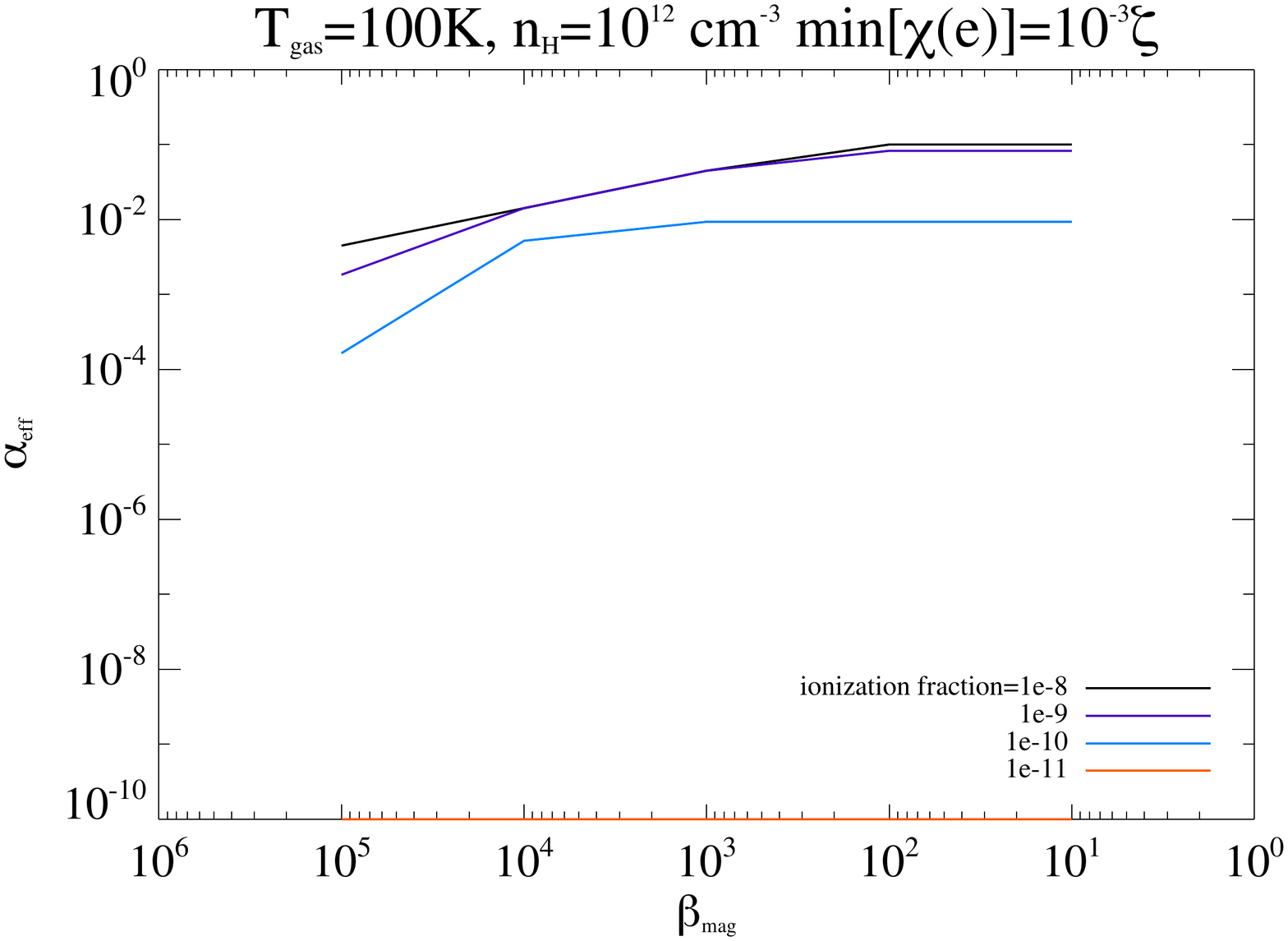}}
\caption{Effective turbulence coefficient $\alpha_{\mathrm{eff}}$ as function of  $\beta_{\mathrm{mag}}$  for gas of density of $n_{\mathrm{<H>}}$=10$^{10}$ cm$^{-3}$ and 10$^{12}$ cm$^{-3}$ at $T$=100~K in a disk midplane. The minimum free electron abundance is 10$^{-3}$ with respect to the total number of negative charges (i.e. all the other charges are on dust grains). The left panels show the ideal-MHD, Ohmic, and ambipolar diffusion contributions to  $\alpha_{\mathrm{eff}}$, while the two right panels show the resulting $\alpha_{\mathrm{eff}}$ including the all mode damping criterion of \cite{Jin1996ApJ...457..798J}. Lines at 10$^{-10}$ on the right panels mean complete no MRI-driven turbulence for total ionisation fraction of 10$^{-11}$  for all values of $\beta_{\mathrm{mag}}$.}
  \label{fig_ohm_Am1}          
\end{figure*}  
% ------------------------------
In this section, we use the model discussed above to propose an empirical non-ideal MHD MRI-driven formulation for $\alpha_{\rm eff}$ that can be used on physico-chemical \pd\ codes like \ProDiMo. \citet{Sano2002ApJ...570..314S} proposed a limitation to the
efficiency of the MRI-driven turbulence when the Elsasser Ohmic number $\Lambda_{\rm Ohm}$
is larger than one
\begin{equation}
\alpha_{\mathrm{Ohm}} = \alpha_{\mathrm{ideal}} \times min\left(1,\frac{\mathrm{v^2_A}}{\eta_{\mathrm{O}}\Omega}\right)=\alpha_{\mathrm{ideal}} \times min\left(1, \sigma_{\mathrm{O}} \mathrm{v^2_A} \Omega \right).\label{alpha_elsasser_ohm}
\end{equation}
The Elsasser Ohmic criterion tests the
coupling between the magnetic field and the charged particles in the
disk. For a constant magnetic field strength in the disk, the coupling
is more efficient at the disk surfaces, where the gas density is low and the
abundances of charged particles are high. \citet{Wardle2012MNRAS.422.2737W} argued that the Hall diffusion can
overcome the resistive damping of the MRI in certain circumstances. Numerical simulations of the
Hall diffusivity give a more complex picture \citep{Lesur2014A&A...566A..56L}, however. In light of the
uncertainties, we decided to not include the effects of the Hall
diffusivity in our estimate of the effective viscosity parameter $\alpha_{\rm eff}$. The adopted functional form for $\alpha_{\mathrm{eff}} $ is
\begin{equation}
\alpha_{\mathrm{eff}} = \alpha_{\mathrm{ideal}}(\beta_{\mathrm{mag}})\times min\left(1,\frac{\mathrm{v^2_A}}{\eta_{\mathrm{O}}\Omega}\right) \times \left[\left(\frac{50}{Am^{1.2}}\right)^2+\left(\frac{8}{Am^{0.3}}+1\right)^2\right]^{-1/2},
\end{equation}
when $\sqrt{\beta_\mathrm{mag}}\Lambda_{\mathrm{Ohm}}>1$ and $\alpha_{\mathrm{eff}}\simeq 0$, otherwise.
It has been shown that, when $\sqrt{\beta_\mathrm{mag}}\Lambda_{\mathrm{Ohm}}<1$, 
all the perturbation modes  are stabilized and thus no MRI-driven turbulence can exist \citep{Jin1996ApJ...457..798J}. Our formula attempts to encompass effects from numerous detailed MHD simulations. As such the effective $\alpha_{\mathrm{eff}}$ should only be considered as a practical simplification to be used in non hydrodynamic codes like {\sc ProDiMo}. With more MHD simulations in the future, the approximation will certainly evolve.

The free parameters of the MRI-model are the value of
$\beta_{\mathrm{mid}}$ and the general disk parameters (see Table~\ref{tab:refmodel}). All other
parameters and variables are derived either from $\beta_{\mathrm{mid}}$ or from the chemistry, gas and dust thermal balance and radiative transfer.
While both Ohm and ambipolar diffusion contributions to the effective turbulence depend on the available charges, the
two effects differ on the other dependencies. The Ohmic Elsasser number is sensitive to the magnetic field strength and the gas temperature, while the ambipolar Elsasser number depends on the absolute gas density.  Consequently, in low-mass disks and at disk upper surfaces, MRI efficiency will be limited predominately by ambipolar diffusion. In the disk midplane, the ionization is weak because even the cosmic ray flux can be attenuated \citep{Rab2017A&A...603A..96R}. Ionization is proportional to the gas density ($\propto \zeta n_{\mathrm{<H>}}$), whereas the electron recombination depends on the square of the density ($\propto k_{\mathrm{rec}} n_{\mathrm{ion}}n_{\mathrm{e}}$). This translates to an inverse density dependency for the number of ions in the midplane. Therefore, $Am$ can also be small in the disk midplane where ionization sources are scarce. Both Elsasser numbers increase with radius with power 1.5. But if the gas density decreases as $r^{-2.5}$ than the ambipolar diffusion Elsasser number will decrease. 

To illustrate those dependencies, we plotted in Fig.~\ref{fig_ohm_Am1} the effective turbulence as the function of the parameter $\beta_{\mathrm{mag}}$ for a gas and dust temperature of 100~K and for two gas densities ($n_{\mathrm{<H>}}$=10$^{10}$ cm$^{-3}$ and 10$^{12}$ cm$^{-3}$). The panels show the effects of the variation of the Ohmic and ambipolar diffusion resistivity on the value of $\alpha_{\mathrm{eff}}$ as well as the value of ideal MHD-MRI $\alpha_{\mathrm{ideal}}$. At low and medium densities and for strong magnetic field strength (i.e. with low values of $\beta_{\mathrm{mag}}$), the effective turbulence is dominated by the ambipolar diffusion term. The Ohmic restriction on $\alpha_{\mathrm{eff}}$ occurs for a large value of $\beta_{\mathrm{mag}}$. The value of $\alpha_{\mathrm{eff}}$ is larger than 10$^{-3}$ down to a total ionization fraction of $\sim$ 10$^{-8}$-10$^{-10}$ depending on the gas density.
At disk surfaces, where the ionization fraction is high and the value of $
\beta_{\mathrm{mag}}$ small, the efficiency of MRI turbulence is probably limited by the ambipolar  diffusion resistivity to a maximum value of $\sim$0.25.

% ----------------------------------------------
\begin{figure*}[!htbp]  
  \centering 
  \includegraphics[angle=0,width=9cm,height=9cm,trim=50 80  80 300, clip]{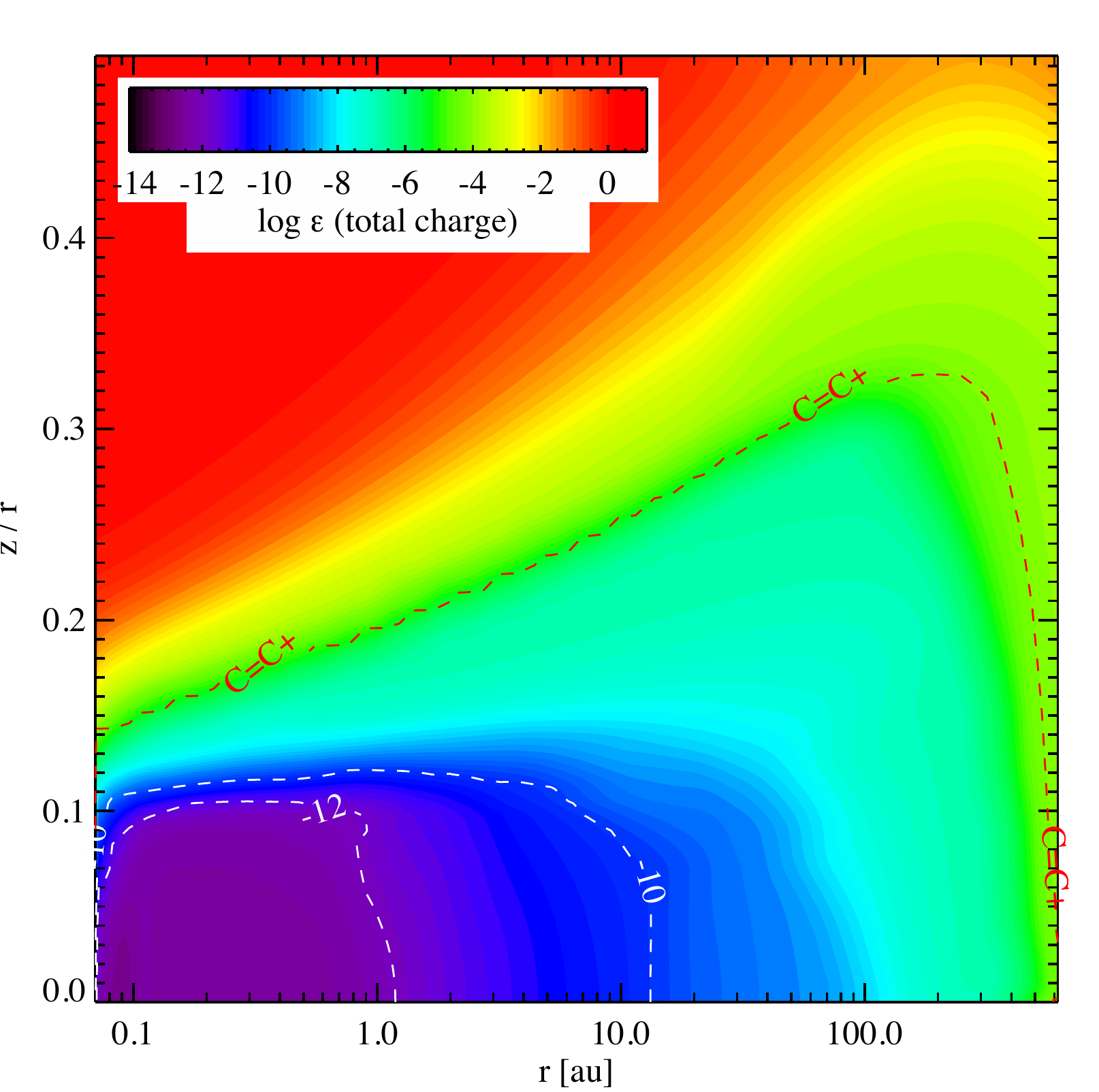}  
  \includegraphics[angle=0,width=9cm,height=9cm,trim=50 80  80 300, clip]{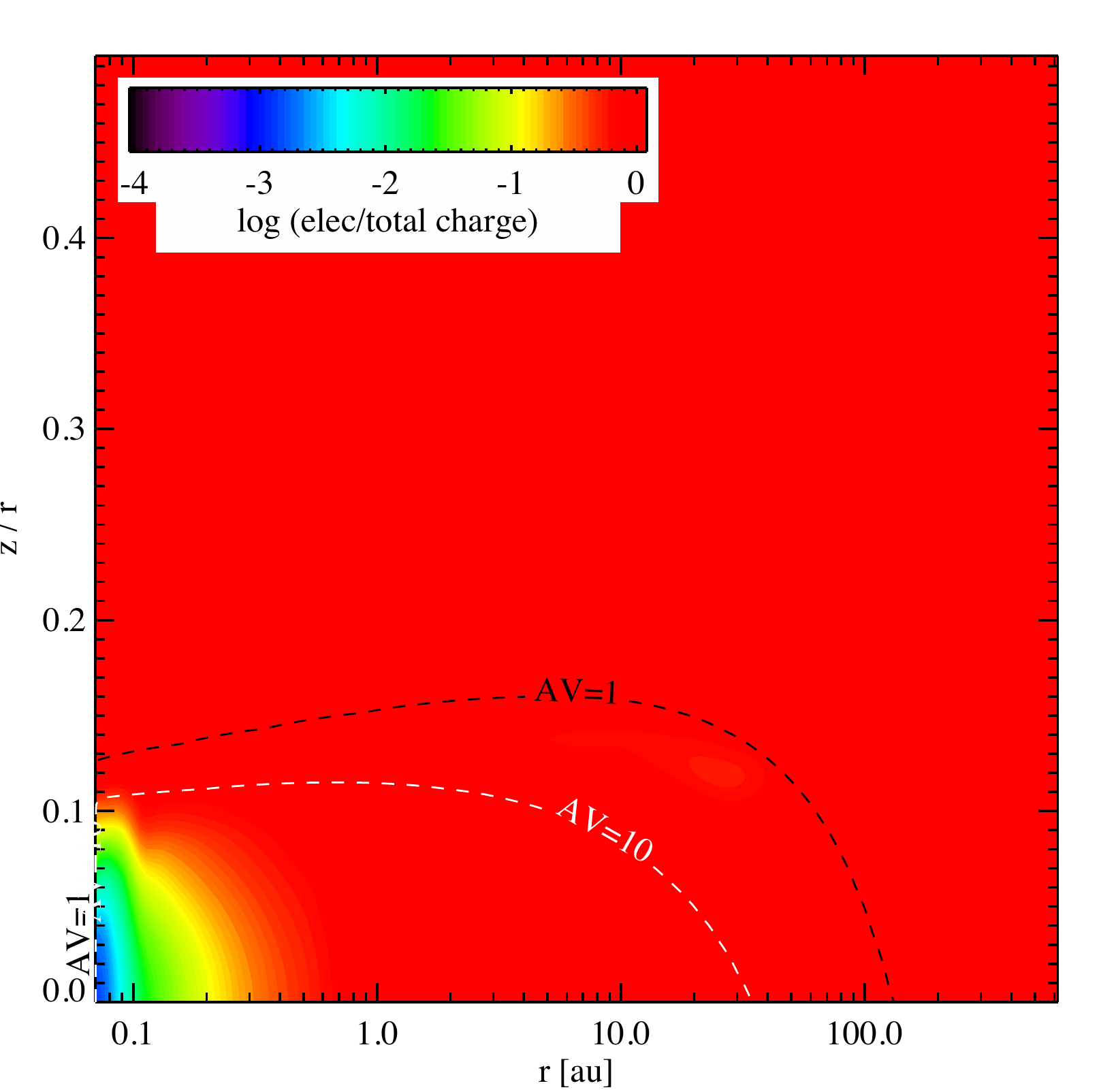}
    \includegraphics[angle=0,width=9cm,height=9cm,trim=50 80  80 300, clip]{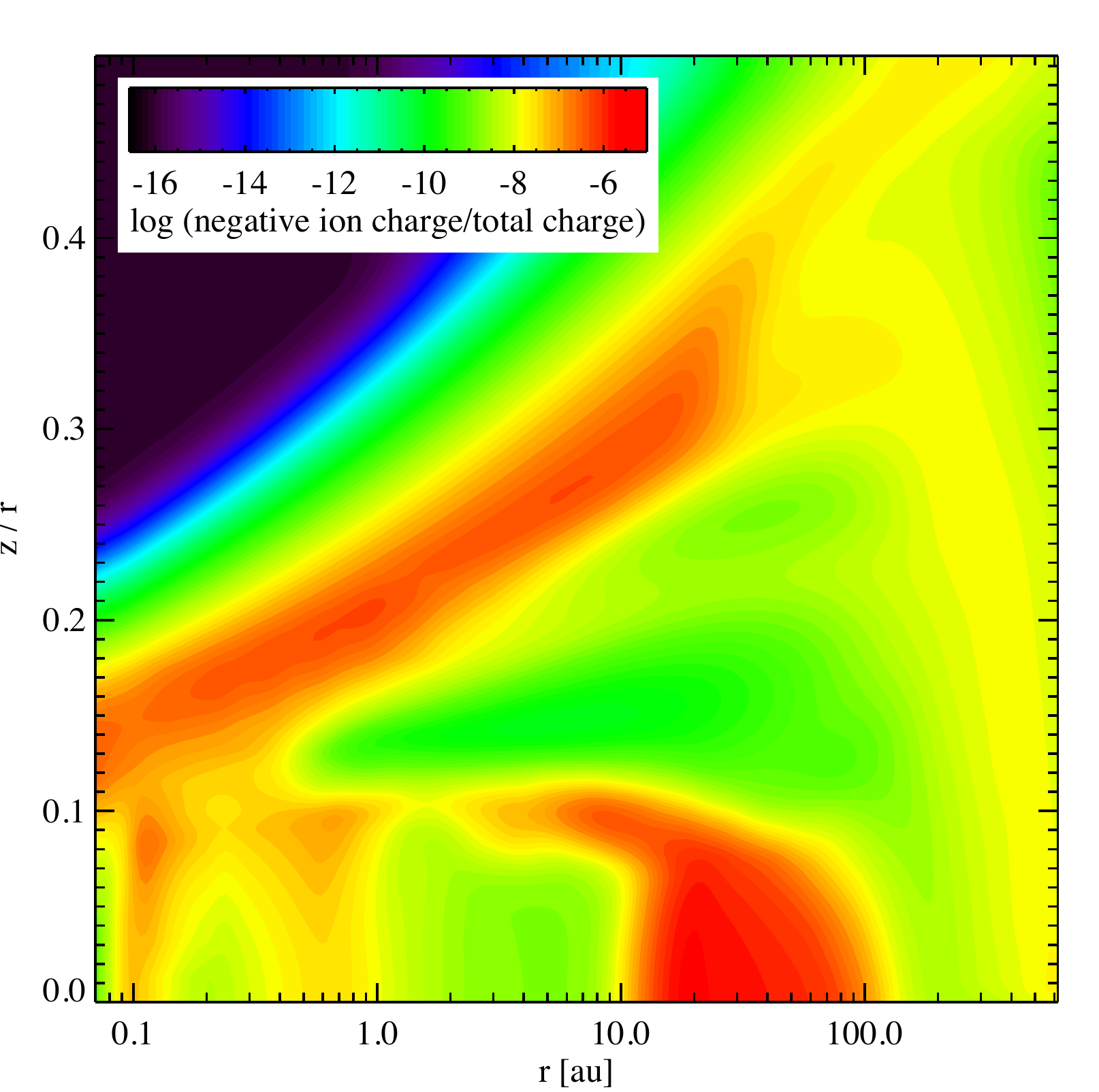}
  \includegraphics[angle=0,width=9cm,height=9cm,trim=50 80  80 300, clip]{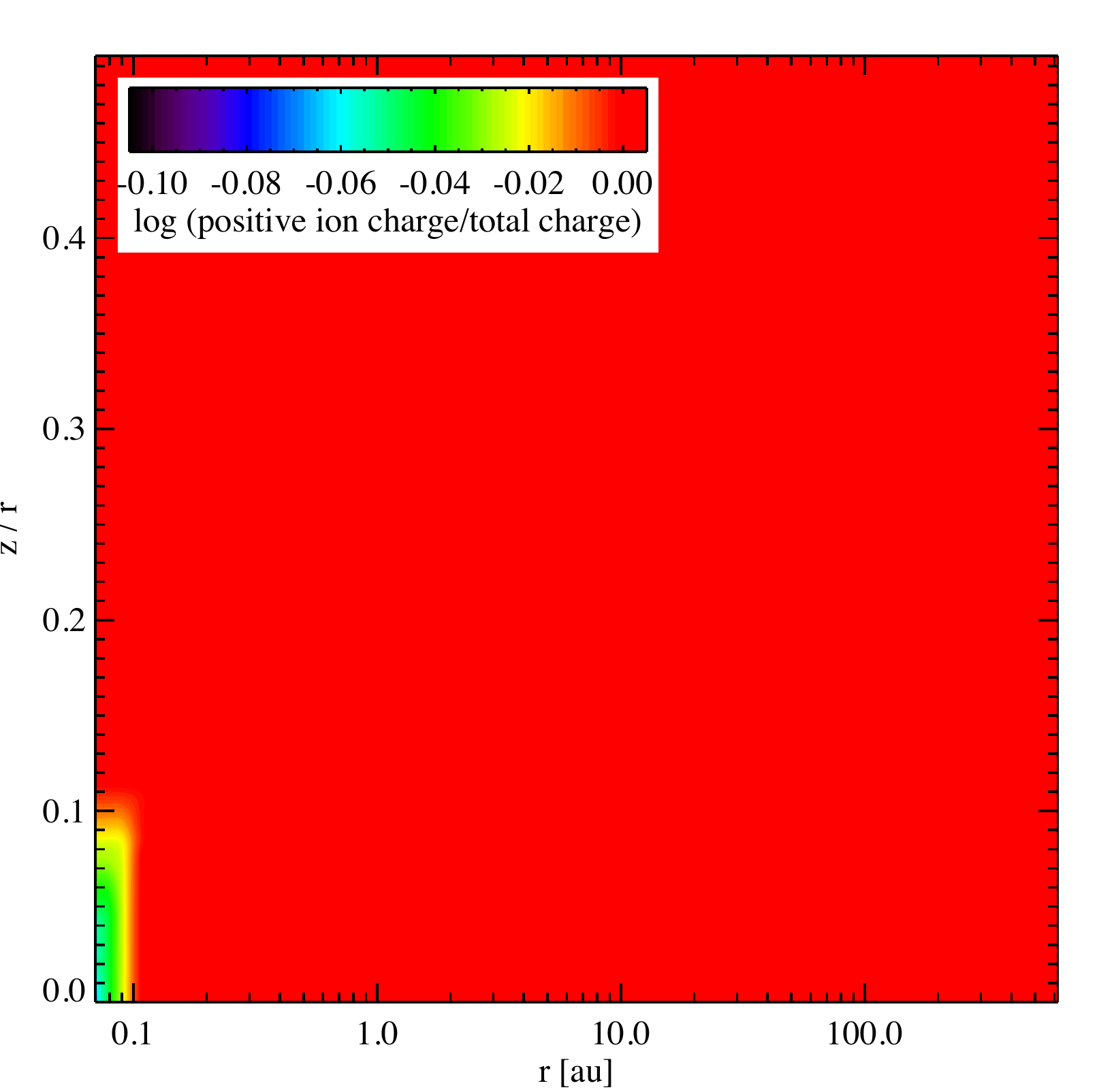}
  \caption{The upper-left panel shows the total charge (positive or negative) abundance. The red-dashed line corresponds to the location in the disk where the abundance of C$^+$ and C are equal. The upper-right panel shows the contribution of free electrons to the total charge. The lower-left panel corresponds to the contribution of the negative ions (H$^-$) and the lower right that of the positively-charged ions. The disk model is  the DIANA typical disk with $\beta_{\rm mid}$=10$^{4}$.}
  \label{fig_disk_results_charges}          
\end{figure*}  
 
% ----------------------------------------------
\section{Disk model}\label{disk_model}

\begin{figure*}[!htbp]  
  \centering 
  \includegraphics[angle=0,width=9cm,height=9cm,trim=50 80  80 300, clip]{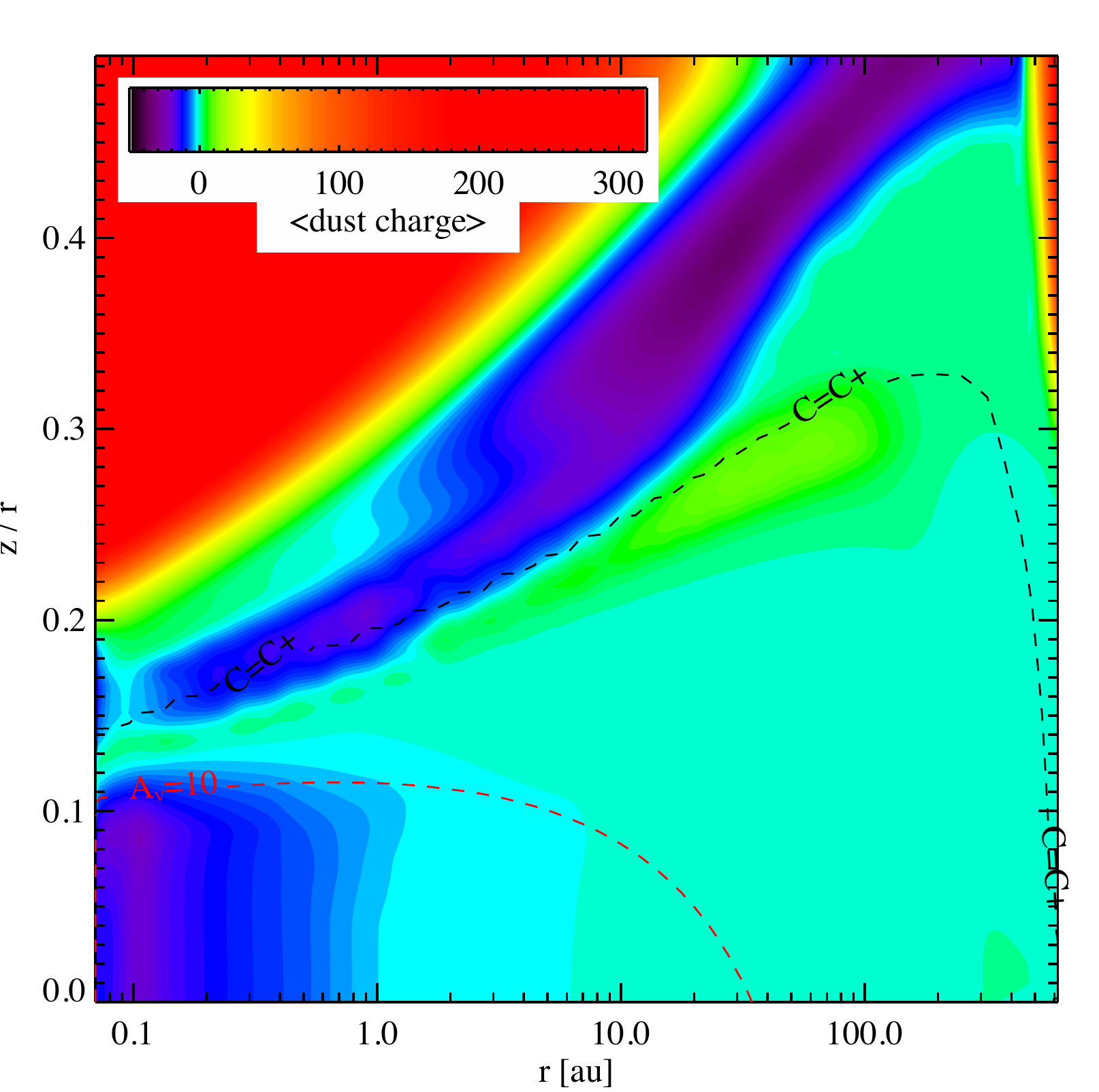}  
  \includegraphics[angle=0,width=9cm,height=9cm,trim=50 80  80 300, clip]{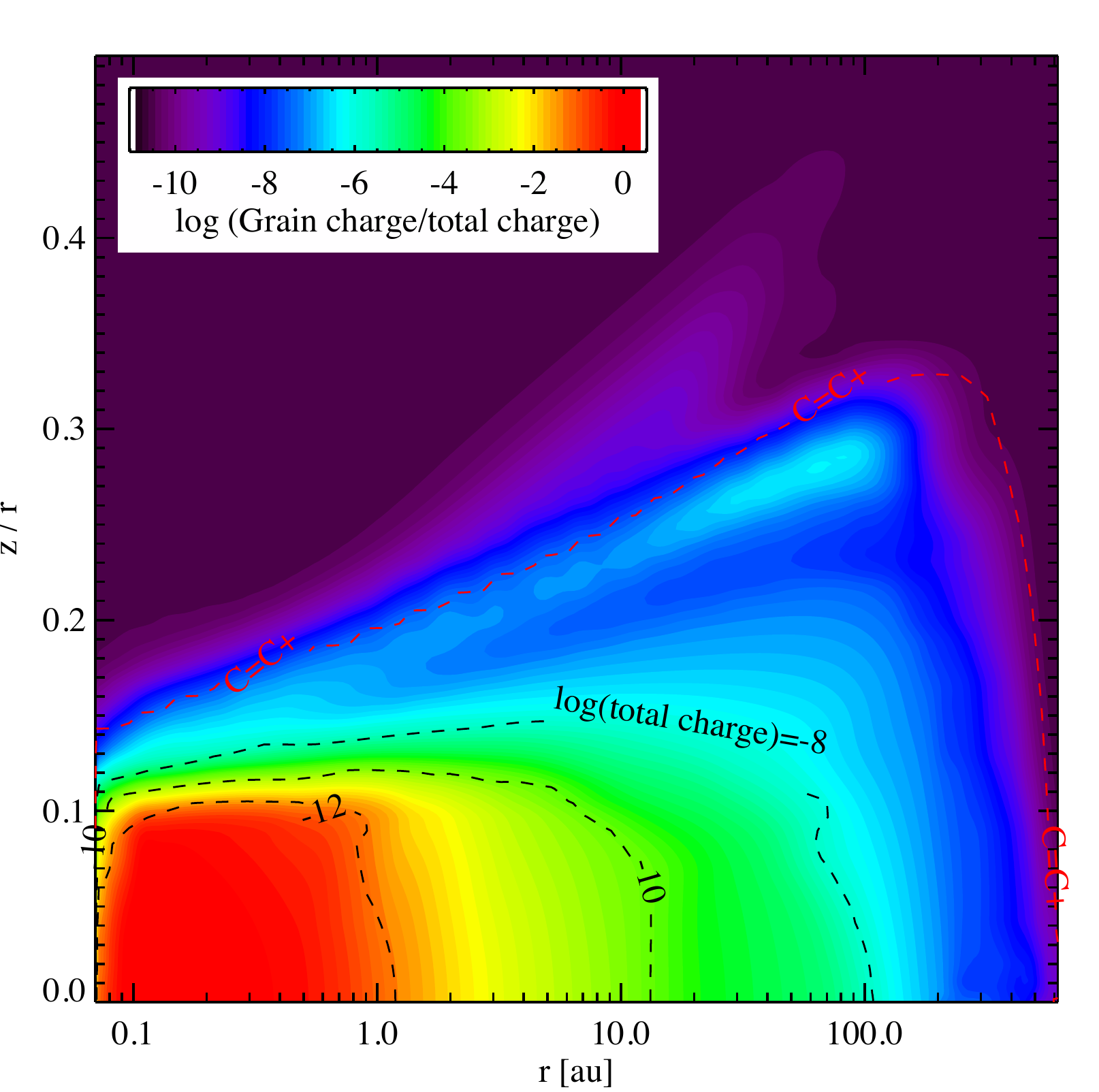} 
    \includegraphics[angle=0,width=9cm,height=9cm,trim=50 80  80 300, clip]{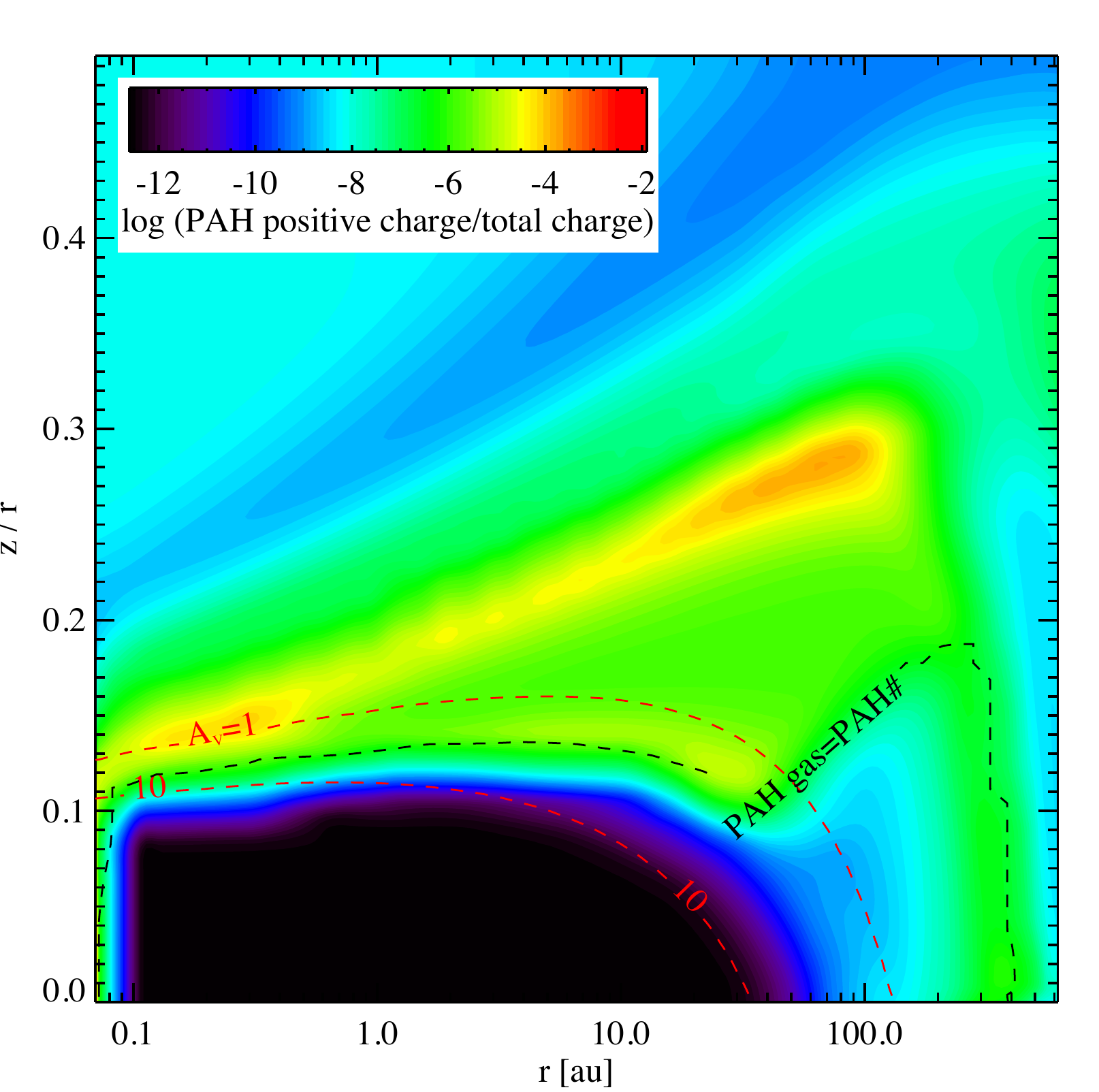}  
   \includegraphics[angle=0,width=9cm,height=9cm,trim=50 80  80 300, clip]{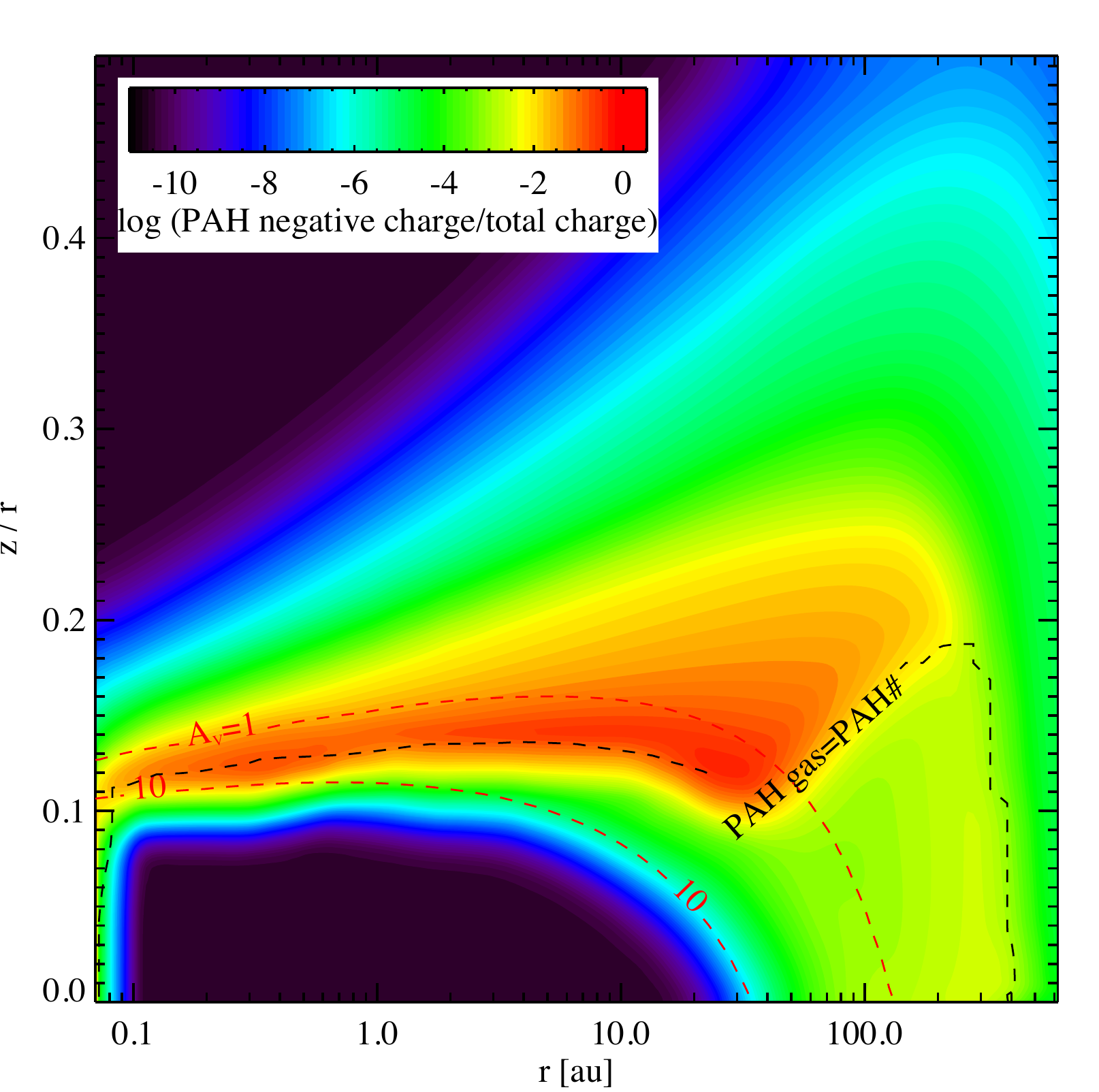}
  \caption{The average charge per grain and the fraction is shown in the upper-left panel while its fractional contribution to
  the total charge is shown in the upper-right panel. The location of the transition from ionized to neutral carbon is overplotted in the upper panels. The contribution of the positively-charged PAHs (PAH$^+$, PAH$^{++}$, and PAH$^{+++}$) is displayed in the lower-right panel while the contribution of the negatively-charged PAHs (PAH$^-$) is displayed in the lower-right panel. The contours show the extinction and the location in the disk where the abundance of gas phase and frozen PAHs are equal. The model is the DIANA typical disk with $\beta_{\rm mid}$=10$^{4}$. Silicate dust grains remains at $T_{\mathrm{dust}}<1500$ K at all disk heights (see Fig.~\ref{fig_disk_temperatures}).}
\label{fig_disk_results_dust_charges}             
\end{figure*}  
% --------------------------------------------
\begin{figure*}[!htbp]  
  \centering 
  \includegraphics[angle=0,width=8cm,height=8cm,trim=50 80  80 300, clip]{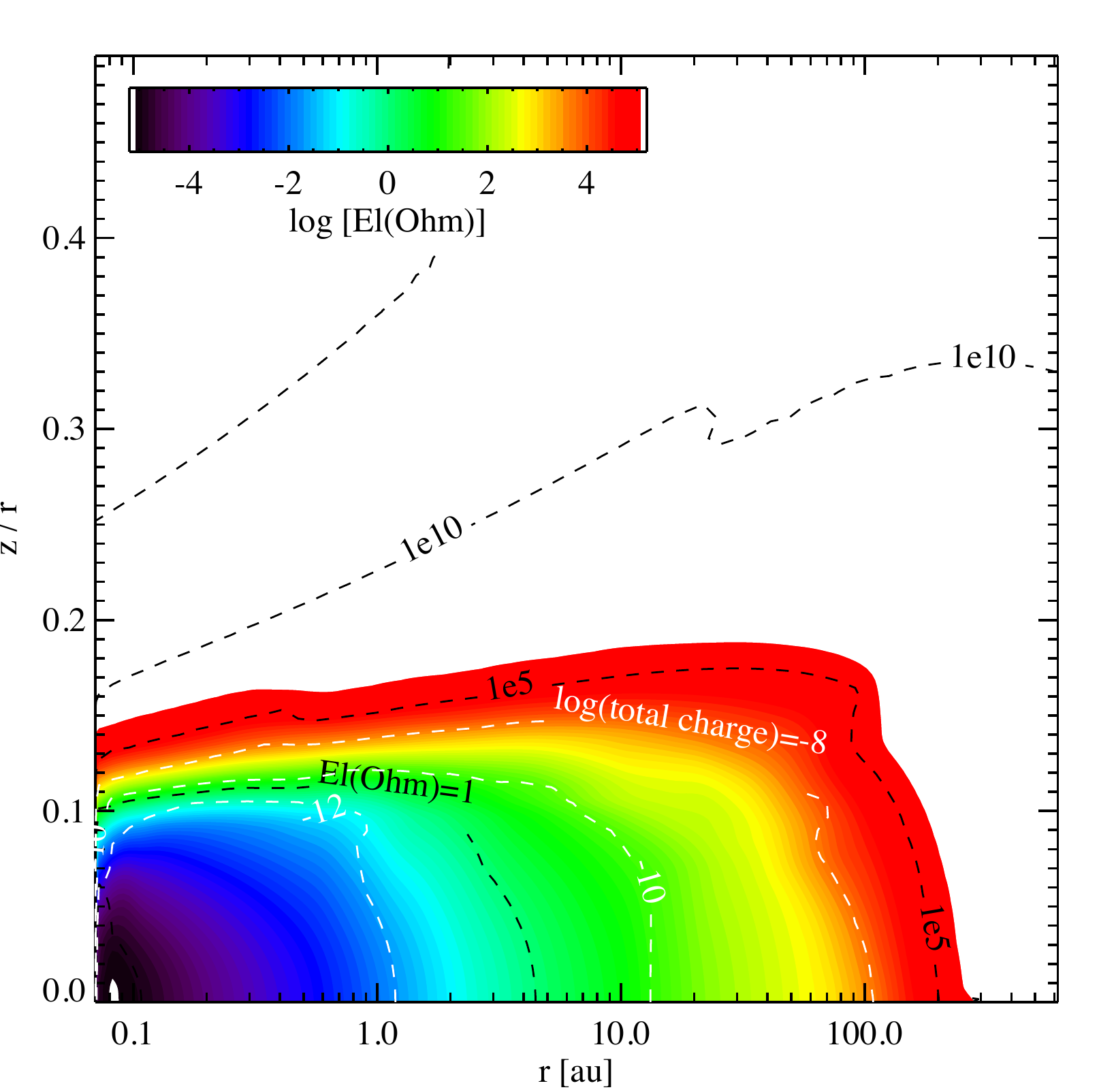}  
  \includegraphics[angle=0,width=8cm,height=8cm,trim=50 80  80 300, clip]{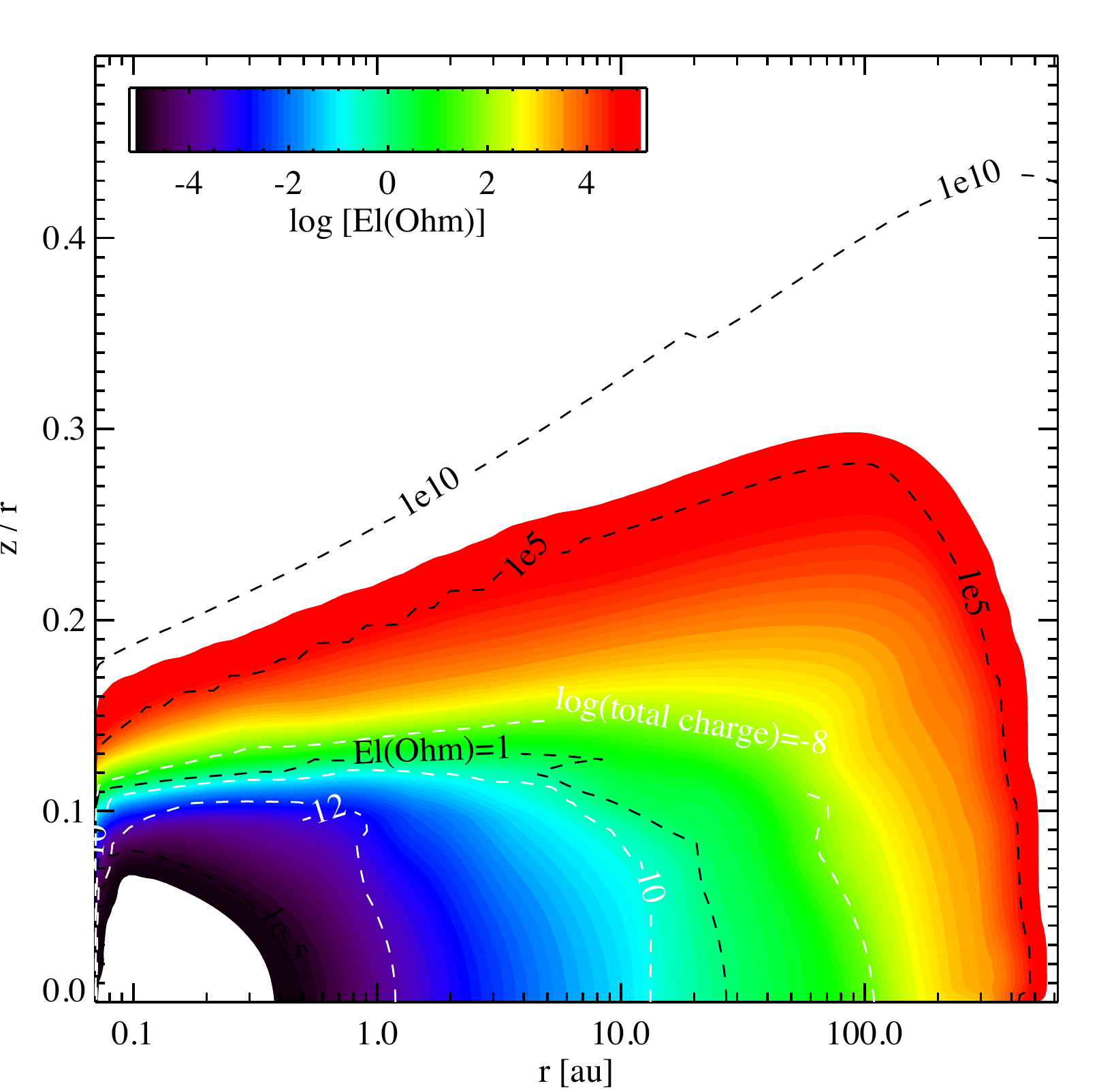}
  \includegraphics[angle=0,width=8cm,height=8cm,trim=50 80  80 300, clip]{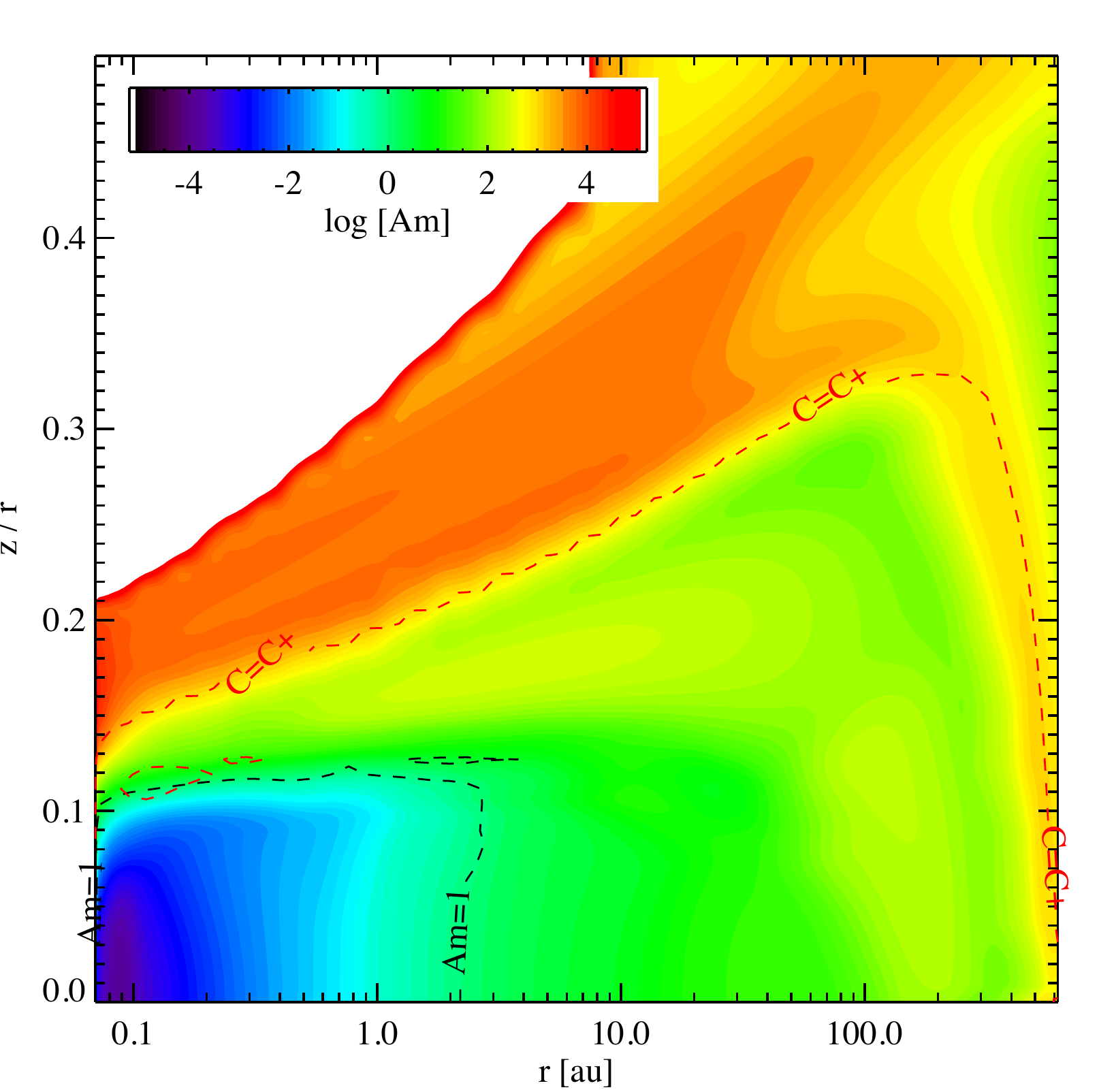}
  \includegraphics[angle=0,width=8cm,height=8cm,trim=50 80  80 300, clip]{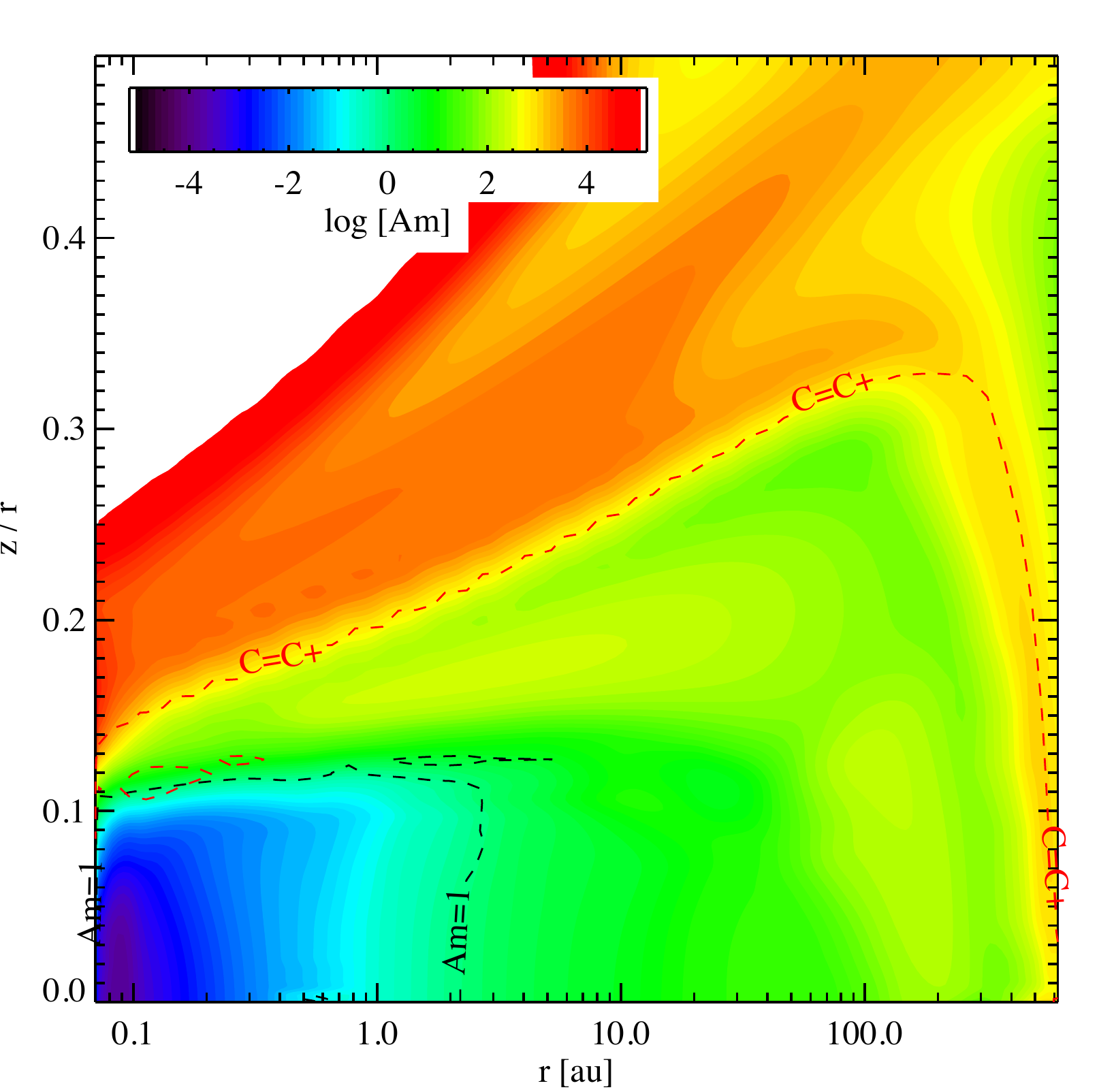}  
  \includegraphics[angle=0,width=8cm,height=8cm,trim=50 80  80 300, clip]{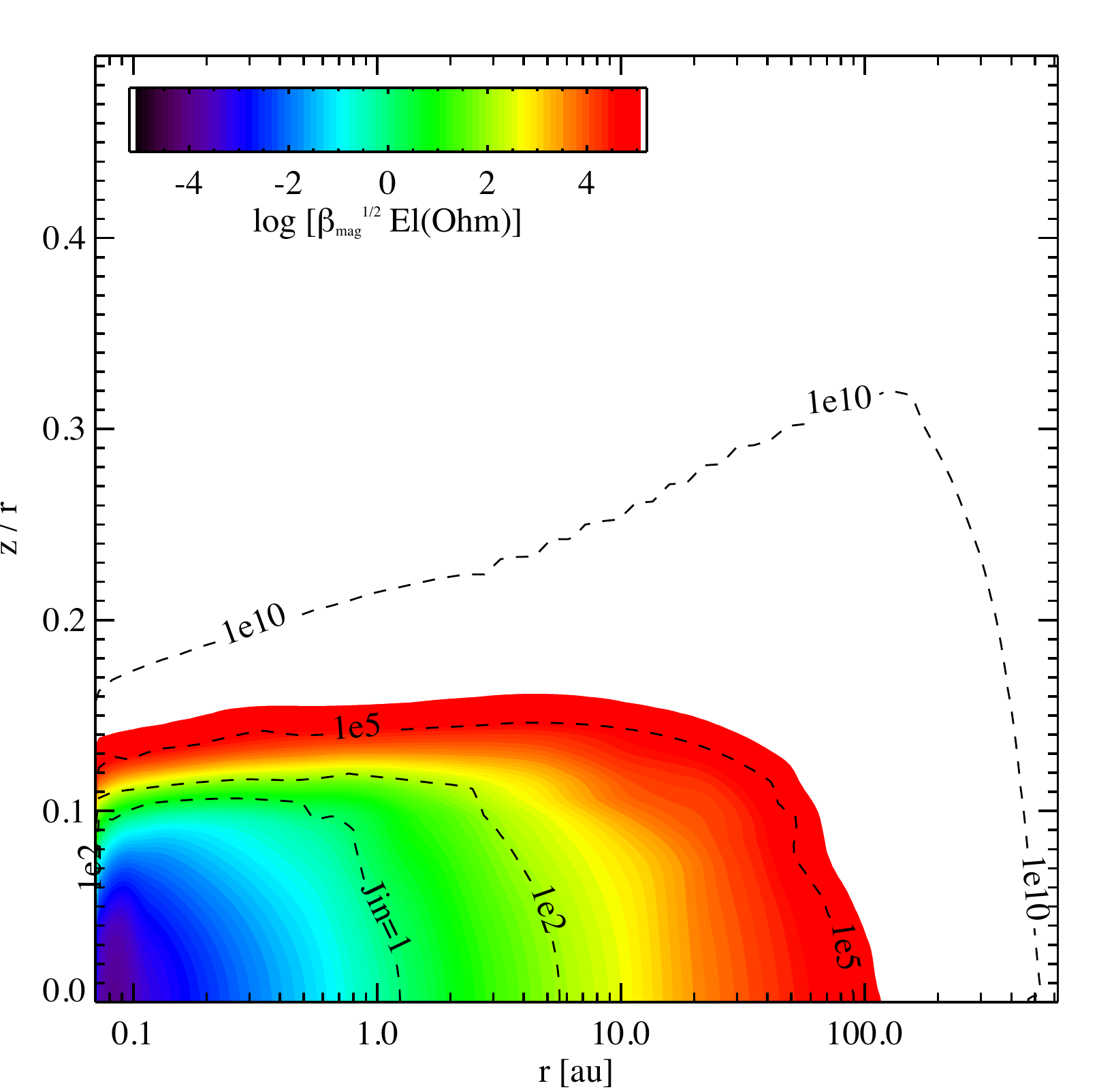}
  \includegraphics[angle=0,width=8cm,height=8cm,trim=50 80  80 300, clip]{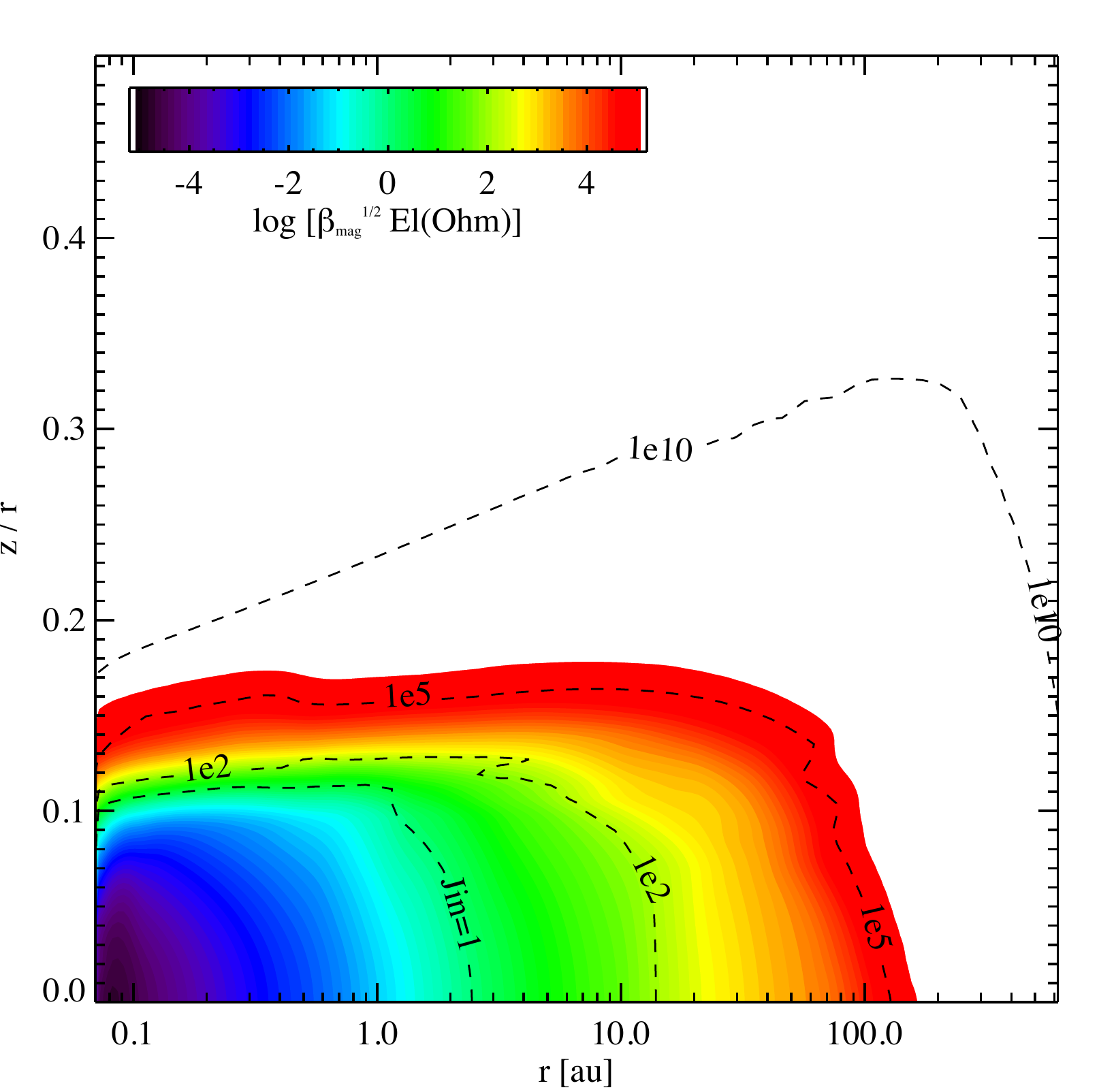}  
  \caption{The Elsasser Ohm number is shown in the upper panels for  the DIANA typical disk and. The white contours corresponds the location of the total charge in the disk. The middle panels show the ambipolar diffusion number in the disk models. The location where the C and C$^+$ abundances are equal are overplotted in red. The criterion for all modes to be damped is shown in the lower panels. The lefts panels are models with $\beta_{\rm mid}$=10$^{4}$ and the right panels are models with 10$^{6}$.}
  \label{fig_disk_results_Ohm}          
\end{figure*}  
% ----------------
\begin{figure*}[!htbp]  
  \centering 
  \includegraphics[angle=0,width=9cm,height=9cm,trim=50 80  80 300, clip]{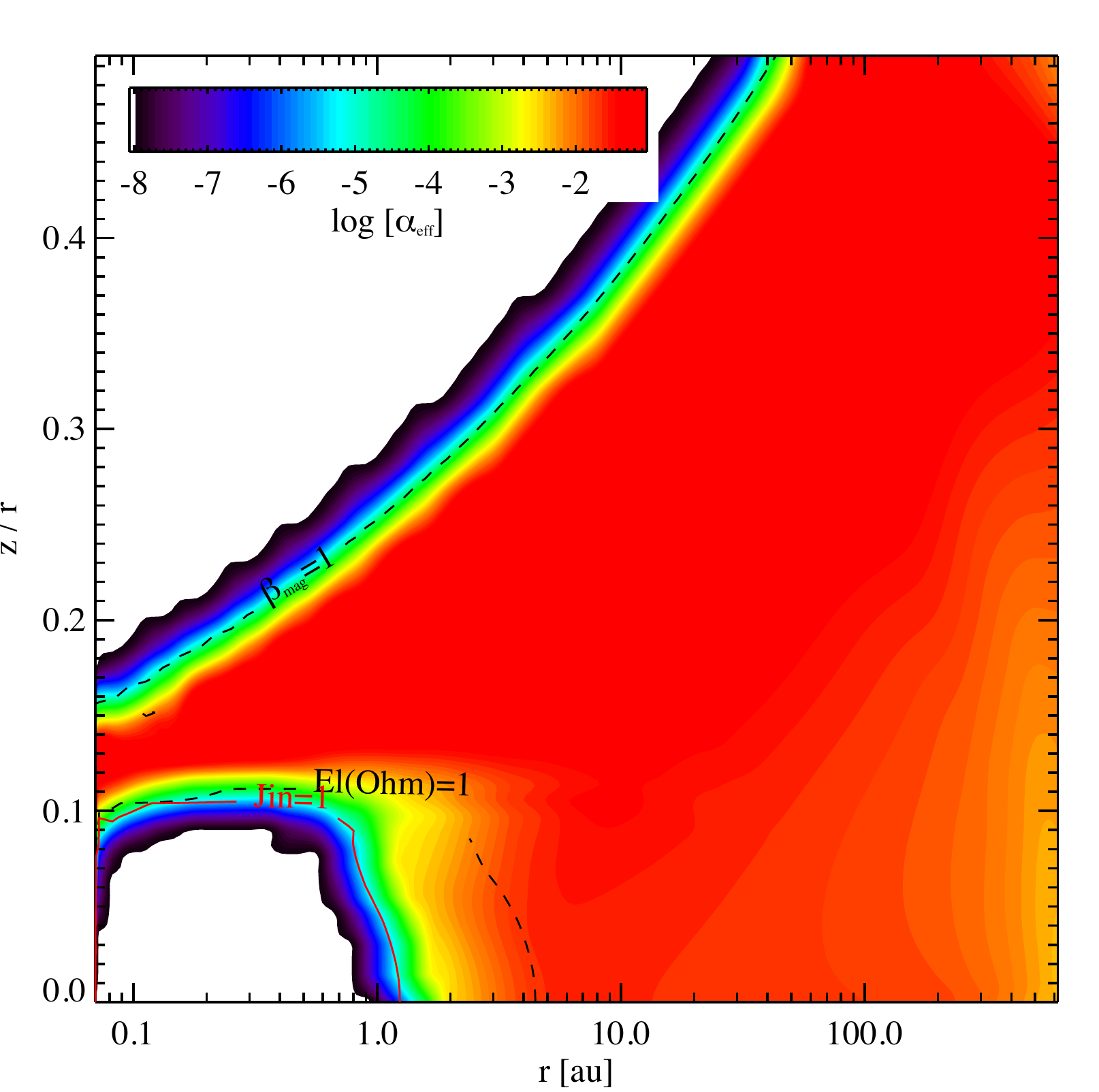}  
  \includegraphics[angle=0,width=9cm,height=9cm,trim=50 80  80 300, clip]{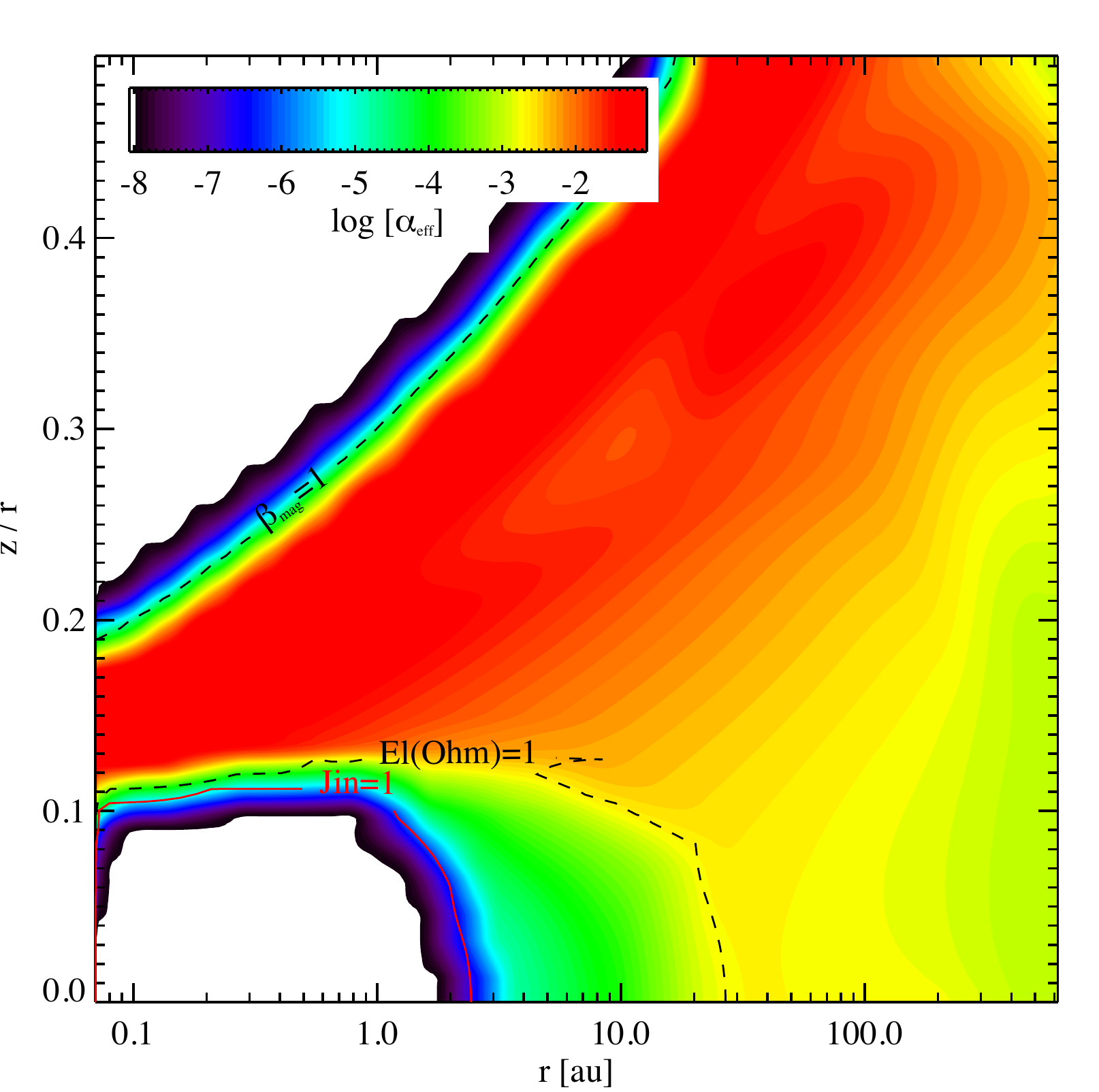}
  \includegraphics[angle=0,width=9cm,height=9cm,trim=50 80  80 300, clip]{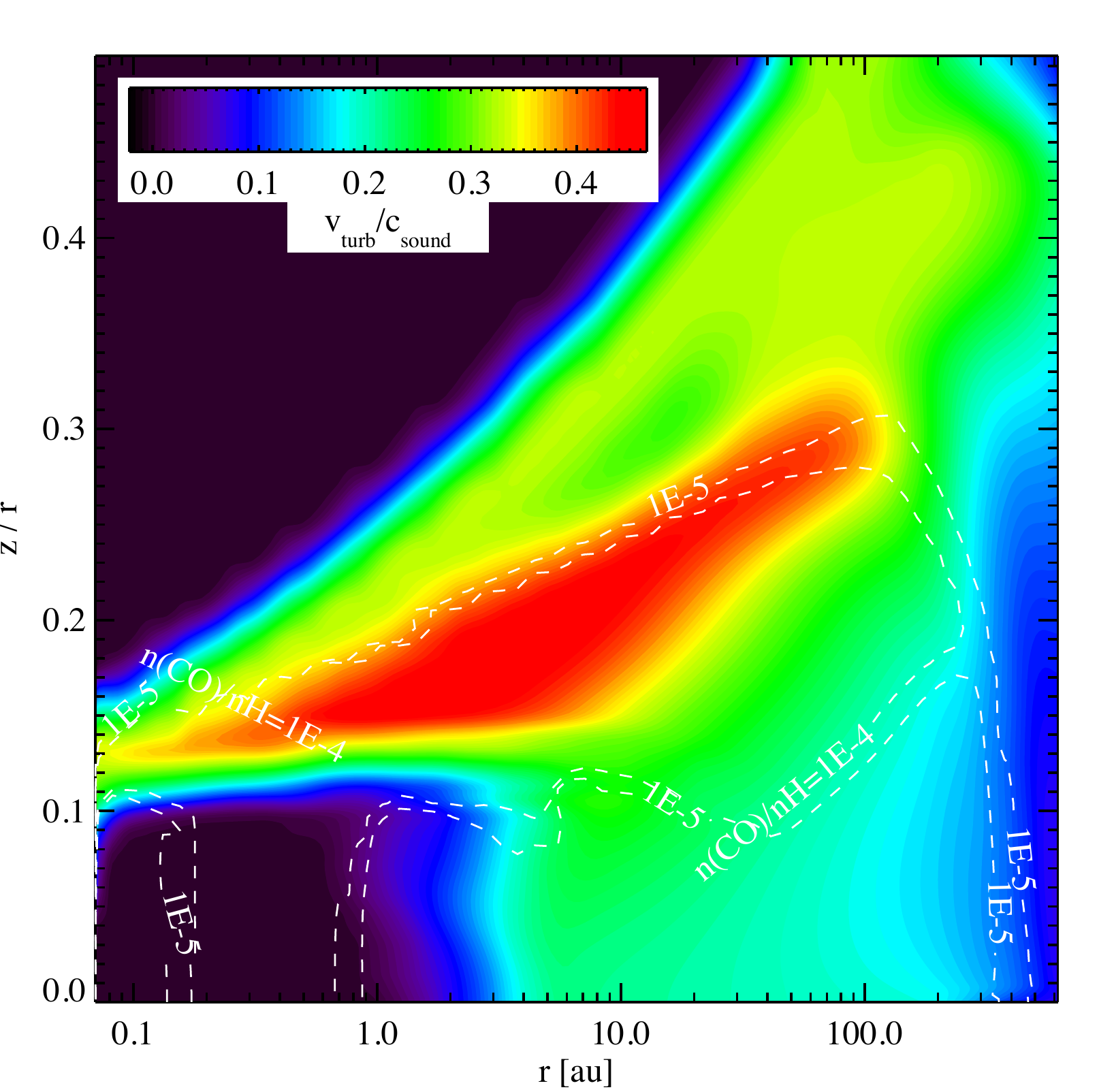}
  \includegraphics[angle=0,width=9cm,height=9cm,trim=50 80  80 300, clip]{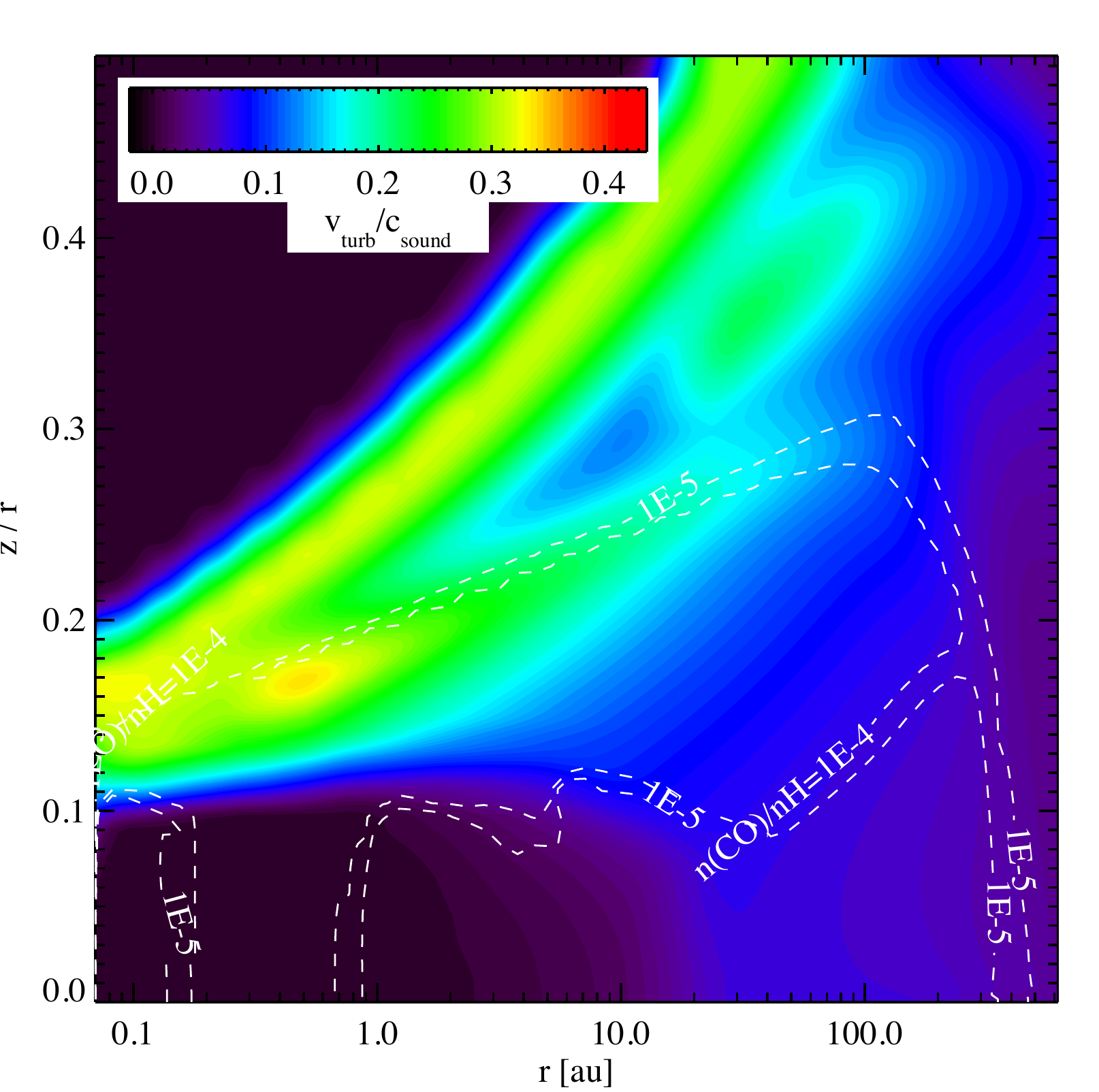}  
  \caption{The effective $\alpha_{\rm eff}$ value and turbulence velocity over sound speed ratio for the DIANA typical disk and $\beta_{\rm mid}$=10$^{4}$ (left panels) and 10$^{6}$ (right) panels. The white contours in the upper panels show the location where the Elsasser Ohm number $\Lambda_{\rm Ohm}$ is unity. The level where there is no MRI in the inner disk midplane, the so-called ''dead-zone'', is shown in red (Jin=1). For $\beta_{\mathrm{mid}}^{1/2} \Lambda_{\mathrm{Ohm}}<1$, MRI is entirely suppressed  and $\alpha_{\mathrm{eff}}$ tends to zero.The contour where $\beta=1$ is also shown in the upper panels. The contours in the low panels indicate the gas-phase CO abundances.}
  \label{fig_disk_results_alpha}          
\end{figure*}  
% ----------------
\begin{figure*}[!htbp]  
  \centering 
  \includegraphics[angle=0,width=9cm,height=10cm,trim=50 80  80 300, clip]{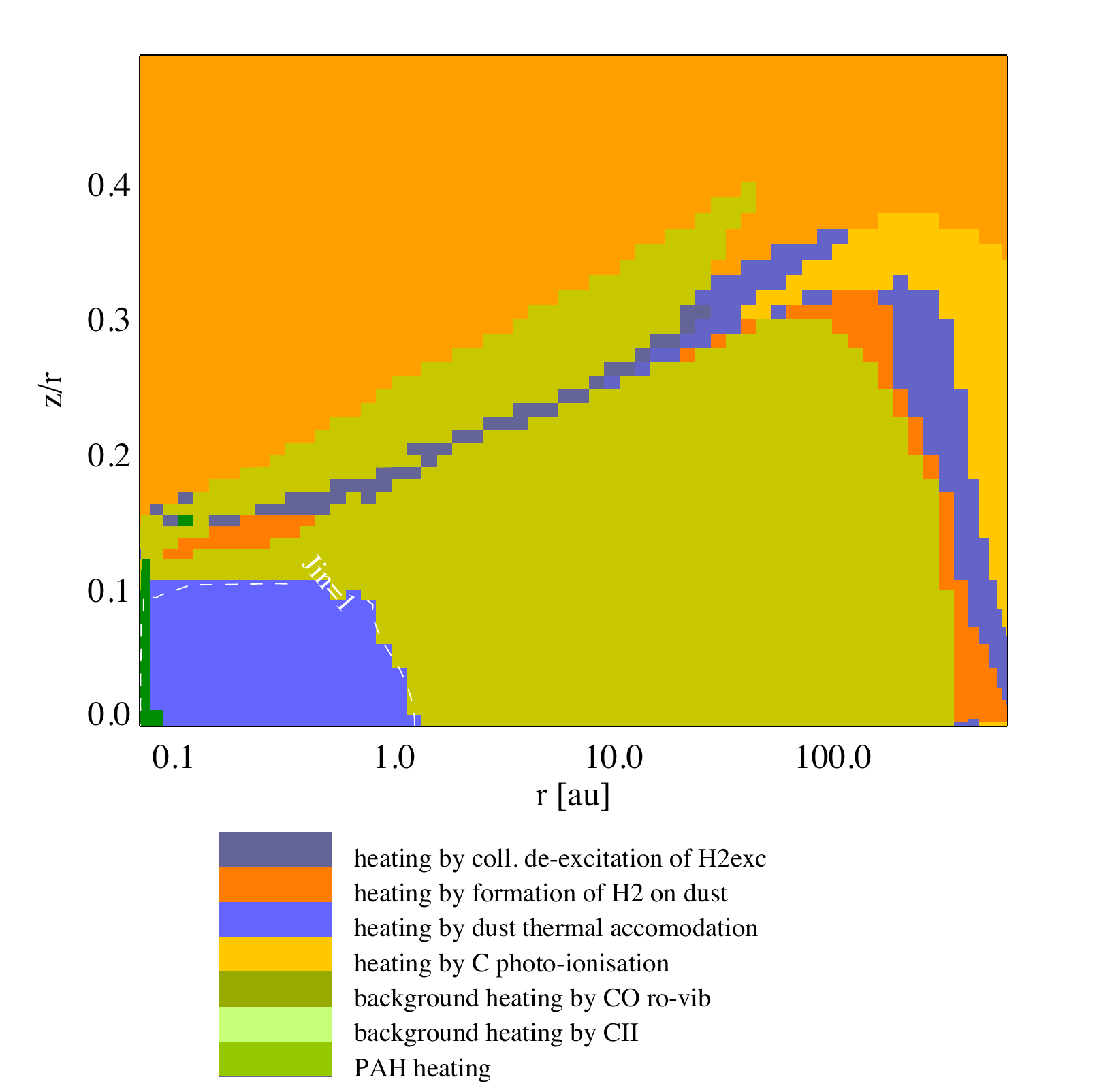}  
  \includegraphics[angle=0,width=9cm,height=10cm,trim=50 80  80 300, clip]{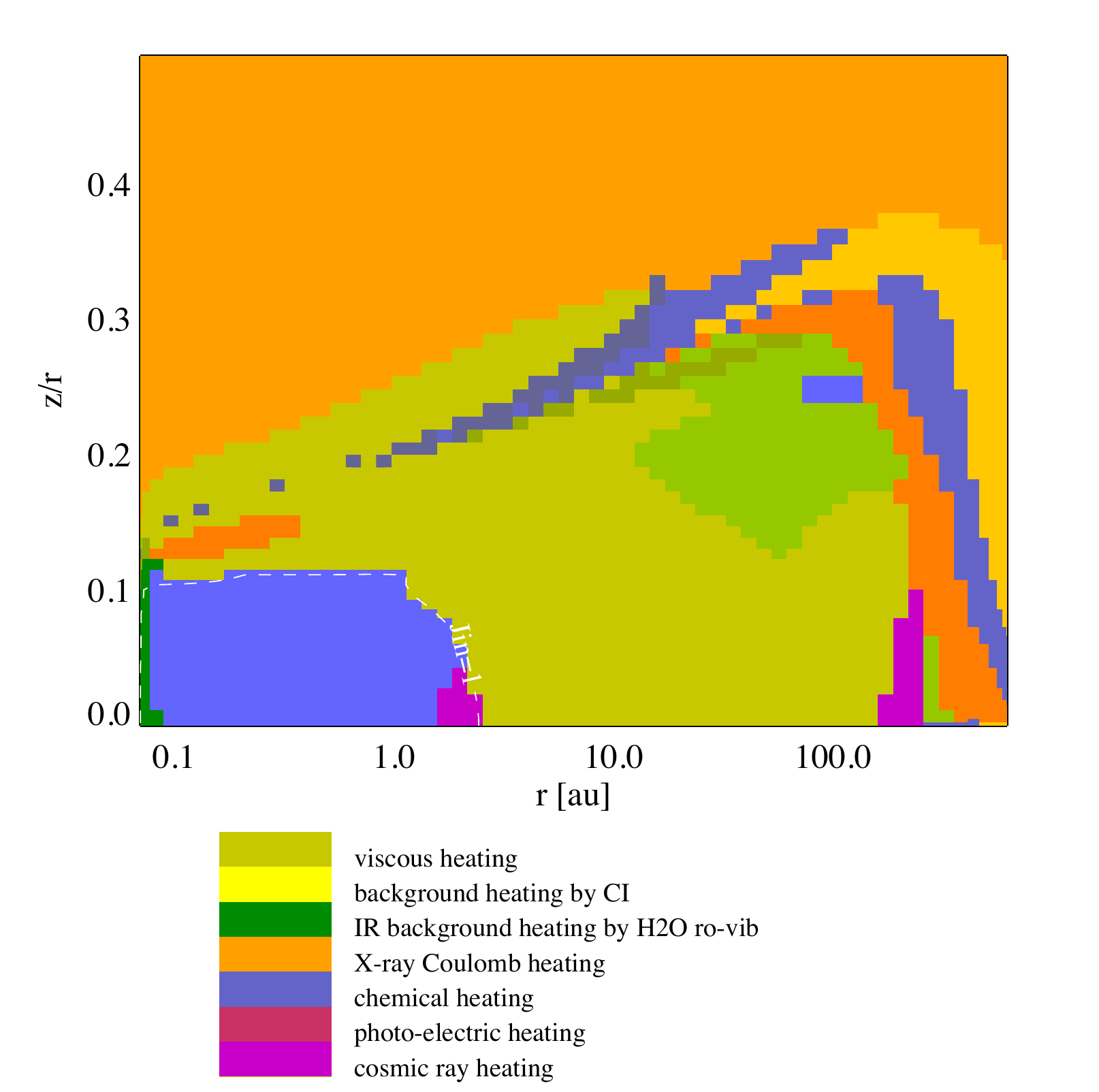}
    \includegraphics[angle=0,width=9cm,height=10cm,trim=50 80  80 300, clip]{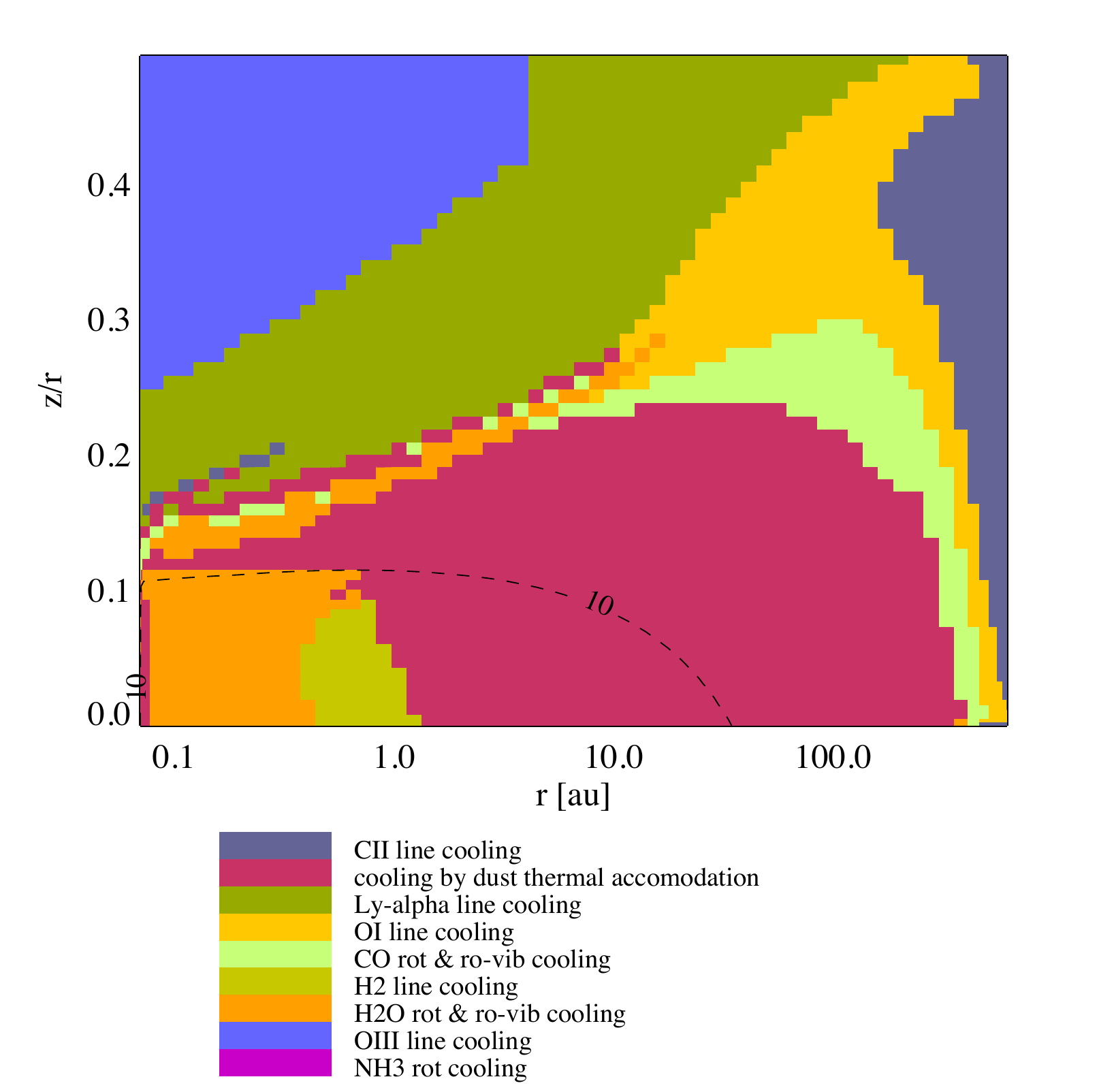}
  \includegraphics[angle=0,width=9cm,height=10cm,trim=50 80  80 300, clip]{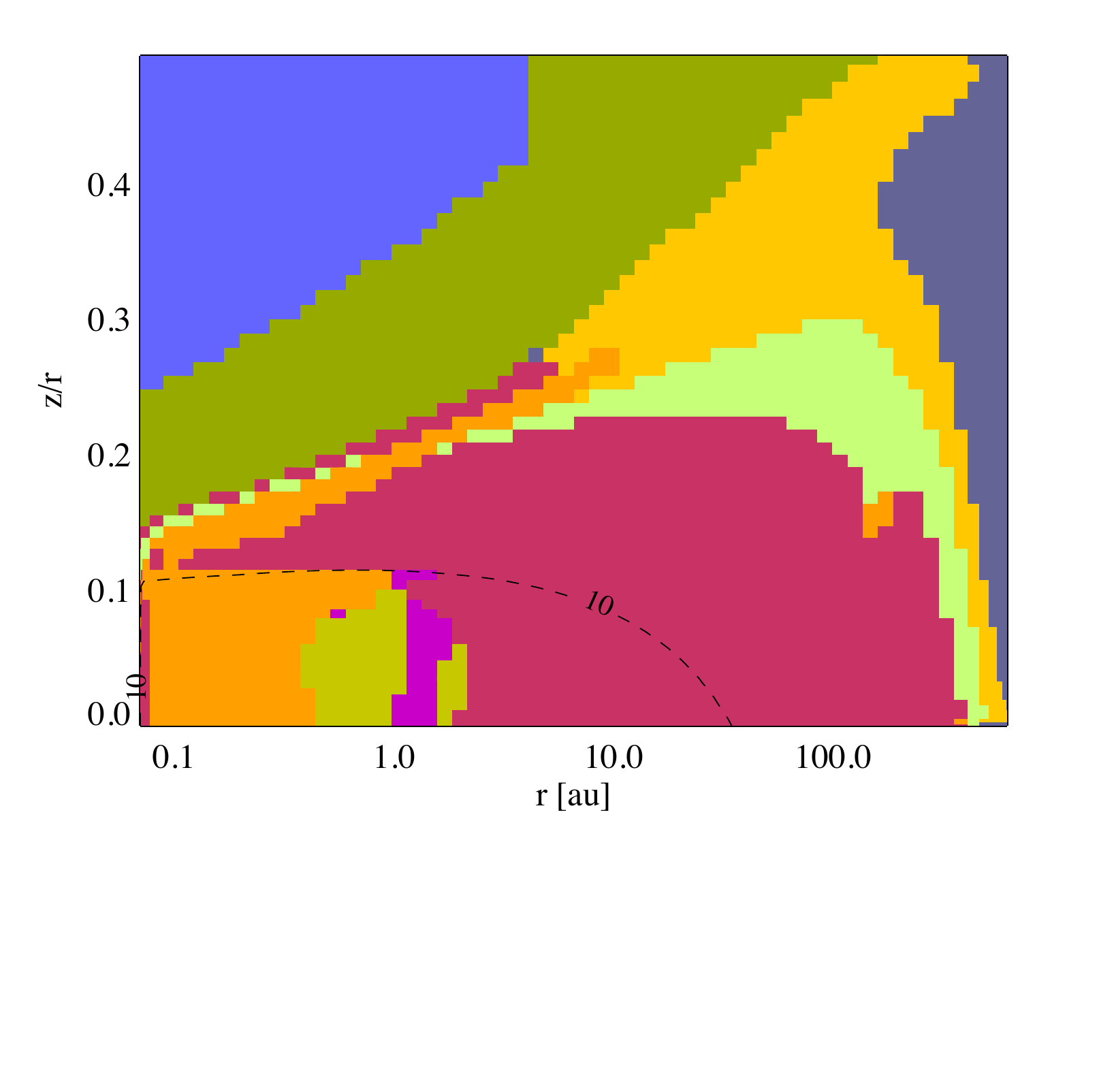}
  \caption{The upper panels show the main heating sources and the lower panels the main cooling sources at the different location in the disk. The left panels are for the $\beta_{\mathrm{mid}}$=10$^4$ disk model, while the right panels correspond to the $\beta_{\mathrm{mid}}$=10$^6$ disk model. In the upper panels, the contour of the Jin's criterion is shown in white contour. In the panels, the black contours correspond to ${\mathrm A_V}$=10.}
  \label{fig_disk_heating_cooling}           
\end{figure*}  
% --------------------------------------------
We chose to model a typical \pd\ with a total gas+solid mass of 10$^{-2}$ M$_\odot$ \citep{Woitke2016A&A...586A.103W}. The disk extends from $r_{\mathrm{in}}$=1~au to $r_{\mathrm{out}}$=600~au with a tapered outer disk. The dust is composed of compact spherical grains
composed of Olivine, amorphous carbon, and Trolite (FeS). The dust grains size distribution follows a power-law with index -3.5 from
$a_{\mathrm{min}}$=0.1~$\mu$m to $a_{\mathrm{min}}$=3000~$\mu$m. The global gas-to-dust mass ratio has the standard value of 100.  Dust grains can settle with a turbulent mixing parameter $\alpha_{\rm settle}$ of 0.01. The PAH abundance depletion factor compared to the interstellar medium value $f_{\mathrm{PAH}}$ is set to 0.01 ($f_{\mathrm{PAH}}=1$ corresponds to an abundance of $3 \times 10^{-7}$). The disk flares with an index of 1.15. The gas scale-height is 1~au at 10~au (10\%). The cosmic ray flux is assumed to be constant throughout the disk at 1.7$\times$10$^{-17}$ s$^{-1}$. Future studies will include varying the X-ray fluxes, lowering the cosmic ray flux and its attenuation in disks \citep{Cleeves2013ApJ...772....5C,Cleeves2014ApJ...794..123C}, and modelling the effects of stellar particles \citep{Rab2017A&A...603A..96R}. The \pd\ parameters are summarized in Table~\ref{tab:refmodel}.

The key parameter of our parametric MRI model is $\beta_{\mathrm{mid}}$. We modelled the disks with $\beta_{\rm mid}$ in the disk midplane from 10$^{2}$ to 10$^{6}$. For comparison, we also ran a model with no (MRI) turbulence, i.e. the line broadening is purely thermal ($\mathrm{v}_{\mathrm{turb}}$=0 km s$^{-1}$) and other models with a constant turbulence width $\mathrm{v}_{\mathrm{turb}}$=0.05, 0.1, 0.15 and 0.2  km s$^{-1}$  and no turbulence heating throughout the disk (passive disk models). The central object is either a TTauri star ($M_*$ = 0.7 M$_\odot$, $T_{\mathrm{eff}}$= 4000 K, $L_*$ = 1.0 $L_\odot$) or a HerbigAe star ($M_*$ = 2.3 M$_\odot$, $T_{\mathrm{eff}}$= 8600 K, $L_*$ = 32 $L_\odot$). The star shows excess UV and strong X-ray emission. The disk parameters are summarized in Table~\ref{tab:refmodel}. 

The disk thermal balance has been modified to include the effects of turbulence heating and turbulence line broadening. 

The release of turbulent energy in the gas at each disk location is
\citep{Hartmann1998apsf.book.....H}
\begin{equation}
E_{\mathrm{acc}}=\frac{9}{4}\alpha_{\rm eff} P_{\mathrm{gas}}\Omega\label{viscous_heating}
\end{equation}
and is included as a gas heating agent in the heating-cooling balance to determine the gas temperature. { \cite{McNally2014ApJ...791...62M} modeled the energy transfer from the turbulence decay to the gas.

The turbulence will effect the gas temperature, which in turn will change the chemistry, the value of
$\beta_{\rm mag}$, the ohmic and ambipolar Elsasser number.
In an analytical analysis in the case of ideal MRI-turbulence, the energy release is
\begin{equation}
E_{\mathrm{acc}} \propto P_{\mathrm{therm}}^{1-\delta}\Omega.
\end{equation}
Adopting $\delta$=0.5,
\begin{equation}
E_{\mathrm{acc}} \propto \sqrt{P_{\mathrm{therm}}}\Omega.
\end{equation}
Assuming that the pressure term can be described by a barotropic law
\begin{equation}
P_{\mathrm{therm}} \propto n^{\gamma},
\end{equation}
the gas heating rate is 
\begin{equation}
E_{\mathrm{acc}} \propto n_{\mathrm{<H>}}^{\gamma /2}\Omega.
\end{equation}
The value of $\gamma$ can vary typically from 1 to 3. It is interesting to note that in our prescription, the turbulence heating efficiency depends on the thermodynamics of the gas. The effects of increasing the gas temperature on the value of
$\alpha_{\mathrm{eff}}$ are not easy to predict. $\alpha_{\mathrm{ideal}}$ is proportional to $T^{-\delta}$, $\Lambda_{\rm Ohm}$ is proportional to $\sqrt{T}$.
The dependence of $Am$ on $T$ is not direct, but assuming electronic recombination rates $\propto T^{-1/2}$, an increase of the gas temperature should result in a decrease of the ion recombination rates and thus a higher value for $Am$. Therefore, as the gas temperature increases due to viscous dissipation, the resistivities decrease but at the same time the ideal-MHD $\alpha_{\rm eff}$ decreases as well.

The turbulent velocity is sub-sonic and is related in our model to $\alpha_{\mathrm{eff}}$ and the sound speed $c_{\mathrm{s}}$ by v$_{\mathrm{turb}}$=$\sqrt{\alpha_{\mathrm{eff}}}$$c_{\mathrm{s}}$
\citep{Hughes2011ApJ...727...85H}. Another relationship such as  v$_{\mathrm{turb}}$=$\alpha_{\mathrm{eff}}$ $c_{\mathrm{s}}$ has been proposed \citep{Simon2013ApJ...775...73S}. 
The local line width becomes $\Delta \mathrm{v} = (\mathrm{v}_{\rm th}^2+\mathrm{v}_{\rm turb}^2)^{1/2}$, where $\mathrm{v}_{\rm th}$ is the thermal broadening. The turbulent velocity will affect the line optical depths as 1/$\Delta \mathrm{v}$. In turn, the cooling term is affected because of the change in the line cooling rates. The MRI-turbulence does not only affect the heating but also the cooling of the gas.
 
% ----------------
\begin{figure*}[!htbp]
  \centering
  \includegraphics[angle=0,width=9.0cm,height=9cm,trim=30 40  80 300, clip]{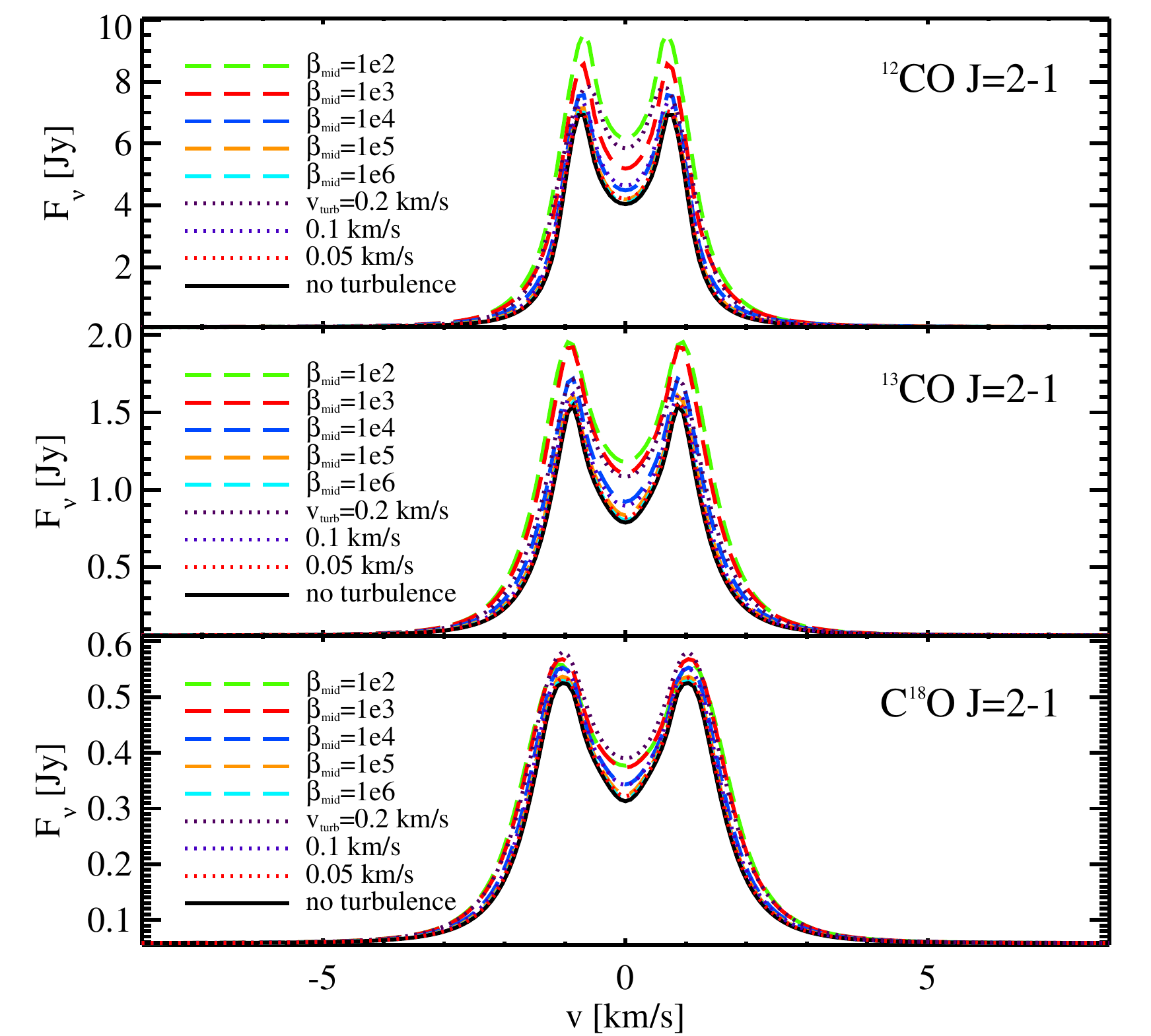}  
  \includegraphics[angle=0,width=9.0cm,height=9cm,trim=30 40  80 300, clip]{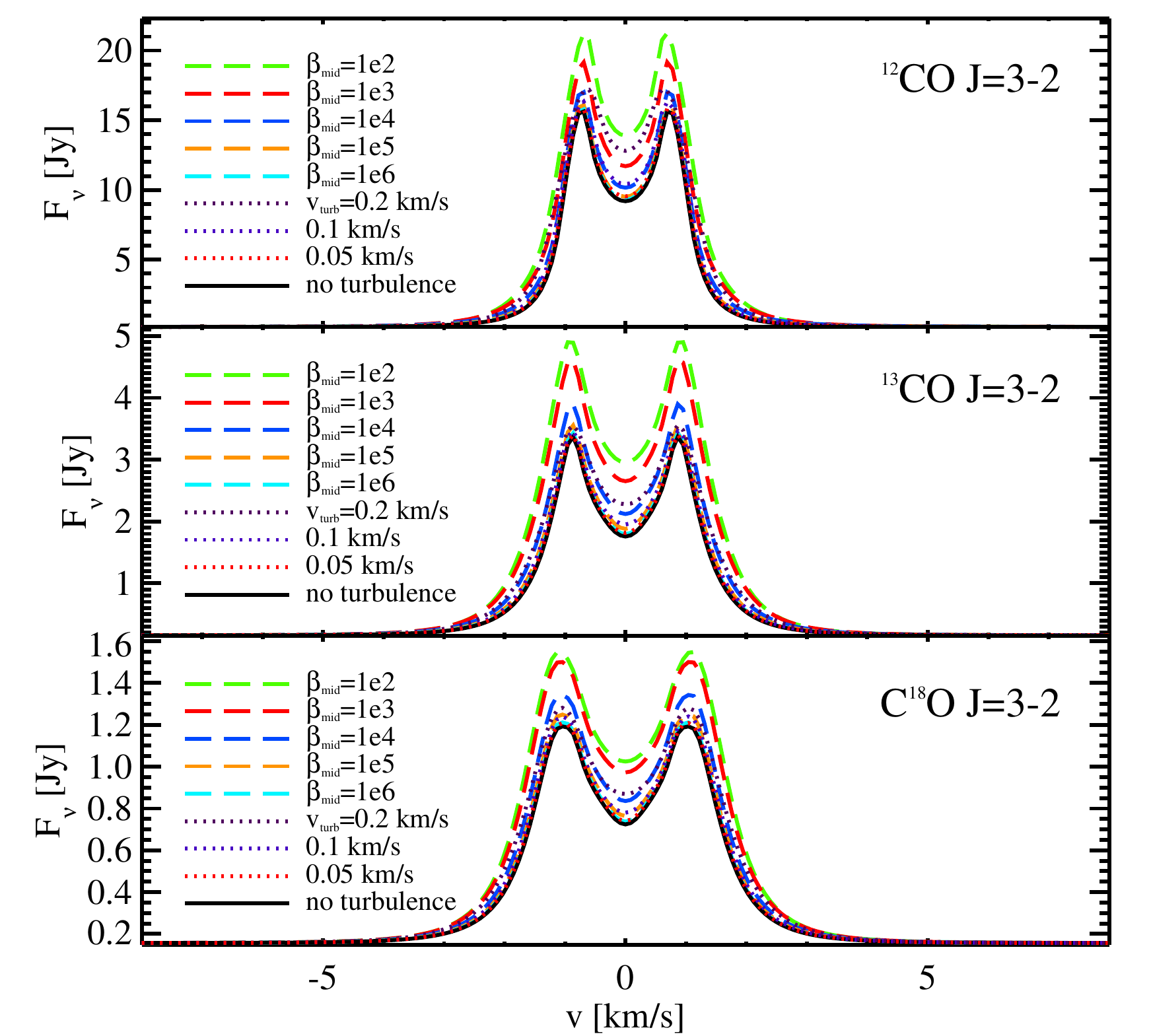}
   \caption{CO and isotopologue line profiles from the typical disk with different values of $\beta_{\rm mid}$ (dashed-lines), as well as from the disk model without turbulence (continuous black lines) and the model with a constant turbulent width of 0.05, 0.1, and 0.2 km s$^{-1}$ (dotted-lines). The disk is seen with an inclination of 45\degr (0\degr means that the disk is seen face-on).}
  \label{fig_CO_disk}           
\end{figure*}  
    
The present model only considers viscous accretion, through the alpha parameter. However, it is known that, in the presence of a poloidal magnetic field, discs can be subject to MHD wind which can also drive accretion (e.g., \cite{Bai2013ApJ...767...30B}). We have purposely neglected this process due to the lack of robust prescriptions for wind-driven accretion.
    
% ------------------------------------------------------------------------
\section{Disk model results}\label{disk_results}

The disk total charge (a.k.a. ionization fraction) is shown in the upper-left panel of Fig.~\ref{fig_disk_results_charges} for a disk model with $\beta_{\rm mid}$=10$^{4}$. The figures show the different disk structures with radius $r$ in au in the horizontal-axis and $z/r$ in the vertical axis.

The disk ionization fraction (or total charge) is governed by the balance between ionization due to UV, X-ray, and cosmic rays in the disk and the recombination and charge exchange reactions. Fig.~\ref{fig_disk_results_charges} shows that the free electrons and the gas-phase cations are the main negative and positive charge carriers except in a small region in the inner disk midplane.  

The dust grain average charge is displayed in the upper-left panel of Fig.~\ref{fig_disk_results_dust_charges}, while the fractional contribution of the charges (positive and negative) on grains to the total charge is exhibited in the upper-right panel. From the upper disk atmosphere to the midplane, the dust grains have first lost hundreds of electrons due to the photoelectric effect. When the UV field decreases, the free electrons (up to a relative abundance of a few 10$^{-4}$) from atomic carbon ionization stick to the grains, rendering them negatively-charged. The total charge reaches the value of the elemental abundance of carbon (1 to 3 $\times 10^{-4}$).
As carbon becomes neutral at lower altitude, the recombination with the free electrons compensates almost exactly for the photoejection of the electrons from the grains and the grain charge fluctuates around zero. The remaining sources of electrons are the atoms with ionization potential lower than 13.6 eV (Fe, Mg, Si, S). Therefore, gas depletion of the metal species (Fe, Mg, Si), in addition to that of Sulfur will determine the electron fraction $\chi({\mathrm{e}})$ in the intermediate disk heights. It should be noted that the dust grains remain at temperatures $T_{\mathrm{dust}}$ well below the sublimation temperatures ($T_{\mathrm{subl.}}\sim$~1500~K) even though the gas temperature can reach up to $\sim$~10,000~K (Fig.~\ref{fig_disk_temperatures}). We have not considered the destruction of dust grains in warm gas due to chemisputtering.
 
In the midplane inner disk region ($r<1$ au with total charge fraction $<$ 10$^{-12}$), the radiation field is too weak to eject the excess electrons on the grains. The grains are negatively charged after the attachment of the free electrons created by the interaction of cosmic rays or X-ray photons with the neutral gas, consistent with the analytical approximation (see Sect.~\ref{charge_analytical_noUV}). The negatively-charged dust grains become the major
dominant negative charge carrier. The cations remain the main positive charge carrier.

In the lower-left panel of Fig.\ref{fig_disk_results_dust_charges}, we see that the contribution of the positively-charged PAHs to the total charge is small. The contribution of the negatively-charged PAHs is relatively high but not dominant in the region above the freeze-out region of the PAHs. The relative contribution of PAHs as charge carriers depends on their abundances in disks. In the typical model, the PAH depletion factor $f_{\mathrm{PAH}}$ is between 0.01 and 1 \citep{Woitke2016A&A...586A.103W}. In the disk midplane, the PAHs are frozen onto the grains.

The Elsasser Ohm number, the ambipolar diffusion number, and the complete mode damping criterion are shown in Fig.~\ref{fig_disk_results_Ohm} for the disk models with $\beta_{\rm mid}$=10$^{4}$ (left panels) and $\beta_{\rm mid}$=10$^{6}$ (right panels). The Elsasser Ohm number is the most sensitive to the value of $\beta_{\rm mag}$. The ambipolar Elsasser number $Am$ distributions are similar for the $\beta_{\rm mid}$=10$^{4}$  and $\beta_{\rm mid}$=10$^{6}$. In unobscured regions, the electron fractional abundance $\chi(\mathrm{e})$ is relatively high in the range 10$^{-8}$--10$^{-4}$. Due to the limit on negative charges on grains (eq. \ref{eq_max_elec_on_grains}), in a gas with total ionization fraction greater than a few 10$^{-12}$, the free electrons will be the main contributor to the Ohm Elsasser conductivity.  In the inner disk midplane, the total charge is low, resulting in low values for the Ohm and ambipolar Elsasser numbers.

One can derive a good approximation to the Elsasser Ohm number in disk regions where $\sigma_{\mathrm{O}}\simeq \sigma_{\mathrm{e,O}}$ and $\chi({\rm e})>10^{-13}$:
\begin{equation}\label{Ohm_Elsasser_approximation}
\Lambda_{\rm Ohm} \simeq 1\ \left(\frac{T}{100}\right)^{1/2}\left(\frac{10^4}{\beta_{\mathrm{mag}}}\right)\left(\frac{r}{\mathrm{au}}\right)^{3/2} \left(\frac{\chi(\mathrm{e})}{10^{-9}}\right).
\end{equation}
The approximated Ohm Elsasser distribution for a disk model with
$\beta_{\mathrm{mid}}=10^4$ is shown in Fig.~\ref{fig_Ohm_analytical}. 

The disk distribution for the effective  turbulent parameter $\alpha_{\rm eff}$ is shown in Fig.~\ref {fig_disk_results_alpha} for $\beta_{\rm mid}$ of 10$^{4}$ (the left panels) and 10$^{6}$ (the right panels). The disk turbulence efficiency structure can be divided into zones. The zone limits are defined by the transitions between the MRI-driven region, the Ohm diffusion limited dead-zone according to the total damping criterion $\mathrm{Jin} \equiv \beta_{\mathrm{mag}}^{1/2} \Lambda_{\mathrm{Ohm}}$=1, the Ohm diffusion restricted MRI region ($\Lambda_{\mathrm{Ohm}}$=1), and the location of the disk where $\beta_{\mathrm{mag}}$=1 in the disk atmospheres. The decrease of $\alpha_{\rm eff}$ with the disk radius stems from the decrease in the gas pressure, hence of $\beta_{\mathrm{mag}}$ (see the middle panels of  Fig.~\ref{fig_Bfield}). The outer disk midplane decrease of $\alpha_{\rm eff}$ stems from the increase of $\beta_{\rm mag}$.
The ambipolar diffusion does not restrict the value of $\alpha_{\rm eff}$ in the disk. At the top disk surfaces, the hydrogen gas is ionized due to the X-rays. As the protons recombined and the gas becomes more neutral, the main ions are the C$^+$ cations. Below the fully ionized carbon layer, the ambipolar diffusion number $Am$ plummets by two orders of magnitude. Despite the drop, $Am$ remains higher than unity for most of the disk. The ambipolar diffusion is also affected by the metal abundances. At large radii, the vertical column density is low such that the gas can be ionized and $Am$ increases again.

The lower panels of Fig.~\ref {fig_disk_results_alpha} show the turbulence over the sound speed velocity ratios. The location of the gas-phase CO is overplotted as contours. For
$\beta_{\rm mid}$=10$^{4}$, CO will be emitted significantly in the turbulent warm molecular disk layers with $\mathrm{v}_{\rm turb}/c_{\rm s}$ between 0.25  and 0.4 . For the $\beta_{\rm mid}$=10$^{6}$ model, $\mathrm{v}_{\rm turb}/c_{\rm s}$ is between 0.1 and 0.3. For completeness, the sound speed $c_{\rm s}$ for the two models are plotted in Fig.~\ref{fig_Bfield}. The CO emission comes mostly from gas with sound speed $\sim$0.3 - 0.5 km s$^{-1}$. 
% ---------------- 
\begin{figure*}[!htbp] 
  \centering
  \includegraphics[angle=0,width=6cm,height=6cm,trim=30 20  80 300, clip]{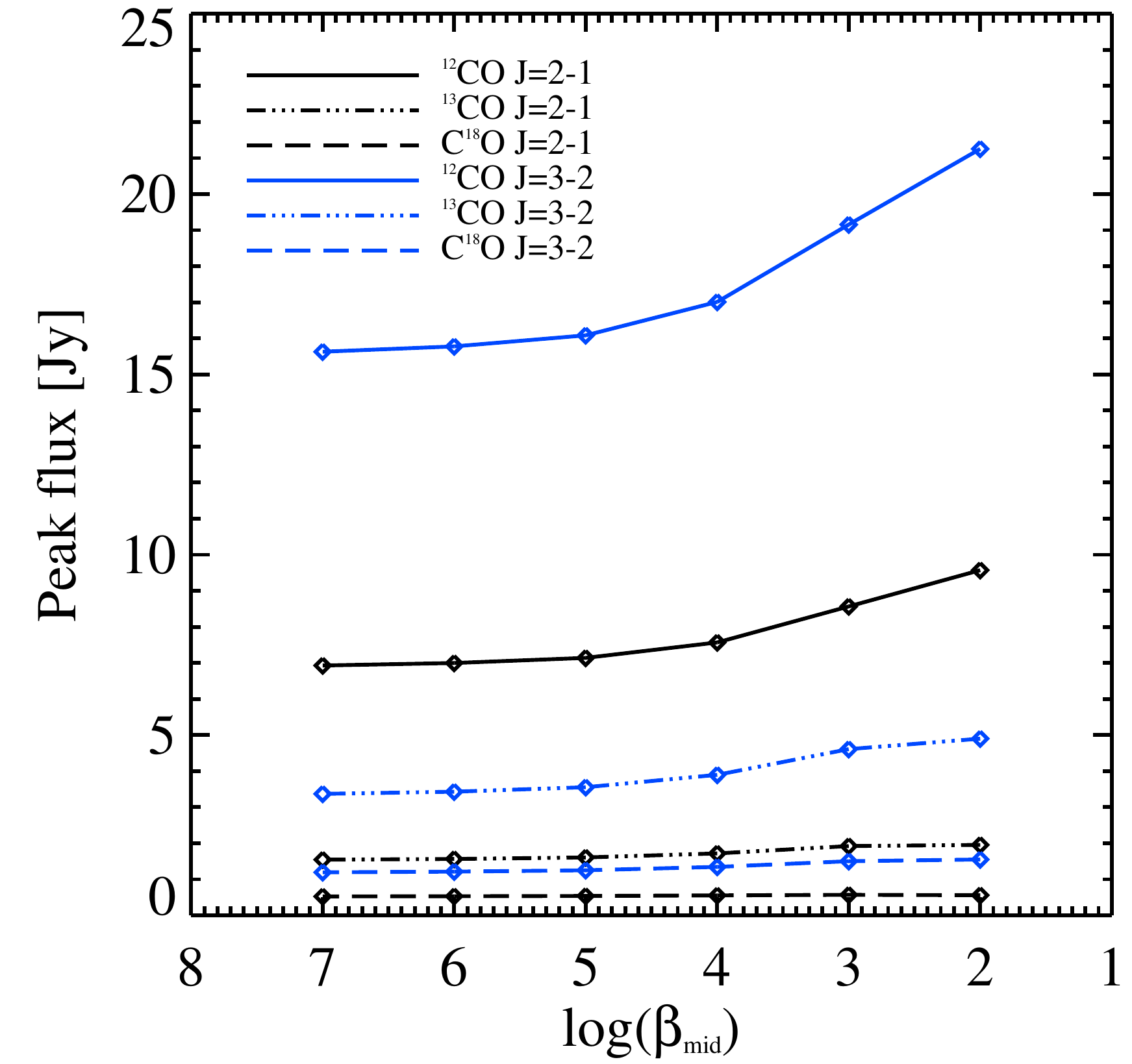}  
  \includegraphics[angle=0,width=6cm,height=6cm,trim=30 20  80 300, clip]{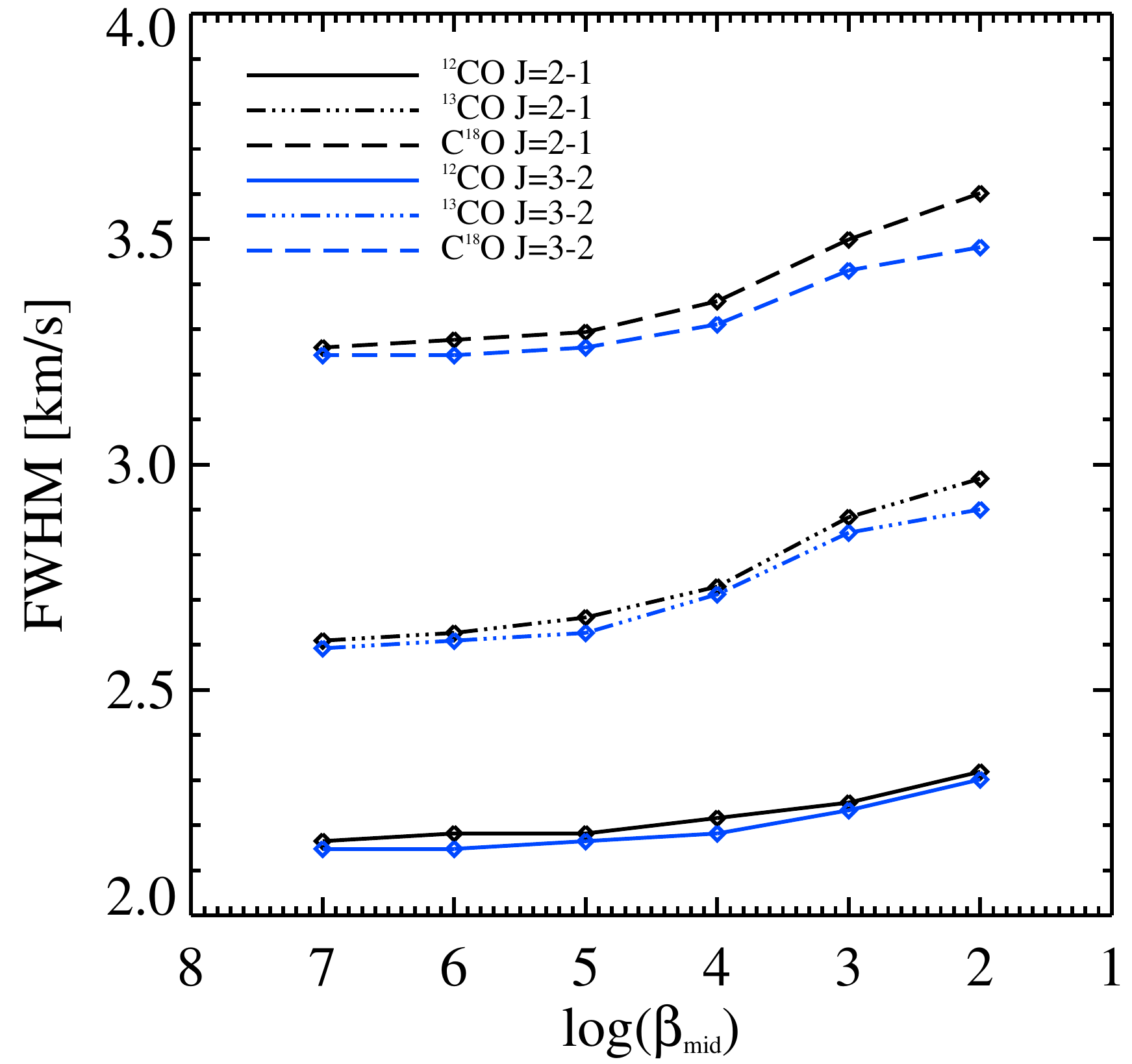}
  \includegraphics[angle=0,width=6cm,height=6cm,trim=30 20  80 300, clip]{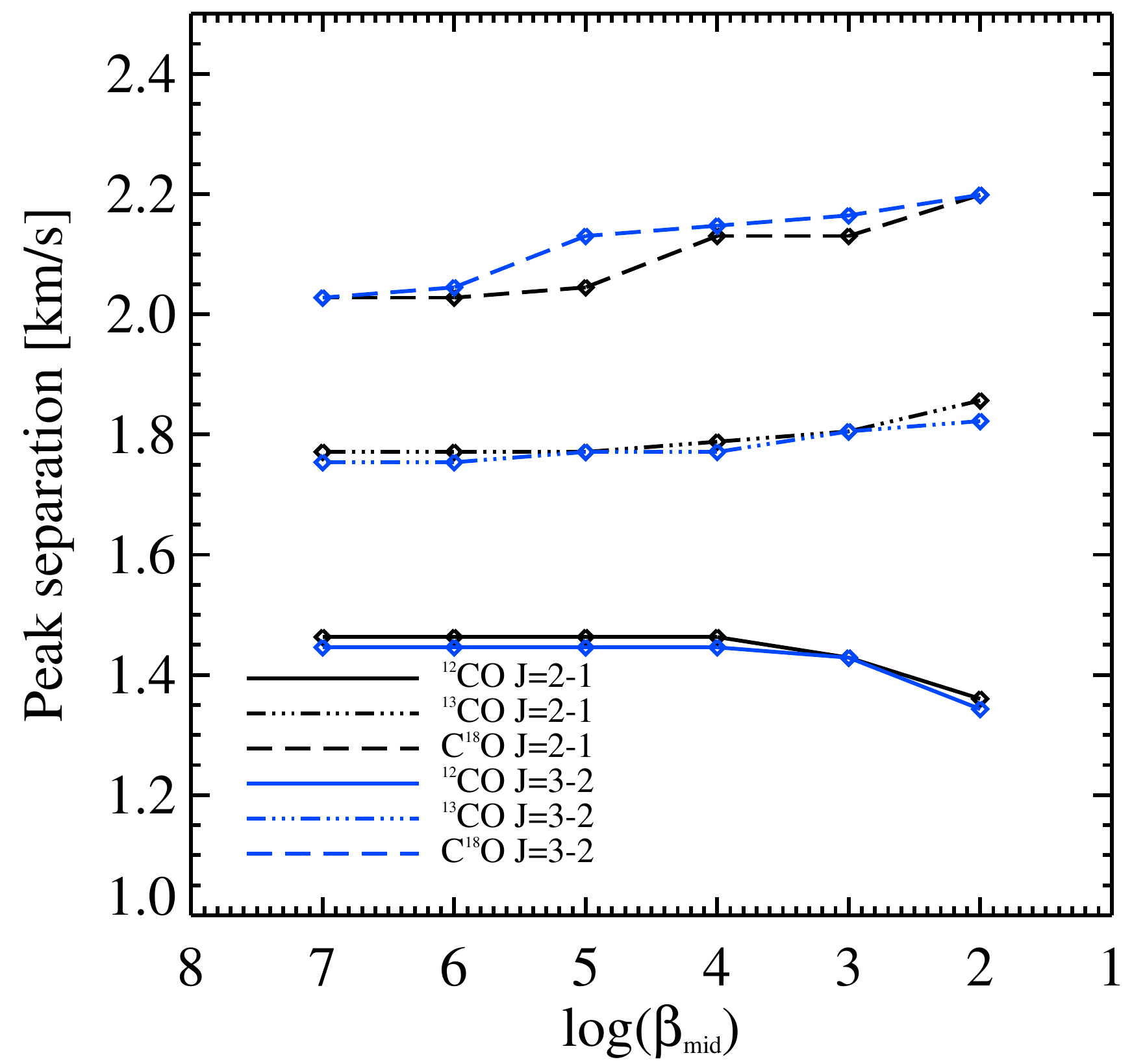}  
  \caption{CO line profile statistics as function of the value of $\beta_{\mathrm{mid}}$. The top left shows the peak fluxes in Jy. The top right panel shows the FWHM in km s$^{-1}$. The lower left panels show the double-peaked profile peak separation in km s$^{-1}$  The value for $\beta_{\mathrm{mid}}$=10$^7$ corresponds to a model with no MRI turbulence in the entire disk.}
  \label{fig_CO_statistics}           
\end{figure*}   
% ------------------- 
Fig.~\ref{fig_disk_heating_cooling} shows the main heating and cooling processes for the disk models with
$\beta_{\mathrm{mid}}$=10$^4$ (left panels) and $\beta_{\mathrm{mid}}$=10$^6$ (right panels). In the dead-zone where $\beta_{\mathrm{mag}}^{1/2} \Lambda_{\mathrm{Ohm}}<1$, the gas heating occurs mainly via the gas thermal accommodation on warm dust grains. Outside the dead-zone, viscous heating is the dominant heating process in both disk models from the midplane to the intermediate layers ($z/r \sim 0.2$) till disk radii of a few hundred au, where the density, the Keplerian rotation, and $\alpha_{\mathrm{eff}}$ are low (see eq.~\ref{viscous_heating}). The outer disk is heated by the energy released by H$_2$ formation on dust grains and by the chemical reactions. The disk atmosphere region where PAHs are not frozen onto dust grains in the $\beta_{\mathrm{mid}}$=10$^{4}$ model can be heated by the photoelectric effect. The main gas heating
processes of passive disks in the atmospheres are hits by cosmic rays and photoelectric effects from dust grains and PAHs \citep{Bergin2007prpl.conf..751B} and chemical heating \citep{Woitke2016A&A...586A.103W}. Towards the midplane the densities are high enough for the gas and the dust to become thermally coupled.
In an active disk, these processes are minor as soon as the magnetic field strength is not too weak (or $\beta_{\mathrm{mid}}>10^{6}$).
In the regions with $\beta_{\mathrm{mag}}<1$, the gas is heated by the X-ray Coulomb heating. The main cooling agents in the dead-zone are the water vapor and molecular hydrogen rovibrational lines. In the viscous heating dominated region, the cooling occurs via thermal accommodation on dust grains. Outside the viscous dominated region, CO isotopologue rotational lines are the main coolants in the warm molecular layers, and  the [OI] and [CII] fine-structure lines are the main cooling lines in the atomic disk atmosphere. In the X-ray dominated region, [OIII] and Lyman $\alpha$ are the main cooling lines. This is consistent with the conclusions of \cite{Najita2017ApJ...847....6N} who modeled the effects of heat generated by turbulence decay and concluded that warm CO and H$_2$O lines can trace the mechanically heated gas.

The actual CO line profiles computed in non-LTE for CO, $^{13}$CO and C$^{18}$O $J$=2-1 and $J$=3-2 are shown in Fig.~\ref{fig_CO_disk} for the TTausi and HerbigAe disks seen with a 45\degr \ inclination. 
The line flux increases with more turbulent gas because of stronger heating and that more extended emitting area per velocity resolution elements.The signature of turbulence broadening on the line profiles is not obvious, although the CO lines should probe the most turbulent regions of \pd s. The CO profiles for the disk model with $\beta_{\rm mid}$=10$^{6}$ and 
that for the model with no turbulence are similar for both the $J$=2-1 and $J$=3-2  transitions. For both the CO $J$=2-1 and CO $J$=3-2 lines computed in models with a fixed value of  $\mathrm{v_{turb}}$, one can constraint $\mathrm{v_{turb}}$ down to 0.05 km s$^{-1}$ (see the red-dotted lines in Fig.~\ref{fig_CO_disk}).
% ---------  
\begin{figure*}[!htbp] 
  \centering
  \includegraphics[angle=0,width=9cm,height=9cm,trim=50 20  80 300, clip]{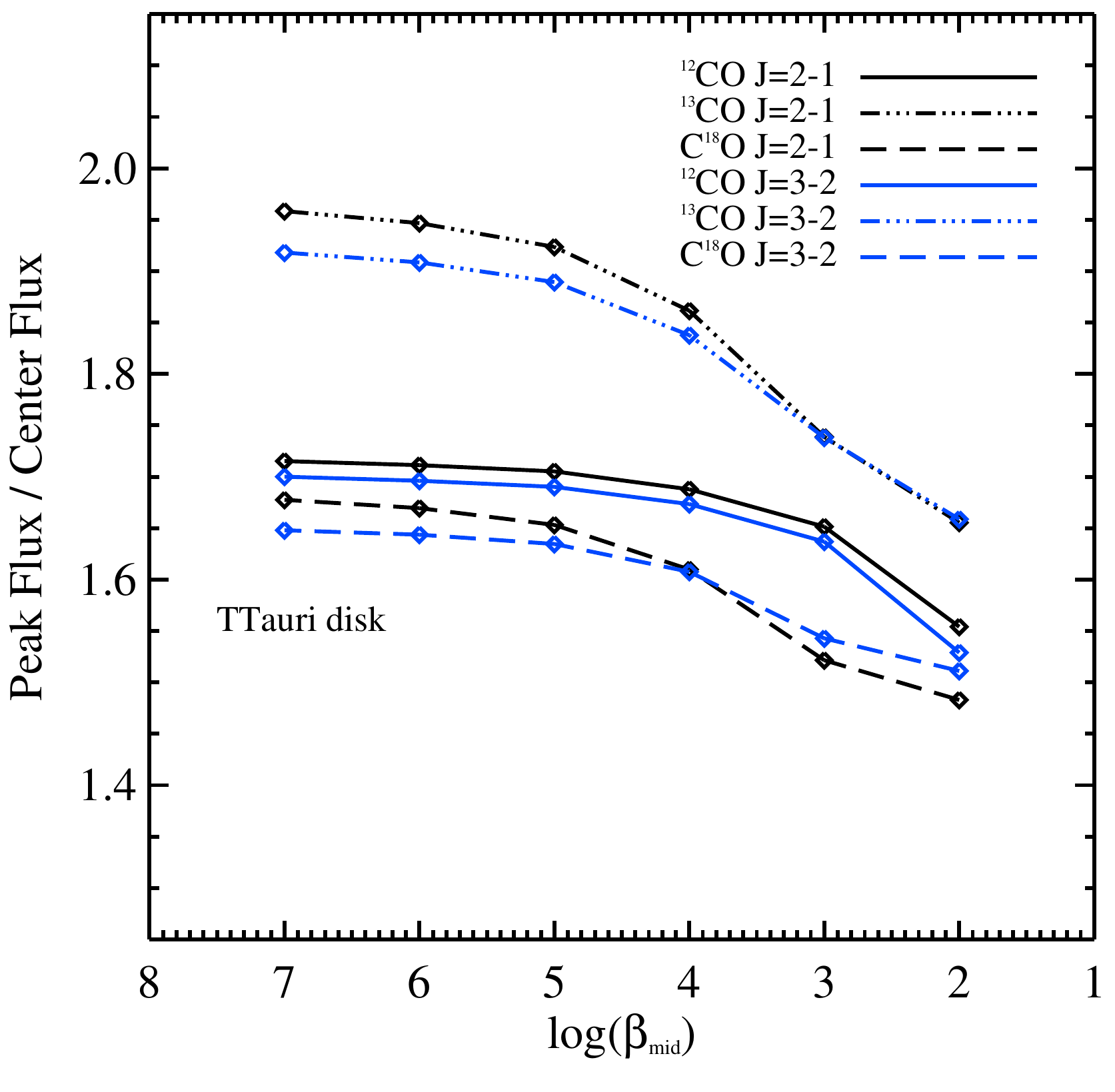}  
  \includegraphics[angle=0,width=9cm,height=9cm,trim=50 20  80 300, clip]{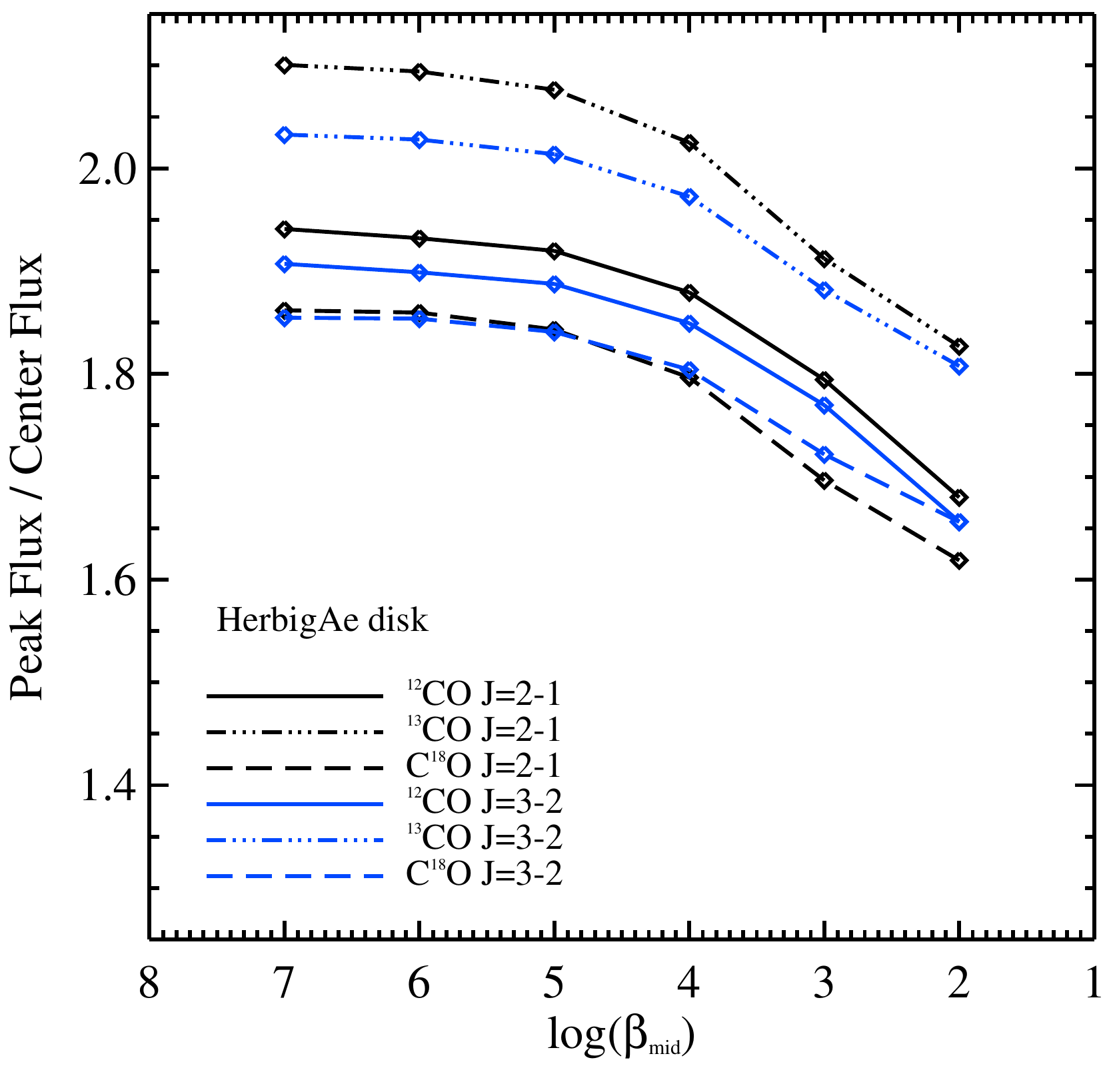}  
  \includegraphics[angle=0,width=9cm,height=9cm,trim=50 20 75 300, clip]{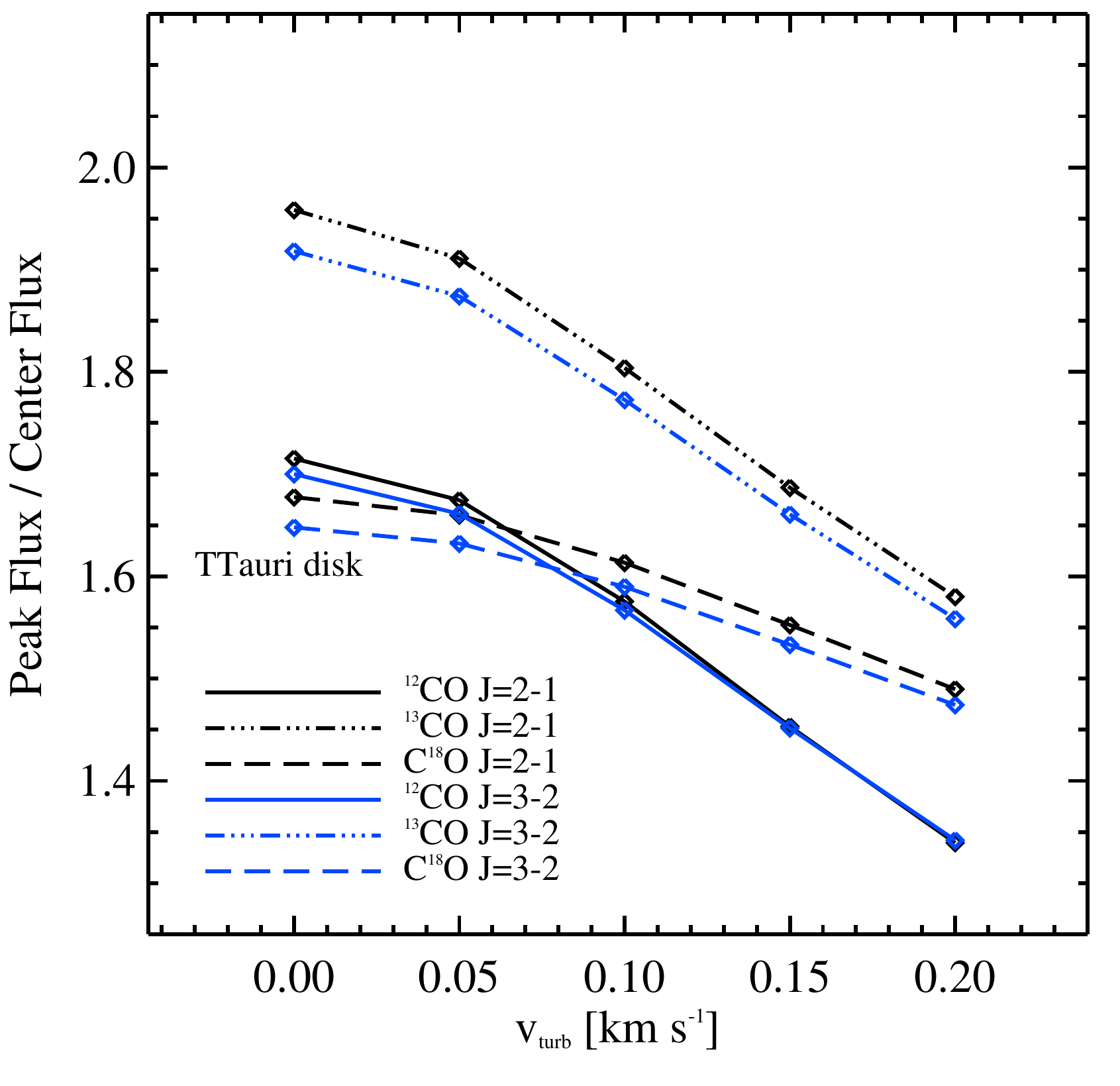} 
  \includegraphics[angle=0,width=9cm,height=9cm,trim=50 20 75 300, clip]{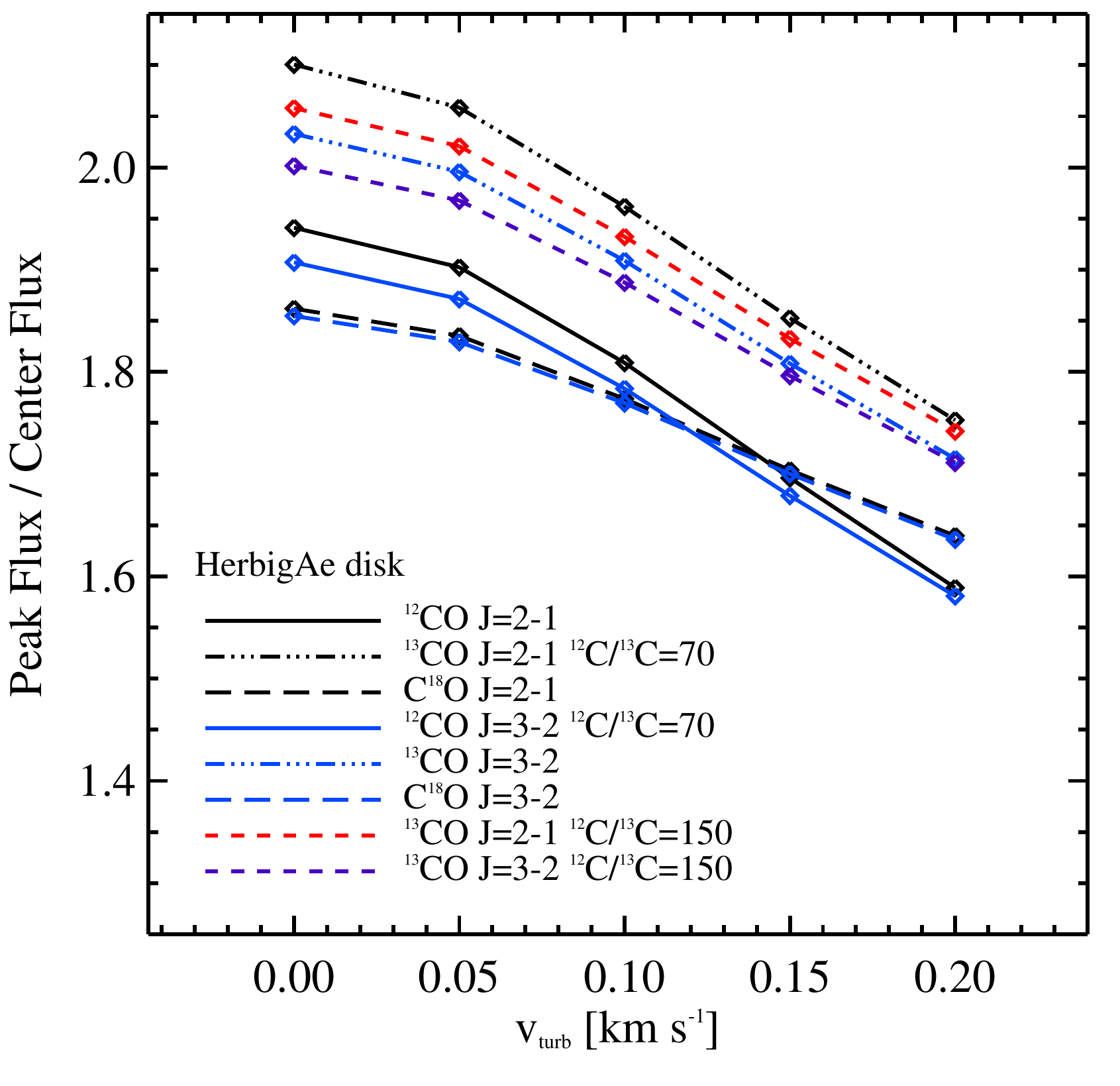} 
  \caption{Peak flux to zero velocity flux ratio for standard TTauri disk model (left panels) and the HerbigAe disk model (right panels) as function of the  $\beta_{\mathrm{mid}}$ parameter (upper panels) and of $\mathrm{v}_{\mathrm{turb}}$ (lower panels). All the models are viewed with an inclination of 45$\degree$ (0$\degree$ \ means that the disk is seen face-on).}
  \label{fig_CO_vturb_stat}           
\end{figure*}   
% --------
Fig.~\ref{fig_CO_statistics} shows the variation of the peak flux, FWHM, and the profile peak separation as function of $\beta_{\mathrm{mid}}$ for the CO lines for the TTauri disk models. The ratio between the peak and line centre flux (peak-to-trough) for the TTauri and HerbigAe disk models are shown in Fig.~\ref{fig_CO_vturb_stat} for the full MRI models (upper panels) and for disk models with a fixed value of the turbulence width (lower panels). The choice of $\beta_{\rm mid}$ influences mostly the ratio between the peak and line centre flux, consistent with the results of the parametric disk model of \cite{Flaherty2015ApJ...813...99F}. The trend of decreasing peak-to-center ratio with increasing disk turbulence velocity shows the same amplitude (0.2 to 0.4) than the trend of a lower ratio for high values of $\beta_{\mathrm{mid}}$. The peak-to-center flux ratios of the $^{13}$CO transitions ($J$=3-2 and $J$=2-1) show the strongest sensitivity to  $\beta_{\mathrm{mid}}$ or $\mathrm{v}_{\mathrm{turb}}$.  A low peak-to-trough ratio indicates a high level of turbulence. The $^{12}$CO  lines are emitted in the region where $\mathrm{v}_{\rm turb}$/$c_{\rm s}$ reaches its maximum. The optically thinner $^{13}$CO lines are emitted in the layers closer to the midplane where $\mathrm{v}_{\rm turb}$/$c_{\rm s}$ is lower.
The peak flux is mostly emitted in the outer disk, where the C$^{18}$O lines can be optically thin. Therefore, the peak fluxes for the C$^{18}$O lines are low and the peak-to-trough ratios for the C$^{18}$O lines are less sensitive to the turbulence broadening than the optically thick $^{12}$CO and $^{13}$CO lines.

One possible parameter that influences the peak-to-trough ratios for the $^{13}$CO lines is the abundance of $^{13}$CO compared to $^{12}$CO.  The lower-right panel in Fig.~\ref{fig_CO_vturb_stat} shows the effect of a global $^{12}$CO/$^{13}$CO ratio of 150 instead of the standard value of 70 \citep{Henkel1994LNP...439...72H,Wilson1999RPPh...62..143W} in the Interstellar Medium (ISM). A higher value for $^{12}$CO/$^{13}$CO in Young Stellar Objects (YSOs) environments compared to the ISM value is consistent with the observations of the $^{12}$CO/$^{13}$CO ratio towards a number of YSOs by \citet{Smith2015ApJ...813..120S}, who found a range of 86 to 165. The higher value for $^{12}$CO/$^{13}$CO may reflect the less efficient $^{13}$CO self-shielding against photodissociation compared to $^{12}$CO \citep{Miotello2016A&A...594A..85M}. Other parameters such as the disk inclination or the central star mass affect the value of the peak-to-trough ratios. 

% ----------------
\section{Discussion}\label{discussion}

Gas and dust in passive disks are heated by the conversion of the
stellar radiation (from the X-ray to the IR) to thermal  
energies
%Kamp2004ApJ...615..991K}
%
\citep{Jonkheid2004A&A...428..511J,Bergin2007prpl.conf..751B,Woitke2009A&A...501..383W,Bruderer2012A&A...541A..91B}. Passive disks show an increase of the temperatures from the midplane to the disk surfaces because of the extinction by dust grains for the UV and optical photons and by the gas for the X-ray photons. 
The warm layers reach temperatures of a few hundred to a few thousand Kelvin. In the inner 10-20~au, the turbulence is low due to the low abundance of charge species. Mass accretion has to occur in the low gas densities, hot and relatively ionized upper disk layers above the dead zone
to compensate for the decrease of mass accretion in the midplane. 
This result contrasts with vertical isothermal and relatively cool disk models where even
extremely high values of $\alpha_{\rm eff}$ are not enough to compensate for the
low accretion in the midplane \citep{Perez-Becker2011ApJ...735....8P}.
Our modelling supports the idea put forward by
\citet{Gammie1996ApJ...457..355G} that accretion would occur mainly through the active surface layers.

Beyond 10-20~au in the midplane, the ionization is high enough such that the MRI-turbulence
can fully develop. Accretion heating becomes important and since the
density is high, gas-grain thermal accommodation is the main source of
gas cooling. The dust grains are thus indirectly partly viscously heated. The 
rate of gas-grain accommodation depends on the density-squared and
thus will become inefficient at large radii. The dust grains act as a thermostat for the gas through the gas-dust thermal accommodation. 
When the gas is heated by the turbulence viscosity, the dust grains receive extra energy from the gas and radiate it. When there  is no turbulence heating, the dust grains transfer some of their energy to the gas.

The value of $\alpha_{\rm eff}$ reaches a maximum of 0.1, which translates to a turbulent width of 0.4 times the sound speed with values between 0.2 and 0.4 $c_{\rm s}$ for a $\beta_{\rm mid}$=10$^{4}$ in the midplane. The computed turbulent line widths assuming $\beta_{\rm mid}$=10$^{4}$ are consistent with the limit found by \citet{Teague2016A&A...592A..49T} in the TW~Hya disk but not with the low value determined by \cite{Flaherty2018ApJ...856..117F}.  \citet{Flaherty2015ApJ...813...99F,Flaherty2017ApJ...843..150F} also found a low value in the \pd\ around the HerbigAe star HD~163296 ($\mathrm{v}_{\rm turb} <  0.03\ c_{\rm s}$). \cite{Simon2018ApJ...865...10S} found that a low ionization rate and a low magnetic field is needed to explain a weak turbulence in the outer regions of protoplanetary disks.

\cite{Teague2016A&A...592A..49T} found a nearly constant ratio between the turbulence width and the local sound speed of $\mathrm{v}_{\rm turb}=0.2\ c_{\rm s}$. The \object{TW Hya} disk is seen at a low inclination so that there is no possibility to use the line peak-to-trough ratio to constrain the turbulence. 

\citet{Flaherty2015ApJ...813...99F} observed a peak-to-trough value of $\sim$2.0 for the $^{12}$CO $J$=3-2 line. In our MRI and fixed $\mathrm{v}_{\rm turb}$ models, the peak-to-trough has a maximum value of 1.9 for the HerbigAe disk model with 45$\degree$ inclination. 
In the ALMA observations presented by \citet{Flaherty2015ApJ...813...99F}, the $^{12}$CO $J$=3-2  and $J$=2-1 profiles show larger peak-to-trough ratios than the $^{13}$CO profiles (see the right panels of Fig.~\ref{fig_CO_vturb_stat}) in contrast to our model results. A higher $^{12}$CO/$^{13}$CO ratio can decrease slightly the peak-to-trough value, which is still not sufficient to explain the observed value (see the lower-right panel of Fig.~\ref{fig_CO_vturb_stat}).
 
An alternative to using high-quality CO lines, the ionization structure of \pd s can be constrained by matching emission lines from molecular ions. \cite{Cleeves2015} argued using emission lines from HCO$^+$ and N$_2$H$^+$ that the TW~Hya disk has a dead-zone extending up to 50--65 au. Their disk model requires a very low cosmic ray ionization rate lower than 10$^{-19}$ s$^{-1}$. Their dead-zone is defined by the criterion $R_{\mathrm{EM}} \leq 3300$, where $R_{\mathrm{EM}}$ is the magnetic Reynolds number \citep{Balbus2001ApJ...552..235B}. 
HCO$^+$ and N$_2$H$^+$ are molecular ions whose theoretical abundances depend sensitively on the treatment of ionization and recombination in the gas \citep{Rab2017A&A...603A..96R} as well as of the $N_2$ self-shielding for N$_2$H$^+$ \cite{Visser2018arXiv180202841V}. In disks models, molecular ions are confined to the upper disk molecular layers. Using molecules other than CO would require a detailed understanding of the chemistry of those species in disks. \cite{Kamp2017A&A...607A..41K} showed that the N$_2$H$^+$ abundance and line fluxes in disks vary significantly when different standard gas-phase chemical networks are used. Finally, although the HCO$^+$ emissions may be strong, N$_2$H$^+$ emissions are faint in disks.
\citet{Guilloteau2012A&A...548A..70G} used CS as a tracer of turbulence because CS is much heavier than CO such that its thermal width is smaller. The low signal-to-noise data allowed them to constrain turbulence to be between 0.4 $c_s$ an 0.5 $c_s$.
ALMA observations of CS in the \object{DM Tau} disk reveals a non-thermal line width between 0.05 and 0.15 km s$^{-1}$ at 300 au \citep{Semenov2018arXiv180607707S}. Sulfur chemistry is complex and CS lines are weak even in massive disks \cite{Dutrey2011A&A...535A.104D}.

The magnetic Reynolds number and the Elsasser Ohmic number are then
related by
\begin{equation}
R_{\mathrm{EM}} = \left(\frac{c_{\mathrm{s}}}{\mathrm{v_A}}\right)^2\Lambda_{\rm Ohm}
\end{equation}
assuming $\mathrm{c}_{\mathrm{s}} = h \Omega$. 
\cite{Flaherty2017ApJ...843..150F} used DCO$^+$ lines in addition to CO and C$^{18}$O to constrain $\alpha$, and found that the limits given from the three tracers are consistent with each other. When turbulence is low, it becomes difficult to distinguish between the CO line profiles of MRI models with $\beta_{\mathrm{mid}}<10^5$, i.e. that CO line profiles can be used to constrain $\alpha$ down to $\sim 5 \times 10^{-3}$. The CO abundances can also vary over the lifetime of the disks \citep{Yu2017arXiv171003657Y}, which may affect the CO line profile. 

The CO rotational lines that are seen with large beams} in the sub-millimeter domain probe a large disk volume but may not be sensitive to the location in the \pd s with the highest $\mathrm{v}_{\rm turb}/c_{\rm s}$ ratio, although high spatial  resolution CO maps can be obtained with ALMA.

Alternatively, the CO fundamental rovibrational line ($\mathrm{v}$=1--0, $\Delta J=\pm 1$) emission area matches the location of the maximum of the $\mathrm{v}_{\rm turb}/c_{\rm s}$ ratio (see the lower panels in Fig.~\ref{fig_disk_results_alpha}).

The knowledge of the gas and dust temperature structure is central to the estimate of the line thermal broadening. Fig.~\ref{fig_disk_temperatures} shows that the disk dust and gas temperature structures may be complex. In particular, the gas temperature structure depends on $\beta_{\rm mag}$. Because of the dead-zone and the efficiency of the gas-dust thermal accommodation at high densities, the inner disk midplane gas temperature does not change much with accretion heating.

% ----------------
The charge in \pd s\  depends on many parameters that were fixed in our models. In the UV-dominated regions, the electrons are provided by the ionization of neutral carbon atoms, while the abundance of the gas-phase metals play an important role in controlling the ionization fraction in the UV-shielded molecular region. Therefore, the ionization level in the UV-shielded region depends strongly on the gas-phase metal abundances. In the disk areas where $T_{\mathrm{dust}}$ is higher than the sublimation temperature of the main silicate dust grains, the gas phase is enriched by a large quantity of silicon, magnesium, iron, sodium, ..., reaching the stellar photospheric abundances. Those elements can still survive at high gas temperatures in form of metal oxides like SiO, MgO, FeO if UV is not present. But since the main UV opacity, i.e. the dust, is no more present, UV photons permeate the dust-free gas and can dissociate efficiently the oxides unless efficient high-temperature formation of those oxides is possible like for water vapor \citep{Thi2005A&A...438..557T}. If the gas chemistry is at equilibrium, one can use the Saha equation to compute the ionization fraction assuming that the collisional ionization controls the ionization fractions \citep{Desch2015ApJ...811..156D}. 
% ---------------------

Negative grains are formed by attachment of an electron or
charge-exchange with an anion (in our simple chemical network there is only one anion H$^-$). Electrons are first created by
ionization of H$_2$ into H$_2^+$ by an energetic ionization event 
(X-ray, Cosmic rays, radioactive decay), and subsequently
H$_3^+$. They can recombine either with an abundant atomic or
molecular ion or with a dust grains. Cations and negatively-charged
grains can interact to neutralize each other.

In the gas phase, the radiative recombinations of metallic ions with electrons are slow because the excess energy has to be radiated away. On the other hand, the excess energy in the ion recombination
on negatively-charged grain surfaces is transferred to the grain, making the recombination rates with grains much higher than with an ion. The grain surfaces act as a heat sink. This may result in neutral gas in the disk midplane and thus a much lower ionization fraction.

Many parameters can influence the ionization structure and hence the efficiency of MRI-driven turbulence in \pd s. The most important are not well constrained: $\beta_{\rm mag}$, the metallicity, the depletion of the metals into refractory solids, cosmic ray flux, and spectrum, stellar particle flux, grain properties such as the size distribution, settling and drift. The effect of those key parameters can compensate each other. For example, a strong magnetic field can compensate for a low cosmic ray flux. Likewise, a strong X-ray flux can compensate for a weak cosmic-ray flux. The direct grain ionization by absorption of X-ray photons was not considered in this paper and would the subject of a subsequent study.
 
% ------
Our disk model assumes dust settling, which can affect the UV penetration and hence the ionization structure. In our modelling of the dust gain distribution, the vertical settling is parametrized by a constant dust turbulence mixing parameter $\alpha_{\rm settle}$ of 0.01 throughout the disk \citep{Woitke2016A&A...586A.103W}. How this dust mixing parameter $\alpha_{\rm settle}$ relates to the effective turbulence parameter $\alpha_{\rm eff}$ is not clear. Detailed modelling of the dust vertical structure of \object{HL~Tau} suggests that the dust vertical mixing is weak with a value of a few 10$^{-4}$ for $\alpha_{\rm settle}$  \citep{Pinte2016ApJ...816...25P}. This corresponds to a disk model with $\beta_{\rm mid}\sim10^6$. 
%{\bf \cite{Riols2018arXiv180500458R} performed local detailed numerical simulations of the dust settling in a turbulent gas. They found that $\alpha_{\rm settle}$ can be very high 10$^{-3}$ - 10$^{-1}$. For high values of $\beta_{\mathrm{mag}}>$~4, they also confirmed that the 1D model of \citet{Dubrulle2005A&A...429....1D} is valid.}

 % -----
In our model, the value of $\alpha_{\rm eff}$ effects both the sound speed (because turbulence is an efficient heating agent and changes the cooling line transfer) and the
turbulence width. The change in the line cooling efficiency is small because the CO lines (one major gas coolant for the outer disk gas)
are highly optically thick. The contribution of the turbulence to the total line width remains small even for high values of $\alpha_{\rm eff}$. A fundamental difference between our models and the parametric disk models used to constrain the turbulence speed is that the value of $\alpha$ is assumed uniform in the parametric disk models.
Future studies are required to model consistently $\alpha_{\rm settle}$ and  $\alpha_{\rm eff}$,  as well as the changes in the disk density structure, in the disk chemistry, and in the grain properties (size, drift, ...) over the disk lifetime.
 
\section{Conclusions}\label{conclusions}

We have implemented a simple parametrized  magnetorotational instability (MRI) driven
turbulence model in the physico-chemical code {\sc ProDiMo} to constrain the gas heating by gas turbulence decay and the potential presence of a dead-zone. The strength of the turbulence in protoplanetary disks is computed consistently with the ohmic and ambipolar diffusion, whose resistivities depend on the abundance of the charge carriers  (free electrons, gas-phase atomic and molecular cations, PAHs, and dust grains). The gas and dust temperatures, chemistry, PAH and grain charges are computed self-consistently together with the continuum and line radiative transfer. The disk models include at the same time active and passive heating processes. The main conclusions are:
\begin{itemize}
\item The free electrons and gas-phase ions are the main charge carriers in most parts of the disk.
\item  The negatively-charged PAHs are the major charge carriers in the disk atmosphere above the PAH freeze-out zone.
\item The dust grains are the dominant negative charge carriers in the inner disk midplane where the ionization fraction is very low.
\item The Ohm resistivity governs the location of the dead-zone in the inner disk midplane.
\item The ambipolar diffusion resistivity does not affect much the efficiency of the MRI-driven turbulence because the density of ions is relatively high in the whole disk. The last two conclusions may change when other disks are considered.
\item In the inner 10-20~au, gas may accrete through the warm layer that sits above the dead-zone. 
\item The gas accretion heating dominates in the disk midplane outside the dead-zone with the dust thermal accommodation being the main gas cooling agent, while in the dead zone thermal accommodation is the main source of gas heating and the gas is cooled by molecular line emissions. In the disk upper layers, the gas is mainly heated by the photoejected electrons from PAHs and dust grains as well as by gas absorption of the dust grain infrared emissions and by H$_2$ formation. 
\item The signatures of the gas turbulence in the CO emission lines are weak and sophisticated modelling of the high signal-to-noise observations is required to derive meaningful estimates of the turbulence parameter $\alpha$. 
\item The CO line profiles from the MRI disk models differ from the profiles of disk models with a single constant value of $\alpha$. Simultaneous fits to multiple transitions of CO lines and to other disk tracers would improve the reliability of the estimate of $\beta_{\mathrm{mid}}$ (or ${\mathrm{v}_{\mathrm{turb}}}$), although the line profiles are shaped by many disk properties that are currently not well constrained.
\end{itemize}
Future works include the study of the influence of several model parameters on the location of the dead-zone. The important parameters include the cosmic ray flux attenuation in the disk, the stellar particle flux, the size and mass of the disk, the stellar properties. Another venue is the study of the disk evolution and the changes in the value of $\alpha_{\mathrm{eff}}$ with time \citep{Bai2016ApJ...821...80B}. Since our model can generate different types of observables (molecular line maps and  continuum emission maps), one can use our model outputs as test cases to improve the reliability of inversion models used to derive $\alpha$ from the observations. Finally we will also explore emission lines from species other than CO as turbulence tracers.

% ---------
\begin{acknowledgements}
  We thank ANR (contract ANR-07-BLAN-0221) and PNPS of CNRS/INSU,
  France for support. IK, WFT and PW acknowledge funding from the EU
  FP7-2011 under Grant Agreement nr. 284405. We acknowledge
  discussions with Ch. Pinte and F. M\'{e}nard. 
  \end{acknowledgements}
%We thank the referee and the editor for the useful comments that helped improve the paper.
\bibliographystyle{aa} % style aa.bst 2
\bibliography{mri_bib,surface_chemistry,disk_observations,references,apex_paper_bib,bringing_water_paper}
  
% ------------------------------------------------------------------
\begin{appendix}

\section{Comparing the contributions of electrons, ions and dust grains to the Ohm conductivity}

The contribution of the electrons to the Ohm conductivity is
\begin{equation}
\sigma_{\mathrm{e,O}}\simeq \frac{e^2}{m_{\mathrm{e}}\langle
  \sigma\mathbf{v}
  \rangle_{\mathrm{e,n}}}\chi(\mathrm{e}),%=\frac{e^2}{m_{\mathrm{e}}\langle
 % \sigma\mathbf{v}
  %\rangle_{\mathrm{e,n}}}\left(\eta_{\mathrm{ion}}+\frac{Z_{\mathrm{d}}n_{\mathrm{d}}}{n_{\mathrm{<H>}}}\right),
\end{equation}
where $\chi($e$)$ is the free electron fractional abundance.
%$\eta_{\mathrm{ion}}$ the total ionization fraction ($\chi(\mathrm{e})=\eta_{\mathrm{ion}}+Z_{\mathrm{d}}n_{\mathrm{d}}/n_{\mathrm{<H>}}$). 
The equation can be rewritten as
\begin{equation}
\sigma_{\mathrm{e,O}}\simeq 3 \times 10^{16} \left(\frac{T}{100}\right)^{-1/2} \chi(\mathrm{e})\ [\text{s}^{-1}].
\end{equation}
One can make a comparison between the contribution of each charged species to
the Ohm Elsasser number. For instance, the ratio between the
contribution from the free electrons to the charged dust grains is of interest:
\begin{equation}
\frac{\sigma_{\mathrm{e,O}}}{\sigma_{\mathrm{dust,O}}}=\left(\frac{Z_{\mathrm{e}}}{Z_{\mathrm{d}}}\right)^2\left(\frac{n_{\mathrm{e}}}{n_{\mathrm{d}}}\right)\left(\frac{m_{\mathrm{d}}\gamma_{\mathrm{d}}}{m_{\mathrm{e}}\gamma_{\mathrm{e}}}\right).
\end{equation}
For the electrons,
\begin{equation}
m_{\mathrm{e}}\gamma_{\mathrm{e}}\simeq 8.28\times10^{-9}T_{100}^{1/2}\frac{m_{\mathrm{e}}}{m_{\mathrm{n}}},
\end{equation}
($m_{\mathrm{e}} \ll m_{\mathrm{n}}$) and for the dust grains
\begin{equation}
m_{\mathrm{d}}\gamma_{\mathrm{d}}\simeq 3\times 10^{-3}a_{\mu{\mathrm{m}}}^2T_{100}^{1/2},
\end{equation}
where the average mass of a grain is 
\begin{equation}
m_{\mathrm{d}} = (4/3)\pi \rho_{\mathrm{d}} a^3= 1.256\times10^{-11} \left(\frac{\mathrm{\mu m^3}}{a^3}\right) \ \mathrm{grams}.
\end{equation}  
Since $m_{\mathrm{d}}>>m_{\mathrm{n}}$ we obtain
\begin{equation}
\frac{\sigma_{\mathrm{e,O}}}{\sigma_{\mathrm{dust,O}}}\simeq 3623.2 \left(\frac{1}{Z_{\mathrm{d}}}\right)^2\left(\frac{n_{\mathrm{e}}}{n_{\mathrm{d}}}\right)\left(\frac{m_{\mathrm{n}}}{m_{\mathrm{e}}}\right)a_{\mu{\mathrm{m}}}^2.
\end{equation}
The average number of dust grains in the disk is
\begin{equation}\label{eq_ndust}
n_{\mathrm{d}} = \frac{2.2 \mathrm{amu}\ n_{\mathrm{<H>}}}{(4/3)\pi \rho_{\mathrm{d}}\ a^3  gd}, 
\end{equation}
\begin{equation}
n_{\mathrm{d}} \approx 2.907\times 10^{-15} n_{\mathrm{<H>}} \left(\frac{\mathrm{\mu m^3}}{ a^3}\right)\left(\frac{100}{gd}\right),\label{nd_formula}
\end{equation}
where $gd$ is the gas-to-dust mass ratio. We have assumed a silicate mass density of 3.0 g cm$^{-3}$. 
Replacing the dust grain number density by the formula above, one obtains
\begin{equation}
\frac{\sigma_{\mathrm{e,O}}}{\sigma_{\mathrm{dust,O}}}\simeq
\frac{10^{18}}{Z_{\mathrm{d}}^2}\left(\frac{n_{\mathrm{e}}}{n_{\mathrm{<H>}}}\right)
\left(\frac{m_{\mathrm{H}}}{m_{\mathrm{e}}}\right)\left(\frac{gd}{100}\right)a_{\mu{\mathrm{m}}}^5
\end{equation}
For a gas-to-dust mass ratio of 100,
\begin{equation}
\frac{\sigma_{\mathrm{e,O}}}{\sigma_{\mathrm{dust,O}}}\simeq
\frac{8.36\times10^{20}}{Z_{\mathrm{d}}^2}\chi (\mathrm{e}) a_{\mu{\mathrm{m}}}^5,\label{Ohm_conductiviity_ratio1}
\end{equation}
where $\chi(e)$ is the relative abundance of free electrons.
If we assume that grains have charge 0 $>$ Z$_{\mathrm{d}} \geq$ -Z$_{\mathrm{max}}$ and using
the maximum possible negative-charge grains (\ref{eq_Zlimit}), the ratio
becomes
\begin{equation}
\frac{\sigma_{\mathrm{e,O}}}{\sigma_{\mathrm{dust,O}}}\geq
5.2\times10^{13} \chi (\mathrm{e}) a_{\mu{\mathrm{m}}}^3.
\end{equation}
Thus for micron-size dust grains, the free electron contribution
to the Ohm Elsasser number always dominates over the dust contribution down
to a free electron abundance of $\sim$1.9 $\times$ 10$^{-14}$. For 0.1 micron grains, the charged 
dust grain contribution to the Ohm Elsasser number becomes important for free electron abundances below
$\sim$1.9 $\times$ 10$^{-11}$.
If one assumes that the dust charge in the UV-obscured region can be approximated by formula \ref{Zestimate_noUV}, the Ohm conductivity ratio is
\begin{equation}
\frac{\sigma_{\mathrm{e,O}}}{\sigma_{\mathrm{dust,O}}}\simeq 1.58 \times 10^{20} \left(\frac{100\mathrm{K}}{T} \right)a_{\mu \mathrm{m}}^4 \chi(\mathrm{e}).
\end{equation}
Note that the ratio is temperature-dependent.

Likewise, one can make a comparison between the contribution of the electrons compared
to that of the ions
\begin{equation}
\frac{\sigma_{\mathrm{e,O}}}{\sigma_{\mathrm{ions,O}}}=\left(\frac{Z_{\mathrm{e}}}{Z_{\mathrm{i}}}\right)^2\left(\frac{n_{\mathrm{e}}}{n_{\mathrm{i}}}\right)\left(\frac{m_{\mathrm{i}}\gamma_{\mathrm{i}}}{m_{\mathrm{e}}\gamma_{\mathrm{e}}}\right)
\end{equation}
Assuming singly-charged ions and that most of the negative charges are
carried by the electrons,
 \begin{equation}
\frac{\sigma_{\mathrm{e,O}}}{\sigma_{\mathrm{ions,O}}}\simeq\frac{m_{\mathrm{i}}\gamma_{\mathrm{i}}}{m_{\mathrm{e}}\gamma_{\mathrm{e}}}\simeq
0.3  \left(\frac{100}{T}\right)^{1/2} \frac{m_{\mathrm{n}}}{m_{\mathrm{e}}} \simeq 920  \left(\frac{100}{T}\right)^{1/2}
\end{equation}
where
\begin{equation}
m_{\mathrm{i}}\gamma_{\mathrm{i}}\simeq 1.9\times10^{-9}
\end{equation}
assuming that $m_{\mathrm{i}}>>m_{\mathrm{n}}$.  In reality, a fraction $\chi(e)$ of the negative charges will be electrons locked in grains.

In UV-dominated regions at disk surfaces, the dust grains will be positively charged with the maximum possible positive charge set by formula
\ref{eq_Zmax}. Formula \ref{Ohm_conductiviity_ratio1} reads
\begin{equation}
\frac{\sigma_{\mathrm{e,O}}}{\sigma_{\mathrm{dust,O}}}\simeq 5.5 \times 10^{13} a_{\mu \mathrm{m}}^4 \chi(\mathrm{e}).
\end{equation}

\section{Effective turbulence coefficient}

\begin{figure*}[!ht]
\centering  
\resizebox{\hsize}{!}{
\includegraphics[angle=0,width=10cm,height=7cm,trim=55 55 60 50, clip]{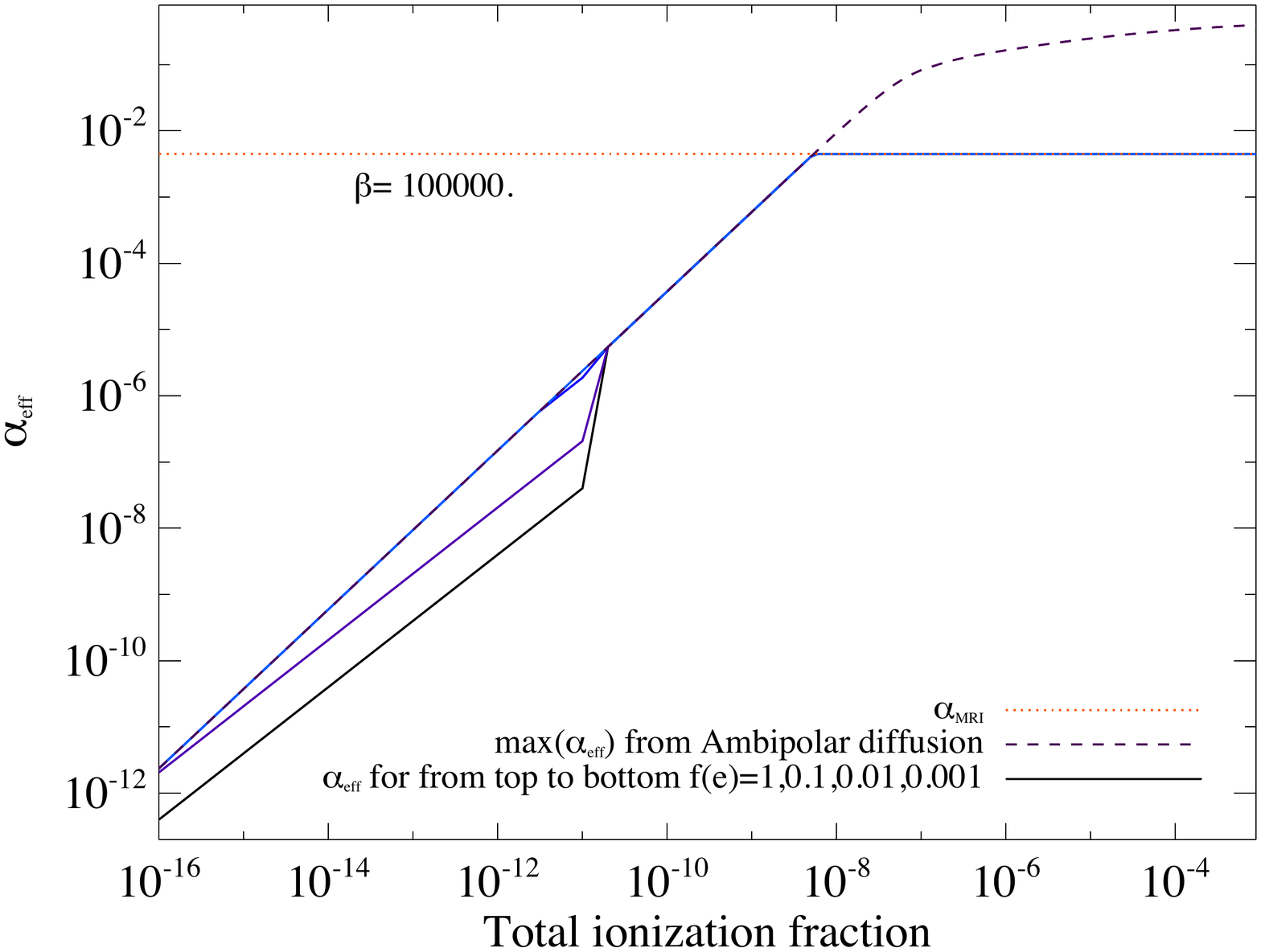}
\includegraphics[angle=0,width=10cm,height=7cm,trim=55 55 60 50, clip]{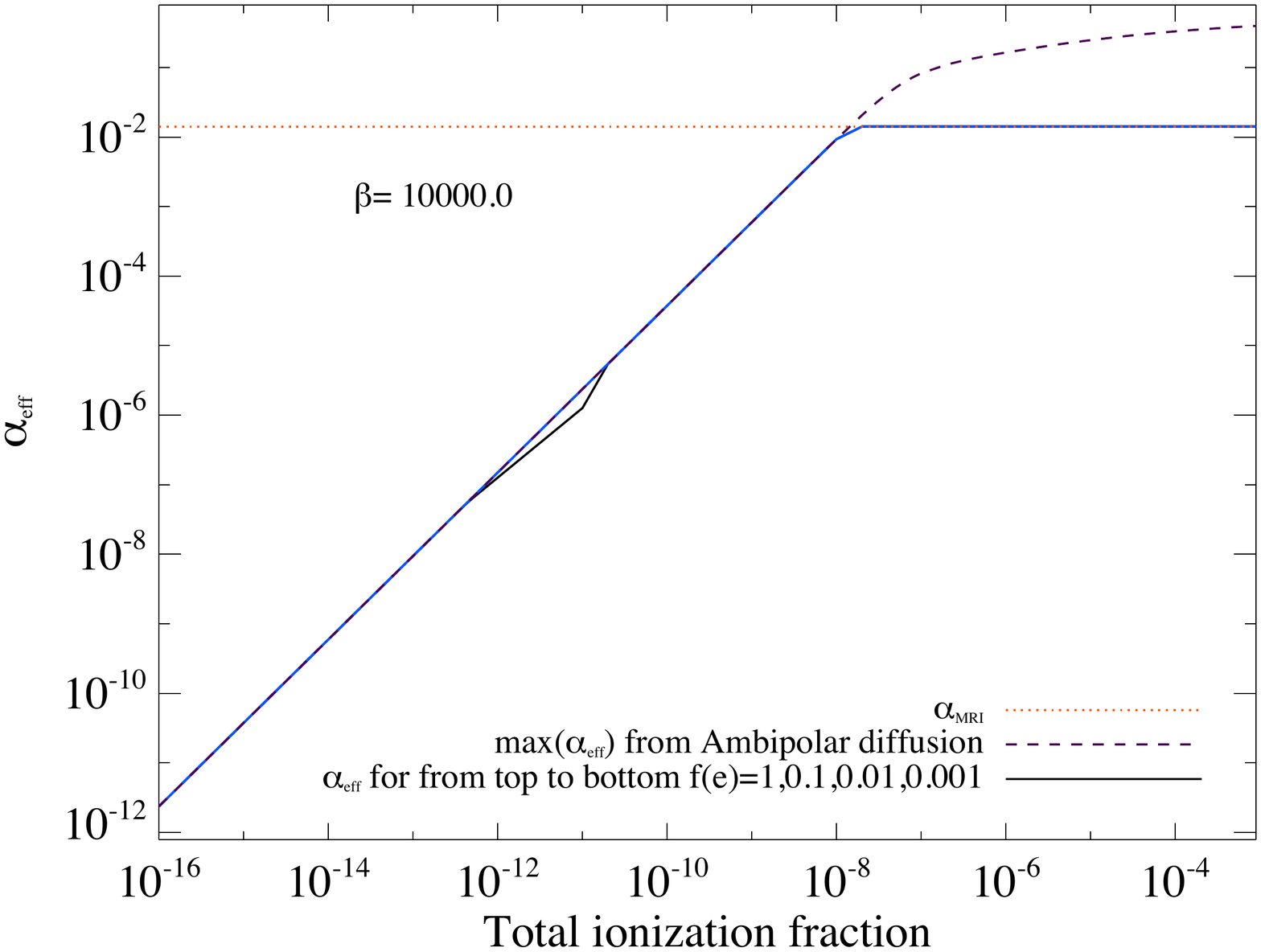}}
\resizebox{\hsize}{!}{
\includegraphics[angle=0,width=10cm,height=7cm,trim=55 55 60 50, clip]{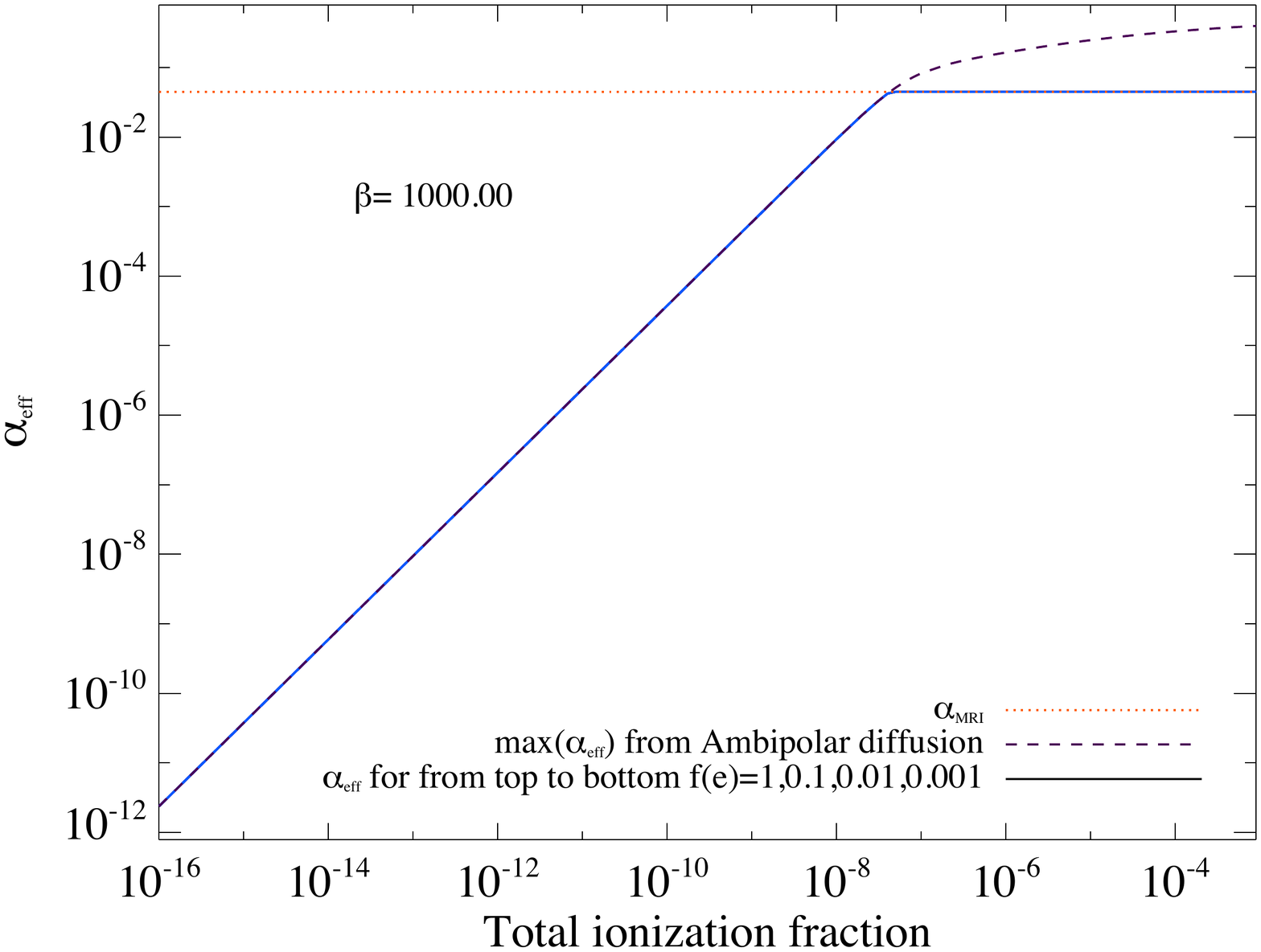}
\includegraphics[angle=0,width=10cm,height=7cm,trim=55 55 60 50, clip]{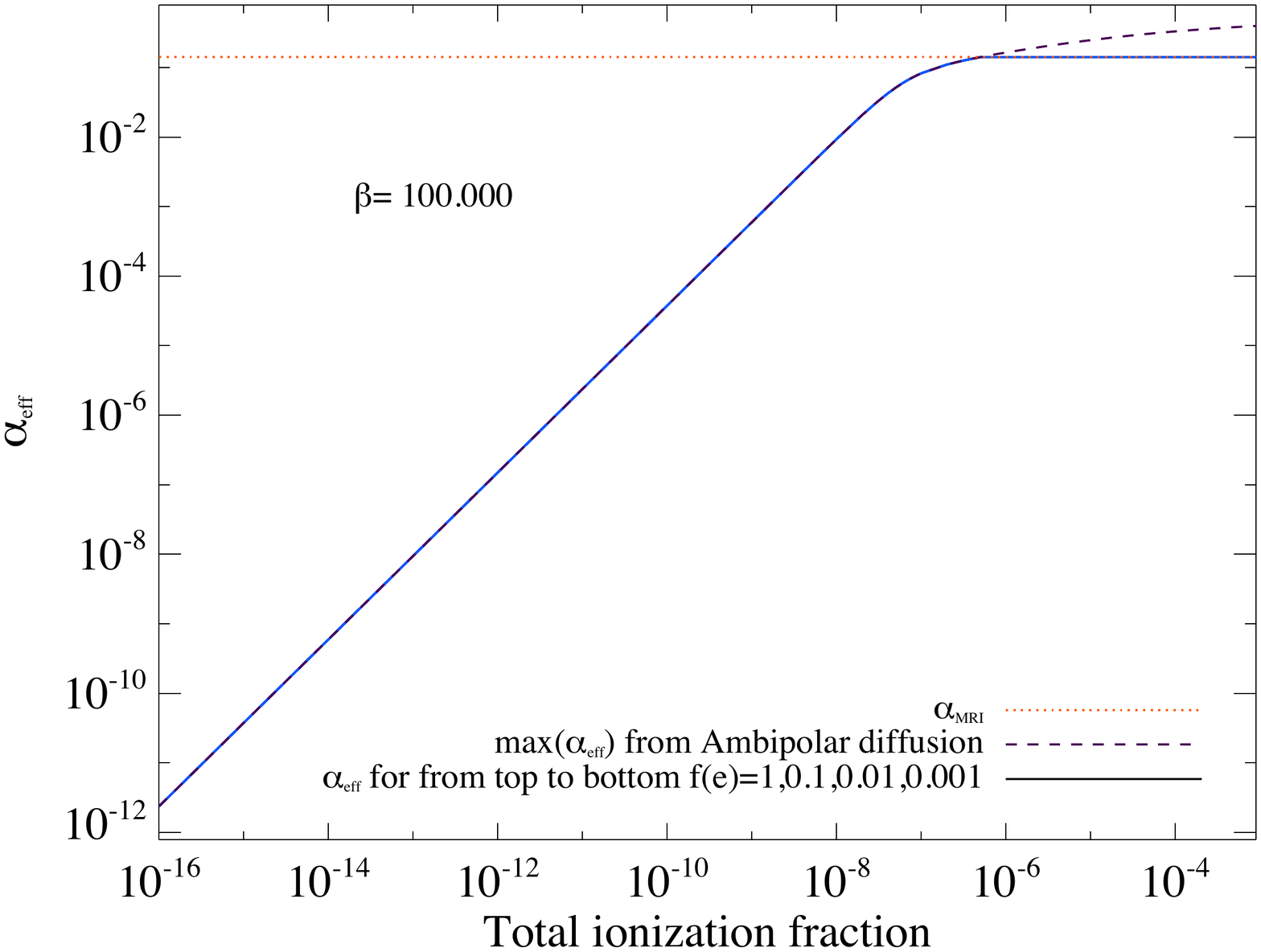}}
\caption{Effective turbulence coefficient $\alpha_{\mathrm{eft}}$ as function of the total ionization fraction for a gas of density $n_{\mathrm{<H>}}$=10$^{10}$ cm$^{-3}$ at $T$=100~K and four values of the $\beta_{\mathrm{mag}}$ parameters (10$^4$, 10$^3$, 10$^2$, 10). The values for $\alpha_{\mathrm{ideal}}$ as plotted in red short-dashed-lines have been derived assuming $\delta$=0.5. The limits imposed by the am bipolar diffusion resistivity are shown in dashed-lines.}
\label{fig_ohm_Am}          
\end{figure*}

Fig.~\ref{fig_ohm_Am} shows the effective turbulence coefficient $\alpha_{\mathrm eff}$ as function of the total ionization fraction for a relatively dense gas at density 10$^{10}$ cm$^{-3}$ and temperature of
100~K. The four panels correspond to different values of $\beta_{\mathrm{mag}}$ (10$^{4}$, 10$^{3}$, 10$^{2}$, 10).

\section{Implementing the combined multi-charge grain physics and gas chemistry}\label{numerical_implementation}

We adopted a simplified implementation of the chemical reaction
network.  In principle, reactions (charge exchange, recombination, ...) have to be included explicitly for each grain
charge (gr$^{Z_{\mathrm{d}}}$= gr$^{Z_{\mathrm{min}}}$, ..., gr$^{3-}$, gr$^{2-}$, gr$^{-1}$, gr$^0$, gr$^{+}$, gr$^{2+}$, ..., gr$^{Z_{\mathrm{max}}}$). However, with $Z_{\mathrm{max}}>10$, the computation cost becomes prohibitive.
Instead, the abundance of grains with a charge $Z_{\mathrm{d}}$ is defined by
$[gr^{Z_{\mathrm{d}}}]= n_{\mathrm{d}} f(Z_{\mathrm{d}})$ where
$Z_{\mathrm{d}}$ is the charge of the grain (integer) and $f(Z_{\mathrm{d}})$ is the normalized discrete distribution (fractional abundance) of grains with 
charge $Z_{\mathrm{d}}$.  The values for $Z_{\mathrm{min}}$ and $Z_{\mathrm{max}}$ are chosen such that $f(Z_{\mathrm{min}})=0$ and $f(Z_{\mathrm{max}})=0$.

Instead of solving for each individual charge, which will increase the number of differential equations in the system by the number of charges, we only considered three grain pseudo charge-species $Z^-$, $Z$, and $Z^+$ with the abundance [$Z^-$], [$Z$], and [$Z^+$] respectively. The gas-phase species react with one of these three pseudo-species only.

The grain charge distribution $f(Z_{\mathrm{d}})$ is
\begin{equation}
\sum_{Z_{\mathrm{min}}}^{Z_{\mathrm{max}}} f(Z_{\mathrm{d}}) =  \sum_{Z_{\mathrm{min}}}^{-1} f(Z_{\mathrm{d}}) + f(0) + \sum^{Z_{\mathrm{max}}}_{1} f(Z_{\mathrm{d}}) = 1.
\end{equation}
The total abundance is
\begin{equation}
\sum_{Z_{\mathrm{min}}}^{Z_{\mathrm{max}}} [gr^{Z_{\mathrm{d}}}] = n_{\mathrm{d}}.
\end{equation}
We defined the pseudo-charge species ($Z^- , Z^- , Z$) such that  
\begin{equation}
[Z^-]+[Z]+[Z^+]=Z_{\mathrm{max}}n_{\mathrm{d}},
\end{equation} 
where we have split the grain population into three categories: the negatively-charged grains
\begin{equation}
[Z^-]=-n_{\mathrm{d}}\sum_{Z_{\mathrm{min}}}^{-1} Z_{\mathrm{d}} f(Z_{\mathrm{d}}),
\end{equation}
the positively-charged grains
\begin{equation}
[Z^+]=n_{\mathrm{d}} \sum^{Z_{\mathrm{max}}}_{1} Z_{\mathrm{d}} f(Z_{\mathrm{d}}).
\end{equation} 
The neutral grains are modelled by the pseudo-species $Z$ of abundance
\begin{equation}
[Z] = n_{\mathrm{d}}Z_{\mathrm{max}}-[Z^-]-[Z^+]
\end{equation}
The average grain charge is
\begin{equation}
<Z_{\mathrm{d}}> = \sum^{Z_{\mathrm{d}}=Z_{\mathrm{max}}}_{Z_{\mathrm{d}}=Z_{\mathrm{min}}} Z_{\mathrm{d}}f(Z_{\mathrm{d}})=([Z^+]-[Z^-])/n_{\mathrm{d}}
\end{equation}
%The fractional abundance $f(Z_{\mathrm{d}})$ is
%normalized ($Z_{\mathrm{min}}=-Z_{\mathrm{max}}$).  
The pseudo-neutral grain species $Z$ includes the actual neutral grains and the positively-charged grains exchanging charges with cations of index $j$ to become more positively-charged and absorbing a photon and ejecting an electron:
\begin{equation}
Z + A^+ \rightarrow Z^+ + A.
\end{equation}
Considering only this reaction, the chemical reaction differential equation is
\begin{equation}
\frac{d[A]}{dt} =
\sum_{0}^{Z_{\mathrm{max}}-1} k_{\mathrm{gr,ion}}^j(Z_{\mathrm{d}})[gr^{Z_{\mathrm{d}}}][A^+] = \sum_{0}^{Z_{\mathrm{max}}-1} k_{\mathrm{gr,ion}}^j(Z_{\mathrm{d}})n_{\mathrm{d}} f(Z_{\mathrm{d}})[A^+].
\end{equation}
Another way to write the reaction differential equation is
\begin{equation}
\frac{d[A]}{dt}=k_{\mathrm{ex,0}}^j[Z][A^+],
\end{equation}
where the positively-grain-charge-averaged charge-exchange rate $k_{\mathrm{ex,i}}^0 $ is
\begin{equation}
k_{\mathrm{ex,0}}^j = \frac{n_{\mathrm{d}}\sum_{0}^{Z_{\mathrm{max}}-1} k_{\mathrm{gr,ion}}^j(Z_{\mathrm{d}})f(Z_{\mathrm{d}})}{n_{\mathrm{d}}Z_{\mathrm{max}}-[Z^-]-[Z^+]},
\end{equation}
or
\begin{equation}
k_{\mathrm{ex,0}}^j = \frac{\sum_{0}^{Z_{\mathrm{max}}-1} k_{\mathrm{gr,ion}}^j(Z_{\mathrm{d}})f(Z_{\mathrm{d}})}{Z_{\mathrm{max}}+\sum_{Z_{\mathrm{min}}}^{-1} Z_{\mathrm{d}} f(Z_{\mathrm{d}}) - \sum_{1}^{Z_{\mathrm{max}}} Z_{\mathrm{d}} f(Z_{\mathrm{d}})}.
\end{equation}
The absorption of a neutral grain or of a positively-charged grain leads to the ejection an electron:
\begin{equation}
Z + h\nu \rightarrow Z^+ + e^-
\end{equation}
The rate for the pseudo species $Z$ is $k_{\mathrm{pe}}^0$, whose derivation follows 
\begin{equation}
\sum_{0}^{Z_{\mathrm{max}}} k_{\mathrm{pe}}(Z_{\mathrm{d}})[gr^{Z_{\mathrm{d}}}] = \sum_{0}^{Z_{\mathrm{max}}-1} k_{\mathrm{pe}}(Z_{\mathrm{d}})n_{\mathrm{d}} f(Z_{\mathrm{d}})=k_{\mathrm{pe}}^0[Z].
\end{equation}
We obtain
\begin{equation}
k_{\mathrm{pe}}^0 = \frac{n_{\mathrm{d}}\sum_{0}^{Z_{\mathrm{max}}-1} k_{\mathrm{pe}}(Z_{\mathrm{d}})f(Z_{\mathrm{d}})}{n_{\mathrm{d}}Z_{\mathrm{max}}-[Z^-]-[Z^+]},
\end{equation}
\begin{equation}
k_{\mathrm{pe}}^0 = \frac{\sum_{0}^{Z_{\mathrm{max}}-1} k_{\mathrm{pe}}(Z_{\mathrm{d}})f(Z_{\mathrm{d}})}{Z_{\mathrm{max}}+\sum_{Z_{\mathrm{min}}}^{-1} Z_{\mathrm{d}} f(Z_{\mathrm{d}}) - \sum_{1}^{Z_{\mathrm{max}}} Z_{\mathrm{d}} f(Z_{\mathrm{d}})}.
\end{equation}
Likewise, the pseudo-neutral grains model the thermal emission of electrons actual neutral and positively-charged grains:
\begin{equation}
Z \rightarrow Z^+ + e^-.
\end{equation} 
The rate of thermionic emission is
\begin{equation}
k_{\mathrm{th}}^0 = \frac{\sum_{0}^{Z_{\mathrm{max}}-1} k_{\mathrm{th}}(Z_{\mathrm{d}})f(Z_{\mathrm{d}})}{Z_{\mathrm{max}}+\sum_{Z_{\mathrm{min}}}^{-1} Z_{\mathrm{d}} f(Z_{\mathrm{d}}) - \sum_{1}^{Z_{\mathrm{max}}-1} Z_{\mathrm{d}} f(Z_{\mathrm{d}})}.
\end{equation}
The pseudo-neutral species $Z$ is also being used to model electron attachment reactions
\begin{equation}
Z + e^- \rightarrow Z^-,
\end{equation}
with the rate
\begin{equation}
k_{\mathrm{e,n}}[gr^{0}]+\sum_{n< 0}^{Z_{\mathrm{min}}+1} k_{\mathrm{e,-}}(n)[gr^{n+}]=k^{0}_{\mathrm{e}}[Z],
\end{equation}
where
\begin{equation}
k^{0}_{\mathrm{e}}=\frac{k_{\mathrm{e,n}}f(0)+\sum_{Z_{\mathrm{min}}+1}^{-1} k_{\mathrm{e,-}}(Z_{\mathrm{d}}) f(Z_{\mathrm{d}})}{Z_{\mathrm{max}}+\sum_{Z_{\mathrm{min}}+1}^{-1} Z_{\mathrm{d}} f(Z_{\mathrm{d}}) - \sum_{1}^{Z_{\mathrm{max}}} Z_{\mathrm{d}} f(Z_{\mathrm{d}})}.
\end{equation}
The electron recombinations of positively-charged grains are modelled by the pseudo-reaction
\begin{equation}
Z^+ + e^- \rightarrow Z.
\end{equation}
Consider the case of the electron recombination with a grain with $n$ positive charges and rate $k_{\mathrm{re}}(n)$, the sum of all the
recombination chemical rate reactions reads
\begin{equation}
\sum_{(Z_{\mathrm{d}}>1} k_{\mathrm{e,+}}((Z_{\mathrm{d}})[gr^{Z_{\mathrm{d}}}]=\sum^{Z_{\mathrm{max}}}_{1} n_{\mathrm{d}}k_{\mathrm{e,+}}(Z_{\mathrm{d}})f(Z_{\mathrm{d}})=k_{\mathrm{e}}^{+}[Z^+]
\end{equation}
We can derive the rate $k^{+}_{\mathrm{e}}$:
\begin{equation}
k^{+}_{\mathrm{e}}=\frac{\sum^{Z_{\mathrm{max}}}_{1} k_{\mathrm{e,+}}(Z_{\mathrm{d}}) f(Z_{\mathrm{d}})}{\sum^{Z_{\mathrm{max}}}_{1} Z_{\mathrm{d}}f(Z_{\mathrm{d}})}.
\end{equation}
The pseudo-negative charge species $Z^-$ models the reactions leading to neutralize the ions A of index $j$:
\begin{equation}
\mathrm{gr}^{\mathrm{Z_d}} + A^+ \rightarrow \mathrm{gr}^{\mathrm{Z_d}+1}  + A.
\end{equation}
The generic reaction is written as
\begin{equation}
Z^- + \mathrm{A}^+ \rightarrow Z + \mathrm{A}
\end{equation}
For example, the pseudo-reaction
\begin{equation}
Z^- + \mathrm{HCO}^+ \rightarrow Z + \mathrm{H} + \mathrm{CO}
\end{equation}
encompasses the dissociation recombination reactions of HCO$^+$ with all negatively-charged grains. 
\begin{equation}
\begin{array}{rcl}
\sum^{-1}_{Z_{\mathrm{min}}} k_{\mathrm{gr,ion}}^j(Z_{\mathrm{d}})[gr^{Z_{\mathrm{d}}}][\mathrm{A}^+] &=& \sum^{-1}_{Z_{\mathrm{min}}} k_{\mathrm{gr,ion}}^j(Z_{\mathrm{d}})n_{\mathrm{d}} f(Z_{\mathrm{d}})[\mathrm{A}^+]\\
& =& k_{\mathrm{ex,-}}^{j}[Z^-][\mathrm{A}^+] ,\\
\end{array}
\end{equation}
where 
\begin{equation}
k_{\mathrm{ex,-}}^j = -\frac{\sum^{-1}_{Z_{\mathrm{min}}} k_{\mathrm{gr,ion}}^j(Z_{\mathrm{d}})f(Z_{\mathrm{d}})}{\sum_{Z_{\mathrm{min}}}^{-1} Z_{\mathrm{d}}f(Z_{\mathrm{d}})}.
\end{equation} 
The pseudo-species $Z^-$ also participate to the photodetachment reaction
\begin{equation}
Z^- + h\nu \rightarrow Z + e^-,    
\end{equation}
with the rate  
\begin{equation}
k_{\mathrm{pe}}^- = -\frac{\sum^{-1}_{Z_{\mathrm{min}}} k_{\mathrm{pe}}(Z_{\mathrm{d}})f(Z_{\mathrm{d}})}{\sum_{Z_{\mathrm{min}}}^{-1} Z_{\mathrm{d}}f(Z_{\mathrm{d}})}.  
\end{equation}  
Finally $Z^-$ can emit electrons due to the thermionic effect  
\begin{equation}
Z^- \rightarrow Z + e^-
\end{equation}
\begin{equation}
k_{\mathrm{th}}^- = -\frac{\sum^{-1}_{Z_{\mathrm{min}}} k_{\mathrm{th}}(Z_{\mathrm{d}})f(Z_{\mathrm{d}})}{\sum_{Z_{\mathrm{min}}}^{-1} Z_{\mathrm{d}}f(Z_{\mathrm{d}})}.
\end{equation}
% ----------
\begin{table}[!th]
\caption{List of pseudo reactions involving the pseudo grain charge species $Z^-$, $Z$, and $Z^+$.}             
\label{PseudoReactions}      
\centering          
\begin{tabular}{lrcllll}     % 7 columns   
\hline
\hline       
\multicolumn{1}{c}{Nr.} & \multicolumn{3}{c}{Pseudo-reaction} & \multicolumn{1}{c}{min. $Z_{\mathrm{d}}$}& \multicolumn{1}{c}{max. $Z_{\mathrm{d}}$}&\multicolumn{1}{c}{Rate}\\
\hline
1 & $Z + A^+$                  & $\rightarrow$ &  $Z^+ + A$    & 0 &$Z_{\mathrm{max}}-1$ &$k_{\mathrm{ex,0}}^j$ \\
2 & $Z + \mathrm{h} \nu$ & $\rightarrow$ & $Z^+ + e^-$  & 0 &$Z_{\mathrm{max}}-1$&$k_{\mathrm{pe}}^0$ \\
3 & $Z$                              & $\rightarrow$ &  $Z^+ + e^-$  & 0 &$Z_{\mathrm{max}}-1$&$k_{\mathrm{th}}^0$ \\
4 & $Z +e^-$                    & $\rightarrow$ & $Z^-$             & $Z_{\mathrm{min}}+1$ & 0 &$k_{\mathrm{e}}^0$ \\
5 & $Z^+ +e^-$                    & $\rightarrow$ & $Z$             & 1 &$Z_{\mathrm{max}}$ &$k_{\mathrm{e}}^+$ \\
6& $Z^- + A^+$                  & $\rightarrow$ &  $Z + A$    & $Z_{\mathrm{min}}$ &-1&$k_{\mathrm{ex,-}}^j$ \\
7 & $Z^- + \mathrm{h} \nu$ & $\rightarrow$ & $Z$  & $Z_{\mathrm{min}}$&-1&$k_{\mathrm{pe}}^-$ \\
8 & $Z^-$                              & $\rightarrow$ &  $Z + e^-$  & $Z_{\mathrm{min}}$ &-1&$k_{\mathrm{th}}^-$ \\
\hline                    
\end{tabular}
\end{table}
% --------------
To illustrate that our concept is valid, we can assume first that the grains can have at most one positive or negative charge, in other words, $Z_{\mathrm{min}}$=-1 and $Z_{\mathrm{max}}$=1. Then the reactions in Table~\ref{PseudoReactions} correspond to the actual reactions as presented in Table \ref{RealReactions}.

\begin{table*}[th]
\caption{List of reactions involving the grain charge species $gr^-$, $gr$, and $gr^+$.}             
\label{RealReactions}      
\centering          
\begin{tabular}{lrcllll}     % 7 columns   
\hline
\hline       
\multicolumn{1}{c}{Nr.} & \multicolumn{3}{c}{Reaction} & \multicolumn{1}{c}{min. $Z_{\mathrm{d}}$}& \multicolumn{1}{c}{max. $Z_{\mathrm{d}}$}&\multicolumn{1}{c}{Rate}\\
\hline
1 & $gr + A^+$                  & $\rightarrow$ &  $gr^+ + A$    & 0 &0 &$k_{\mathrm{ex,0}}^j=k_{\mathrm{gr,ion}}^j(0)$ \\
2 & $gr + \mathrm{h} \nu$ & $\rightarrow$ & $gr^+ + e^-$  & 0 &0&$k_{\mathrm{pe}}^0=k_{\mathrm{pe}}(0)$ \\
3 & $gr$                              & $\rightarrow$ &  $gr^+ + e^-$  & 0 &0 &$k_{\mathrm{th}}^0=k_{\mathrm{th}(0)}$ \\
4 & $gr +e^-$                    & $\rightarrow$ & $gr^-$             & 0 & 0 &$k_{\mathrm{e}}^0=k_{\mathrm{e,n}}$ \\
5 & $gr^+ +e^-$                    & $\rightarrow$ & $gr$             & 1 &1 &$k_{\mathrm{e}}^+=k_{\mathrm{e,+}}(1)$ \\
6& $gr^- + A^+$                  & $\rightarrow$ &  $gr + A$    & -1 &-1&$k_{\mathrm{ex,-}}^j=k_{\mathrm{gr,ion}}^j(-1)$ \\
7 & $gr^- + \mathrm{h} \nu$ & $\rightarrow$ & $gr$  & -1 &-1&$k_{\mathrm{pe}}^-=k_{\mathrm{pe}}(-1)$ \\
8 & $gr^-$                              & $\rightarrow$ &  $gr + e^-$  & -1 &-1&$k_{\mathrm{th}}^-=k_{\mathrm{th}}(-1)$ \\
\hline                    
\end{tabular}
\end{table*}
% --------------
The definition of the rates above depends on the charge distribution $f$(Z$_{\mathrm{d}}$), which has to be known or solved simultaneously with the chemical reactions.
In a steady-state treatment, the chemistry and the grain charge distribution are solved alternatively. 

Using an estimate of the electron and ion abundances, the grain charge distribution $f(Z_{\mathrm{d}})$ is calculated by iteration. The average grain charge $<Z_{\mathrm{d}}>$ is determined by balancing the photoemission
(photoejection and photodetachment) rates and the thermionic rate with
the recombination and charge exchange rates. Then all the chemical rates between a gas-phase species and the pseudo-species $Z^-$, $Z$, and $Z^+$ can be derived. The electron abundance is computed from the global gas neutrality considering the charges on
gas-phase species, PAHs, and dust grains. The grain charge distribution $f(Z_{\mathrm{d}})$ is subsequently re-evaluated knowing the abundances of the cations and electrons.  This iterative process ensures that the grain charges and the gas-charge chemistry are
computed self-consistently. Especially the global neutrality of the medium is enforced. We assume that a dust grain can have charges from
$Z_{\mathrm{min}}$ and $Z_{\mathrm{max}}$, which are numerical free parameters. The results do not depend on the exact values for $Z_{\mathrm{min}}$ or $Z_{\mathrm{max}}$ provided
that $Z_{\mathrm{min}}\leq  Z_{\mathrm{d}} - 10$ and $Z_{\mathrm{max}}\geq  Z_{\mathrm{d}} + 10$. We assumed in this work for the numerical reason
that $Z_{\mathrm{min}}=-Z_{\mathrm{max}}$. If the chemistry is solved time-dependently, the grain charge fractional distribution has to be solved together with the chemical reaction differential equations. Solving a time-dependent grain charge distribution entails many challenges, whose resolution goes beyond the scoop of this paper.

\section{Disk model parameters}

The \pd\ parameters are summarized in Table~\ref{tab:refmodel}.
% --------------------------------------------------------------------------------------------
\begin{table}
\begin{center}
\caption{Model parameters, and values for the reference model. The meaning of the symbols are explained in \cite{Woitke2009A&A...501..383W,Woitke2016A&A...586A.103W}.}
\vspace*{-0.5mm}
\label{tab:refmodel}
\resizebox{88mm}{!}{\begin{tabular}{lcc}
\\[-3.8ex]
\toprule
 quantity & symbol & value\\
 \noalign{\smallskip}
\hline 
\noalign{\smallskip}
stellar mass                      & $M_{\star}$      & $0.7, 2.3\,M_\odot$\\
effective temperature             & $T_{\star}$      & $4000, 8600\,$K\\
stellar luminosity                & $L_{\star}$      & $1, 32\,L_\odot$\\
UV excess                         & $f_{\rm UV}$     & $0.01$\\
UV powerlaw index                 & $p_{\rm UV}$     & $1.3$\\
X-ray luminosity                  & $L_X$           & $10^{30}\rm erg/s$\\
X-ray emission temperature        & $T_{X,\rm fit}$  & $2\times10^7$\,K\\
\noalign{\smallskip}
\hline
\noalign{\smallskip}
strength of interstellar UV       & $\chi^{\rm ISM}$ & 1\\
strength of interstellar IR       & $\chi^{\rm ISM}_{\rm IR}$ & 0\\
cosmic ray H$_2$ ionisation rate  & $\zeta_{\rm CR}$   
                                  & $\!\!1.7\times 10^{-17}$~s$^{-1}\!\!\!$\\
\noalign{\smallskip}
\hline
\noalign{\smallskip}
disk mass$^{(1)}$                 & $M_{\rm disk}$   & $0.01\,M_\odot$\\
dust/gas mass ratio$^{(1)}$       & $\delta$        & 0.01\\
inner disk radius                 & $r_{\rm in}$     & 0.07\,AU\\
tapering-off radius               & $r_{\rm tap}$    & 100\,AU\\
column density power index        & $\epsilon$      & 1\\
reference scale height         & $H_{\rm g}(100\,{\rm AU})$ & 10\,AU\\
flaring power index               & $\beta$         & 1.15\\ 
\noalign{\smallskip}
\hline
\noalign{\smallskip}
minimum dust particle radius      & $a_{\rm min}$         & $0.05\,\mu$m\\
maximum dust particle radius      & $a_{\rm max}$         & $3\,$mm\\
dust size dist.\ power index      & $a_{\rm pow}$         & 3.5\\
turbulent mixing parameter        & $\alpha_{\rm settle}$ & 0.01\\
max.\ hollow volume ratio         & $\mathrm{v}_{\rm hollow}^{\rm max}$   & 80\%\\
dust composition                  & $\rm Mg_{0.7}Fe_{0.3}SiO_3$ & 60\%\\
(volume fractions)                & amorph.\,carbon            & 15\%\\
                                  & porosity                   & 25\%\\
\noalign{\smallskip}
\hline
\noalign{\smallskip}
PAH abundance rel. to ISM         & $f_{\rm PAH}$        & 0.01\\
chemical heating efficiency       & $\gamma^{\rm chem}$  & 0.2\\
\noalign{\smallskip}
\hline   
\noalign{\smallskip}
distance                          & $d$ & 140\,pc\\
disk inclination                  & $i$ & 45\degr\\
\noalign{\smallskip}
\hline   
\noalign{\smallskip}
$^{12}$C/$^{13}$C     & &  70  (150)  $^{(2)}$\\
$^{18}$O/$^{16}$O     & & 540\\
\noalign{\smallskip}
\hline   
\noalign{\smallskip}
Thermal over magnetic &  $\beta_{\rm mid}$ & 10$^{2}$--10${^6}$\\
 in the midplane & & \\
\bottomrule
\end{tabular}}
\end{center}
%\vspace*{-3mm}
\resizebox{90mm}{!}{
\begin{minipage}{100mm}{$^{(1)}$:  The chemical
      heating efficiency $\gamma^{\rm chem}$ is an efficiency by which
      exothermic chemical reactions are assumed to heat the gas. A detailed discussion on the disk parameters and their effects on the disk thermal and chemical structure can be found in \citet{Woitke2016A&A...586A.103W}. $^{(2)}$ 70 is the adopted standard value \citep{Henkel1994LNP...439...72H,Wilson1999RPPh...62..143W}. The choice of 100 corresponds to an average value between the observed ratios by \citet{Smith2015ApJ...813..120S} towards YSOs.}
%      \citep[][see Appendix A.8 therein]{Woitke2011} for details.}
\end{minipage}}
%\vspace*{-1mm}
\end{table}  

\section{Disk vertical component of the magnetic field}\label{Bfield}

\begin{figure*}[!htbp] 
  \centering
  \includegraphics[angle=0,width=8cm,height=8cm,trim=50 80  80 300, clip]{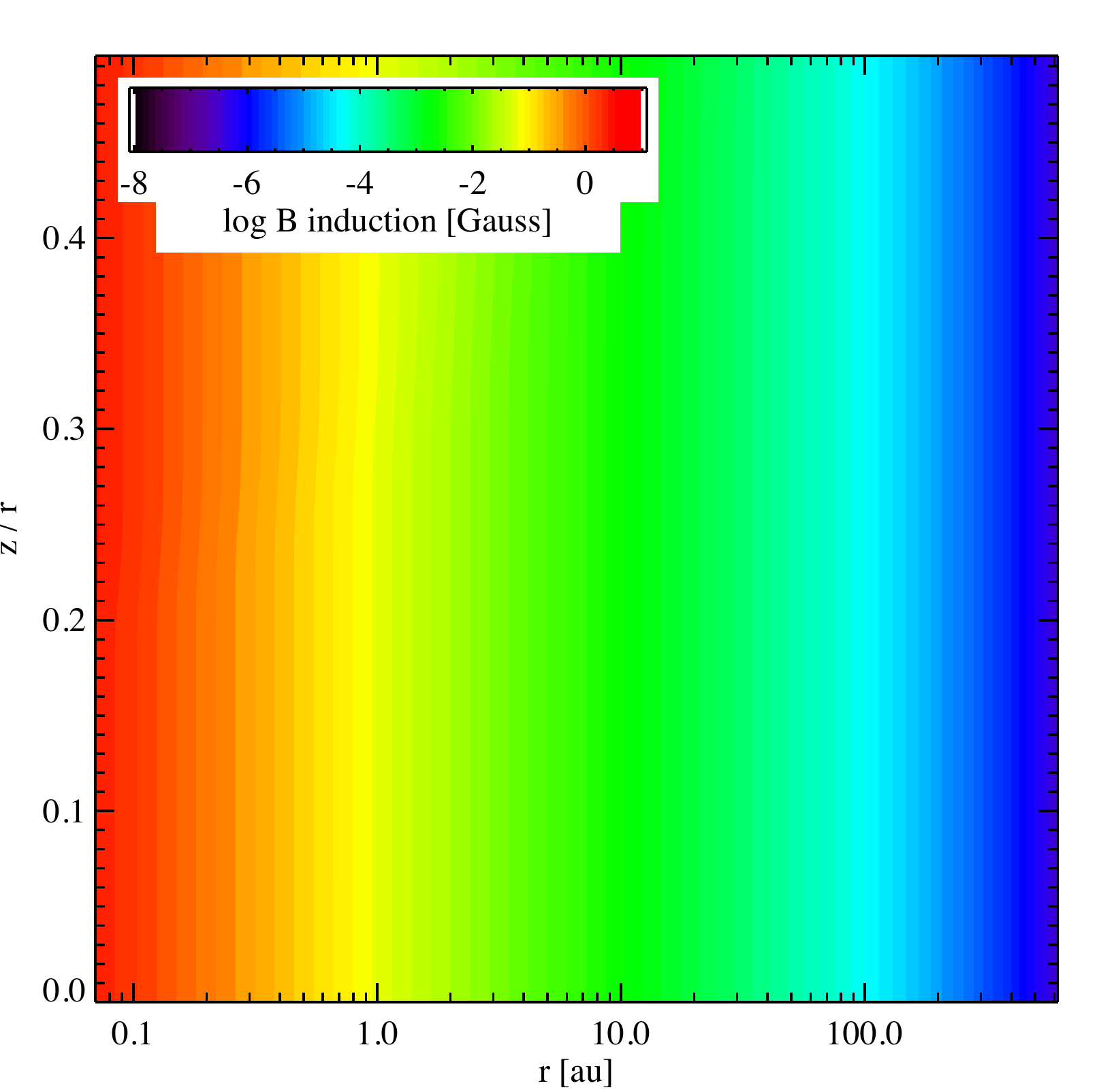}  
  \includegraphics[angle=0,width=8cm,height=8cm,trim=50 80  80 300, clip]{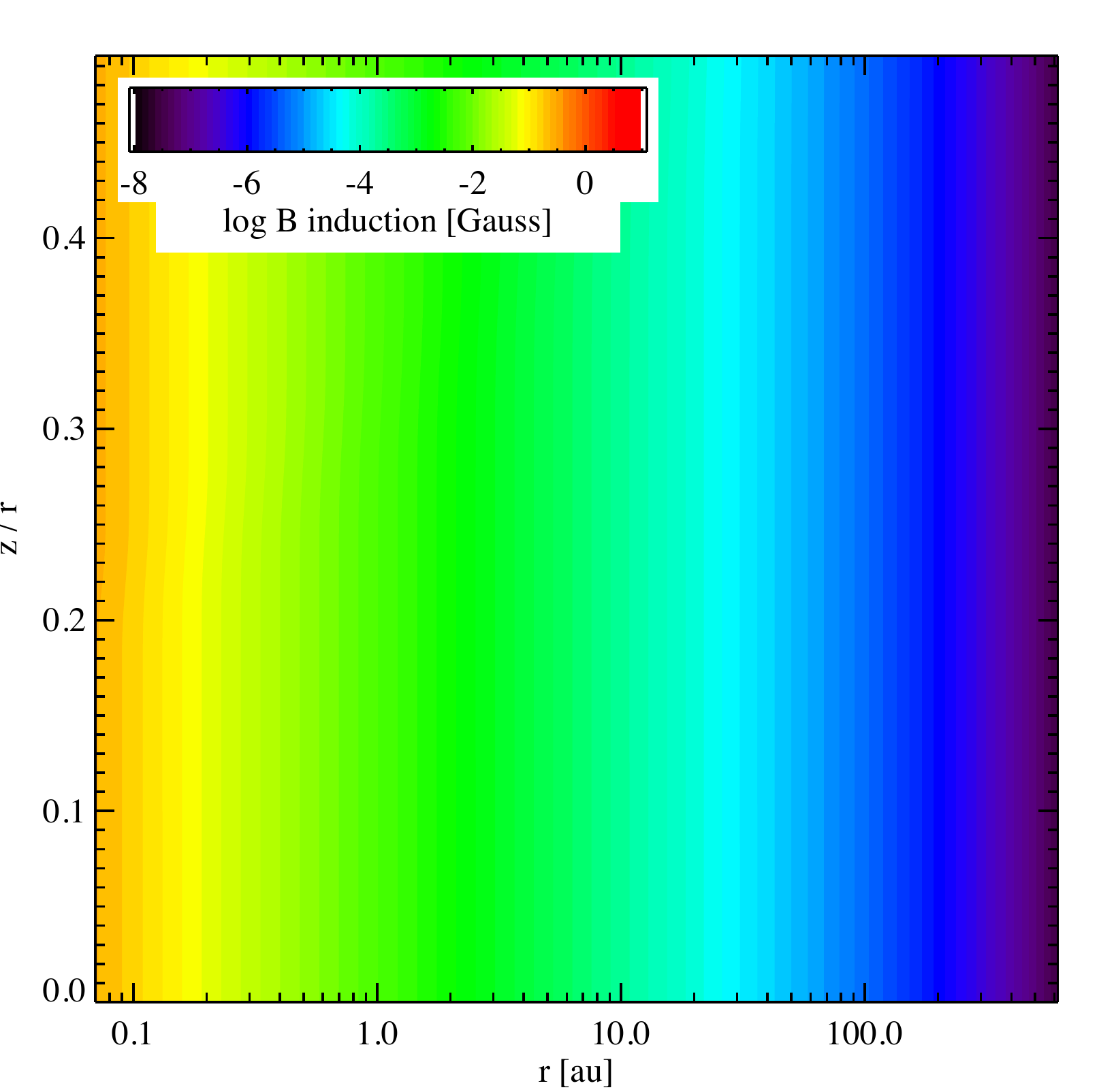}
  \includegraphics[angle=0,width=8cm,height=8cm,trim=50 80  80 300, clip]{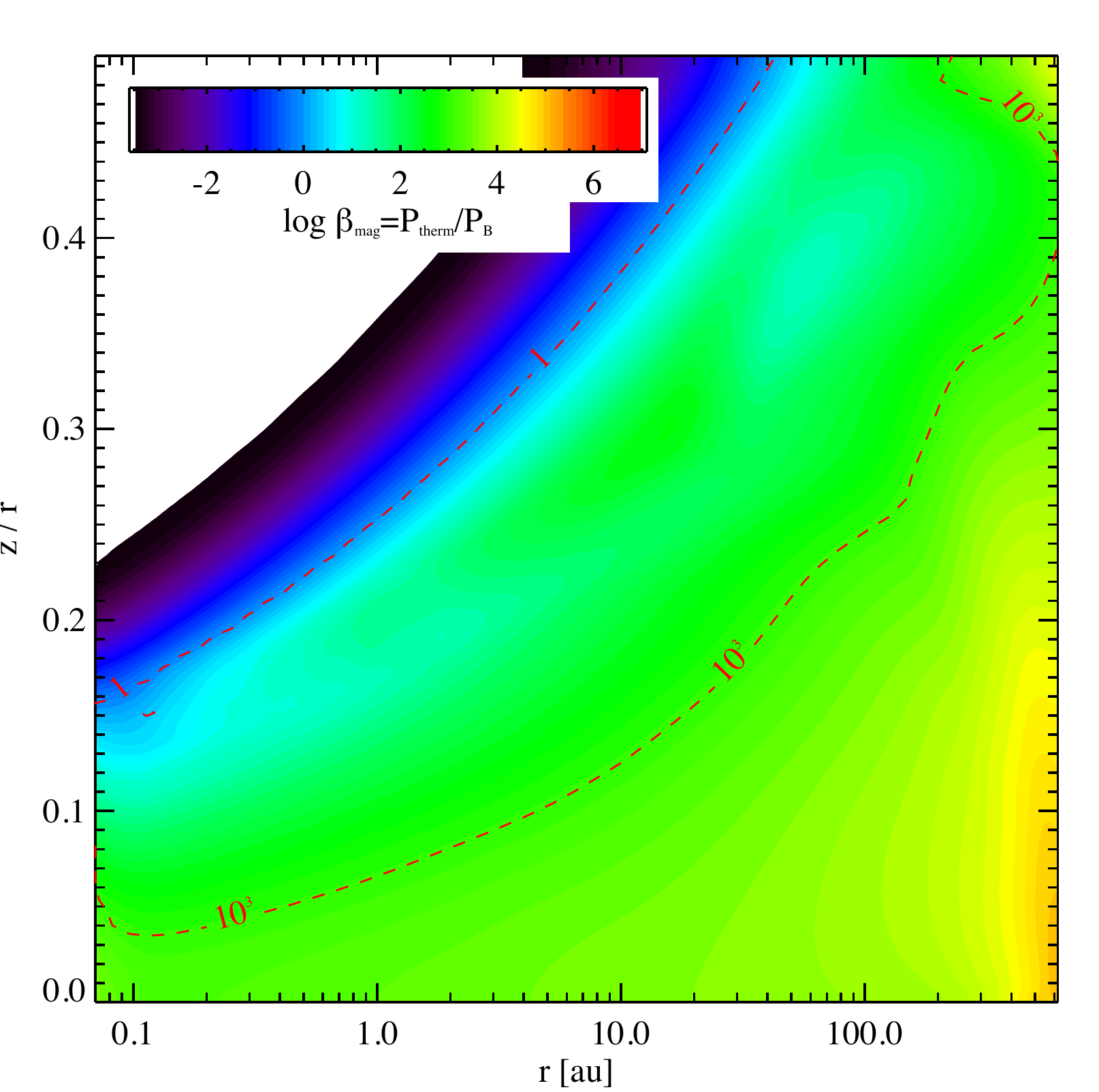}
  \includegraphics[angle=0,width=8cm,height=8cm,trim=50 80  80 300, clip]{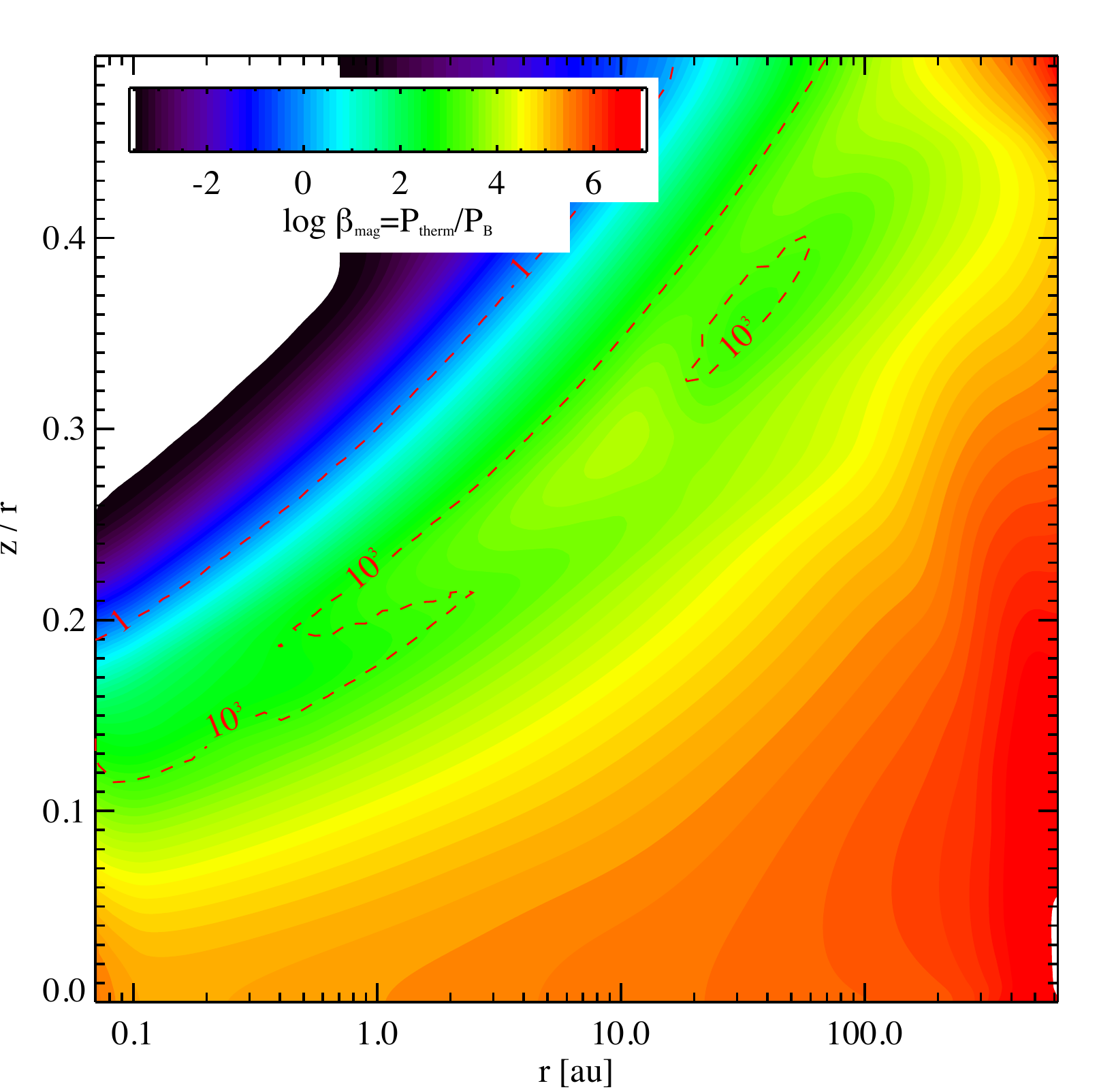}
   \includegraphics[angle=0,width=8cm,height=8cm,trim=50 80  80 300, clip]{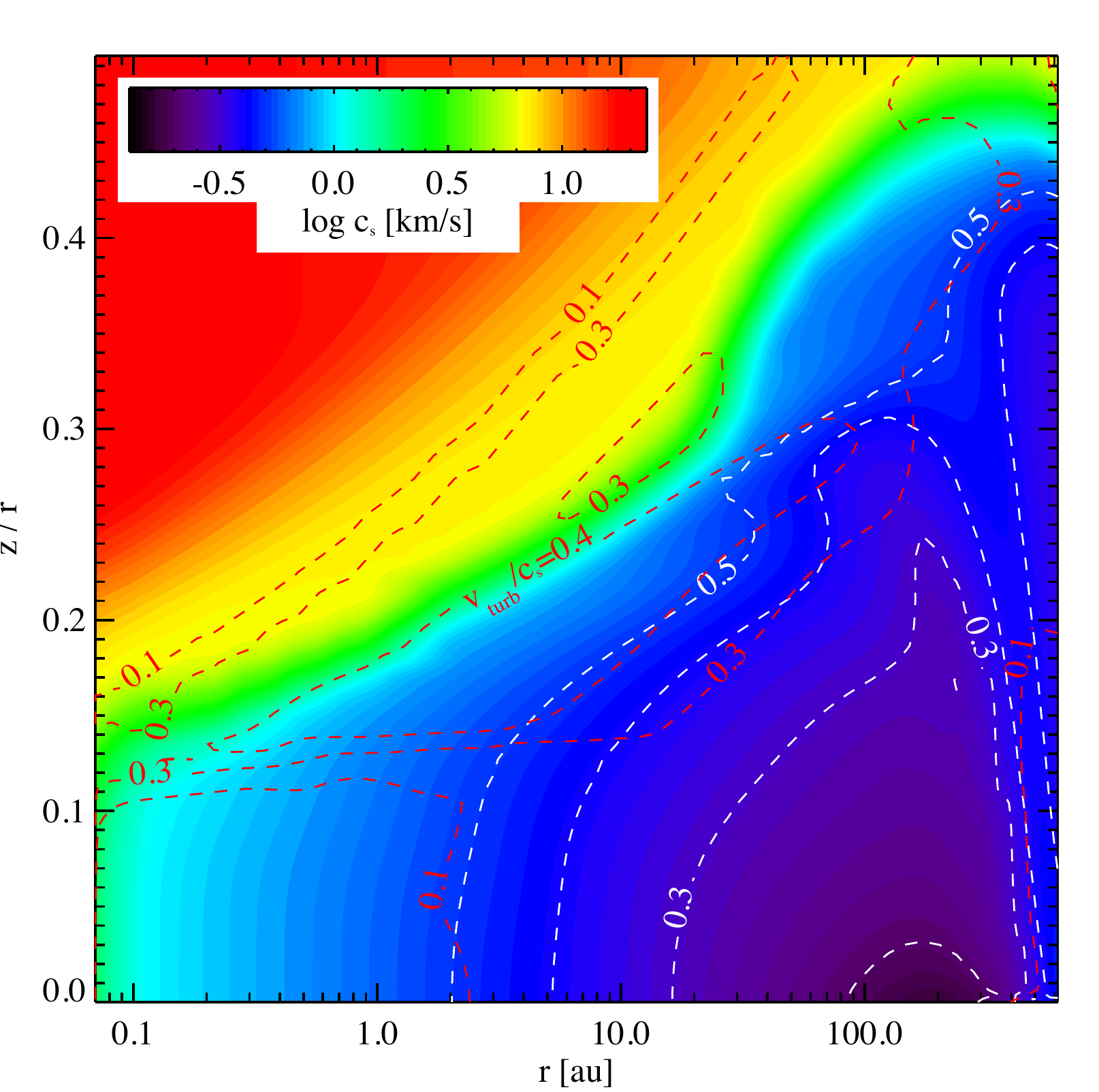}
    \includegraphics[angle=0,width=8cm,height=8cm,trim=50 80  80 300, clip]{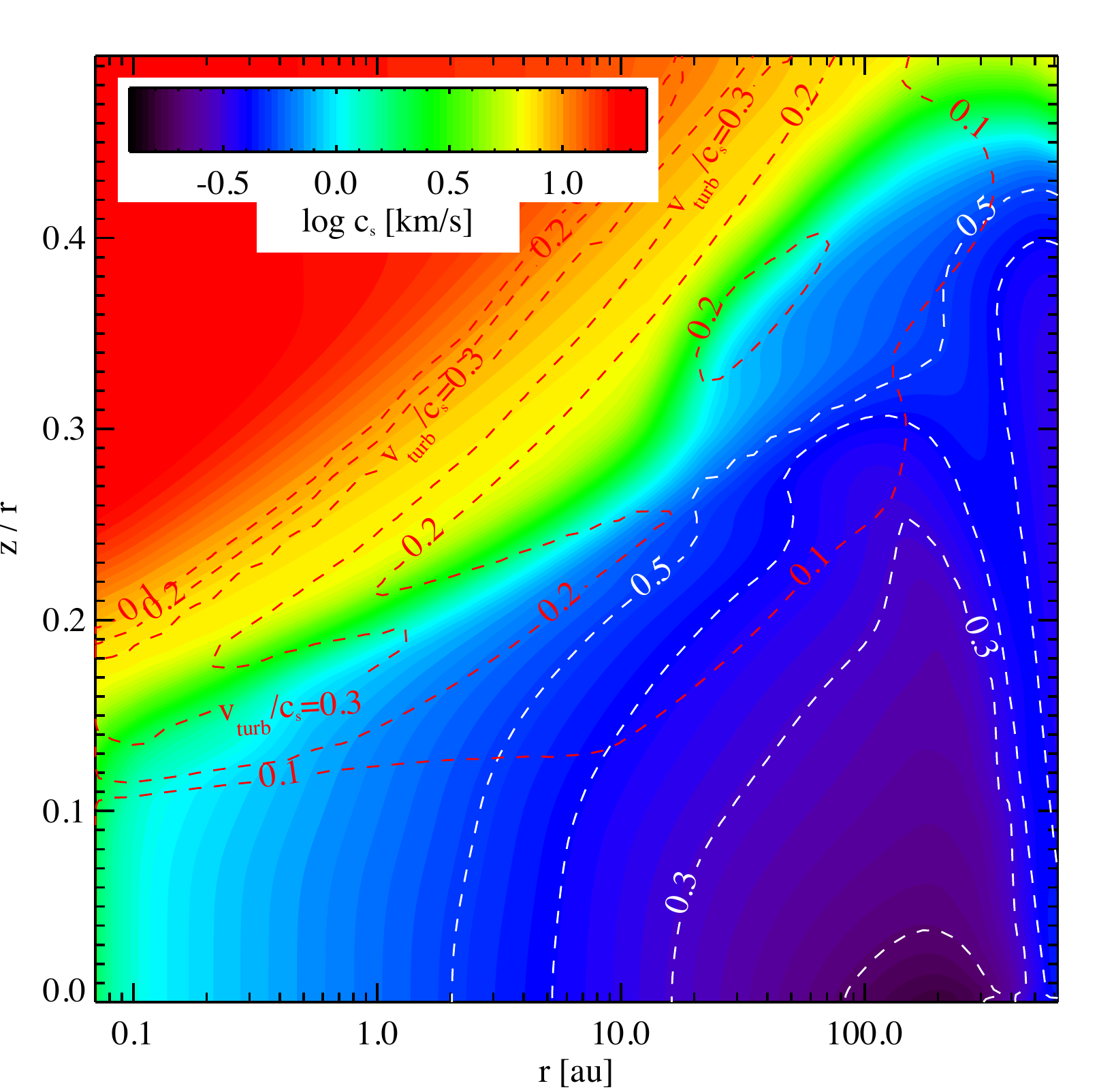}
  \caption{The resulting vertical component of magnetic field (upper panels) and of the distribution of $\beta_{\mathrm{mag}}$ (middle panels) for disk model with  $\beta_{\mathrm{mid}}$=10$^{4}$ (left panel) and $\beta_{\mathrm{mid}}$=10$^{6}$ (right panel).  Notice that $\beta_{\mathrm{mag}}=\beta_{\mathrm{mid}}$ at all radius $r$ at $z=0$. 
 The contour with $\beta_{\mathrm{mag}}=1$ defines the disk surface. The sound speed structures are shown in the lower panels. The red contours in the lower panels show $\mathrm{v_{turb}}/\mathrm{c_s}$.}
  \label{fig_Bfield}           
\end{figure*}

The vertical component of magnetic field, the distribution of $\beta_{\mathrm{mag}}$, and the sound speed distribution are shown for the disk model with $\beta_{\mathrm{mid}}$=10$^{4}$ (left panels) and $\beta_{\mathrm{mid}}$=10$^{6}$ (right panels)in Fig.~\ref{fig_Bfield}. 

% ---------------- 
% ----------------

\section{Disk model temperature structure}\label{disk_temperatures}

\begin{figure*}[!htbp]
  \centering 
  \includegraphics[angle=0,width=8cm,height=8cm,trim=50 80  80 300, clip]{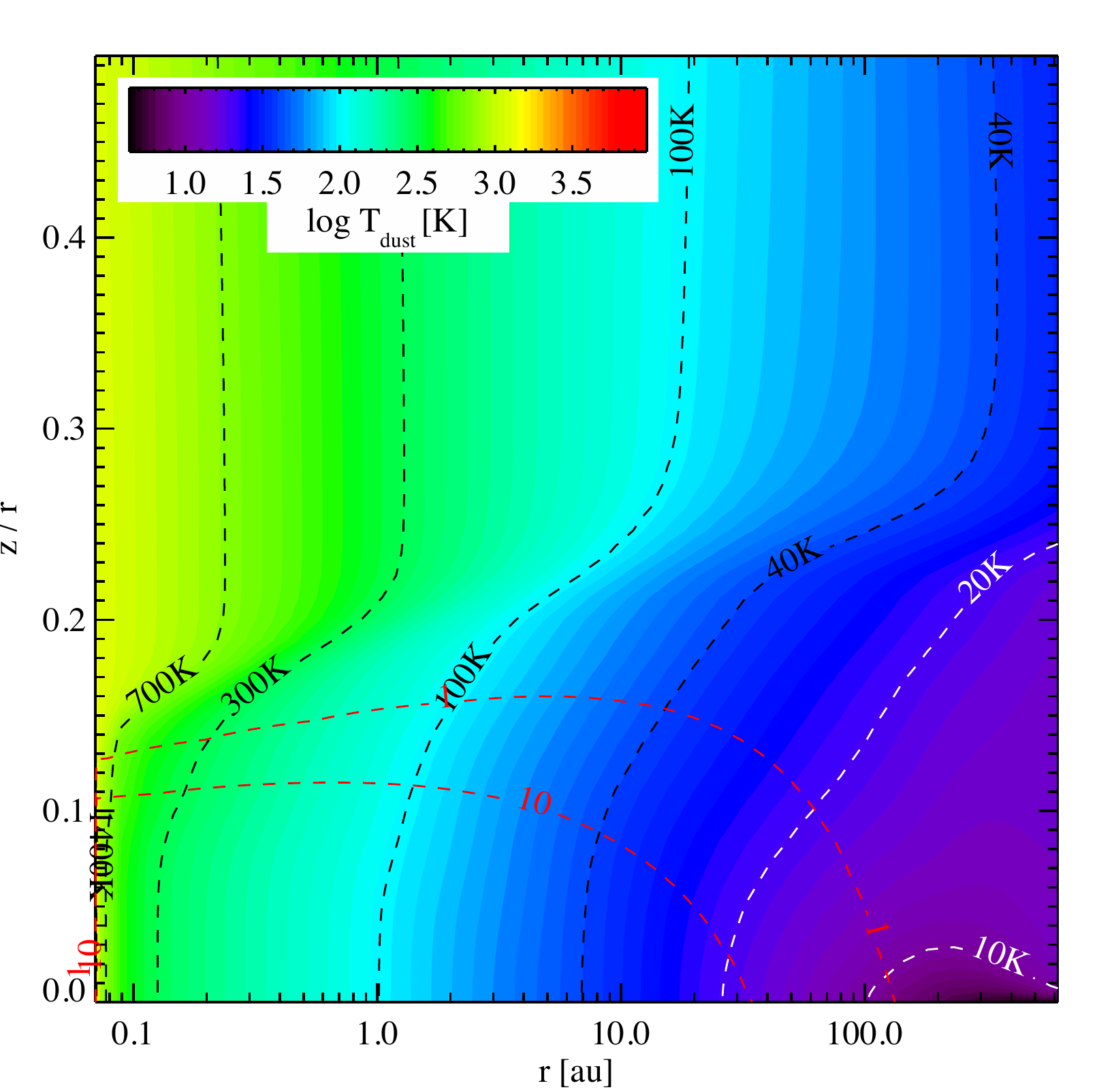}  
  \includegraphics[angle=0,width=8cm,height=8cm,trim=50 80  80 300, clip]{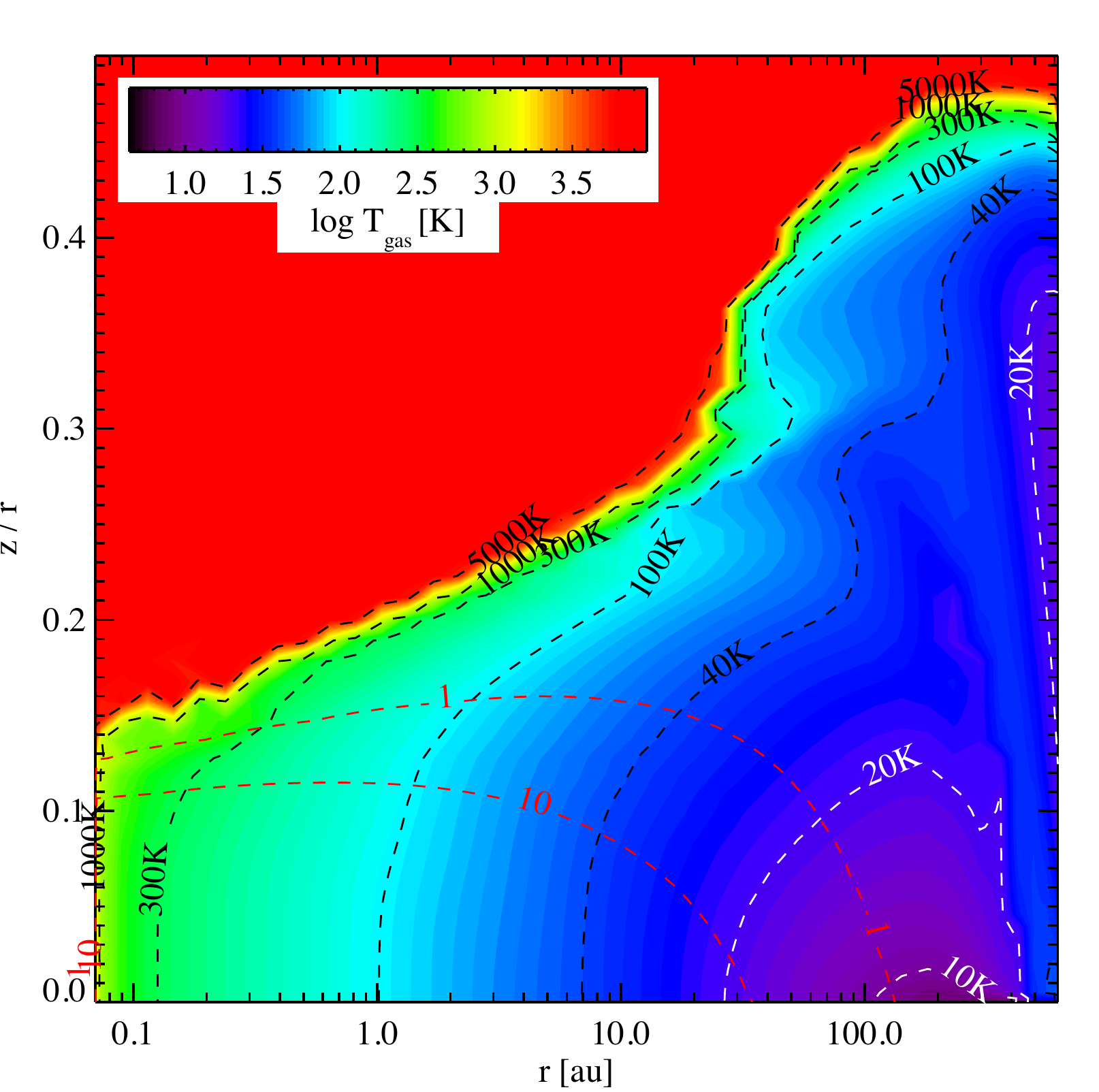}
  \includegraphics[angle=0,width=8cm,height=8cm,trim=50 80  80 300, clip]{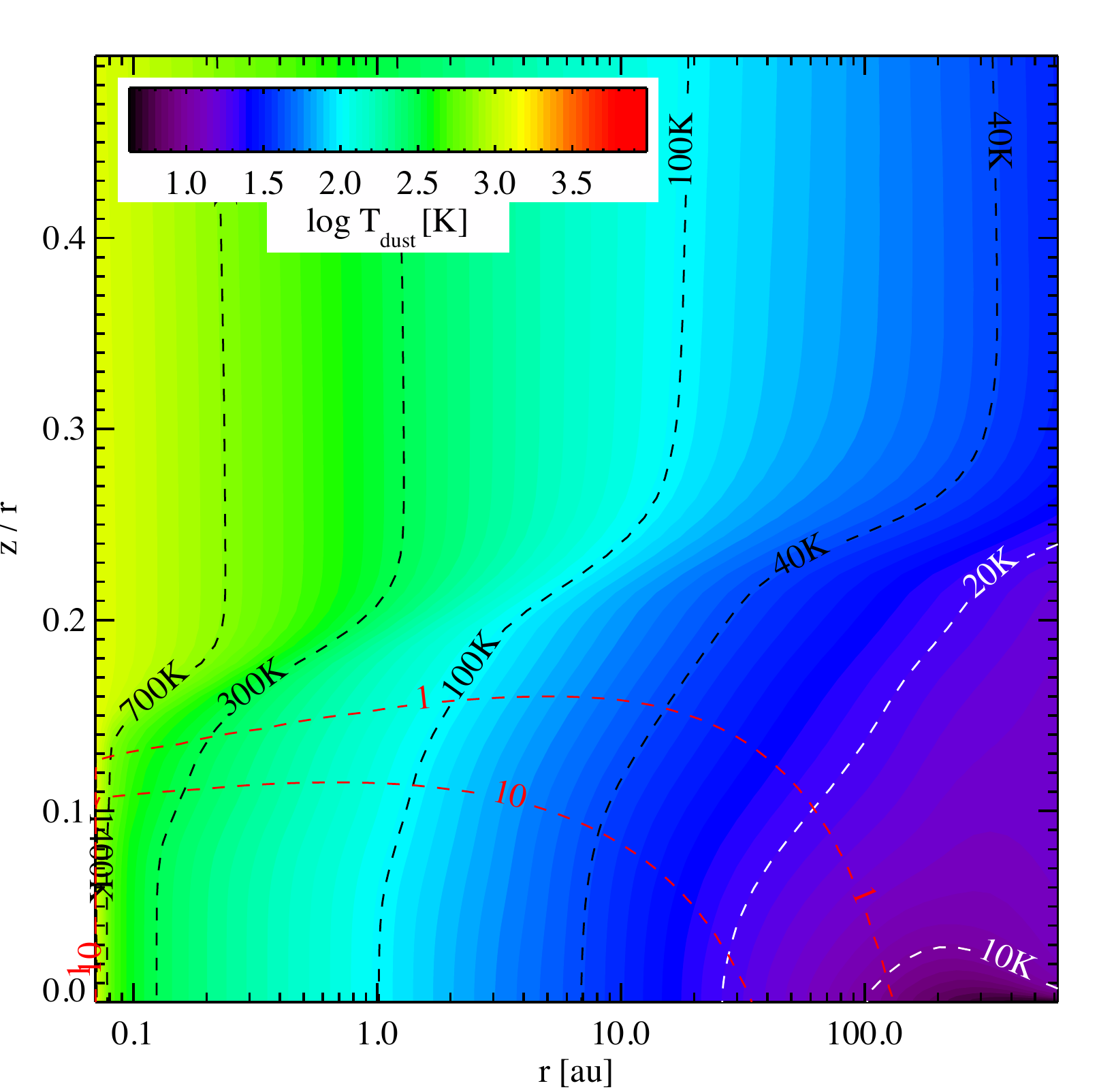}  
  \includegraphics[angle=0,width=8cm,height=8cm,trim=50 80  80 300, clip]{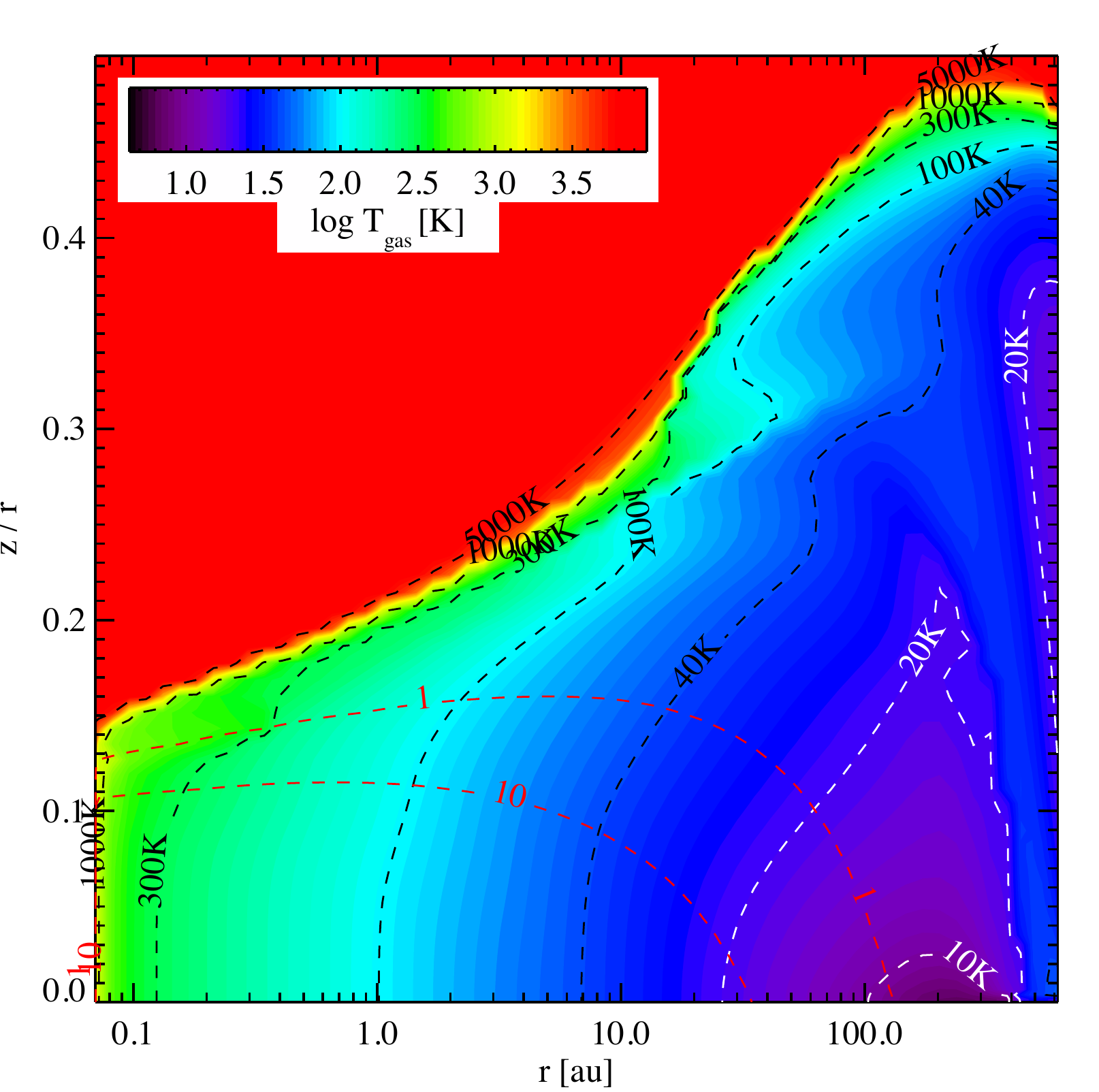}  
  \caption{Disk dust (left panel) and gas (right panel) temperature structure. The top panels correspond to the $\beta_{\rm mid}$=10$^{4}$ model and the lower panels to the $\beta_{\rm mid}$=10$^{6}$ model.}
  \label{fig_disk_temperatures}           
\end{figure*}

The disk dust and gas thermal structure are shown in Fig.~\ref{fig_disk_temperatures}. 

% ----------------
\section{Analytical Ohm Elsasser number}\label{Ohm_analytical}

\begin{figure}[!htbp] 
  \centering
  \includegraphics[angle=0,width=8cm,height=8cm,trim=50 80  80 300, clip]{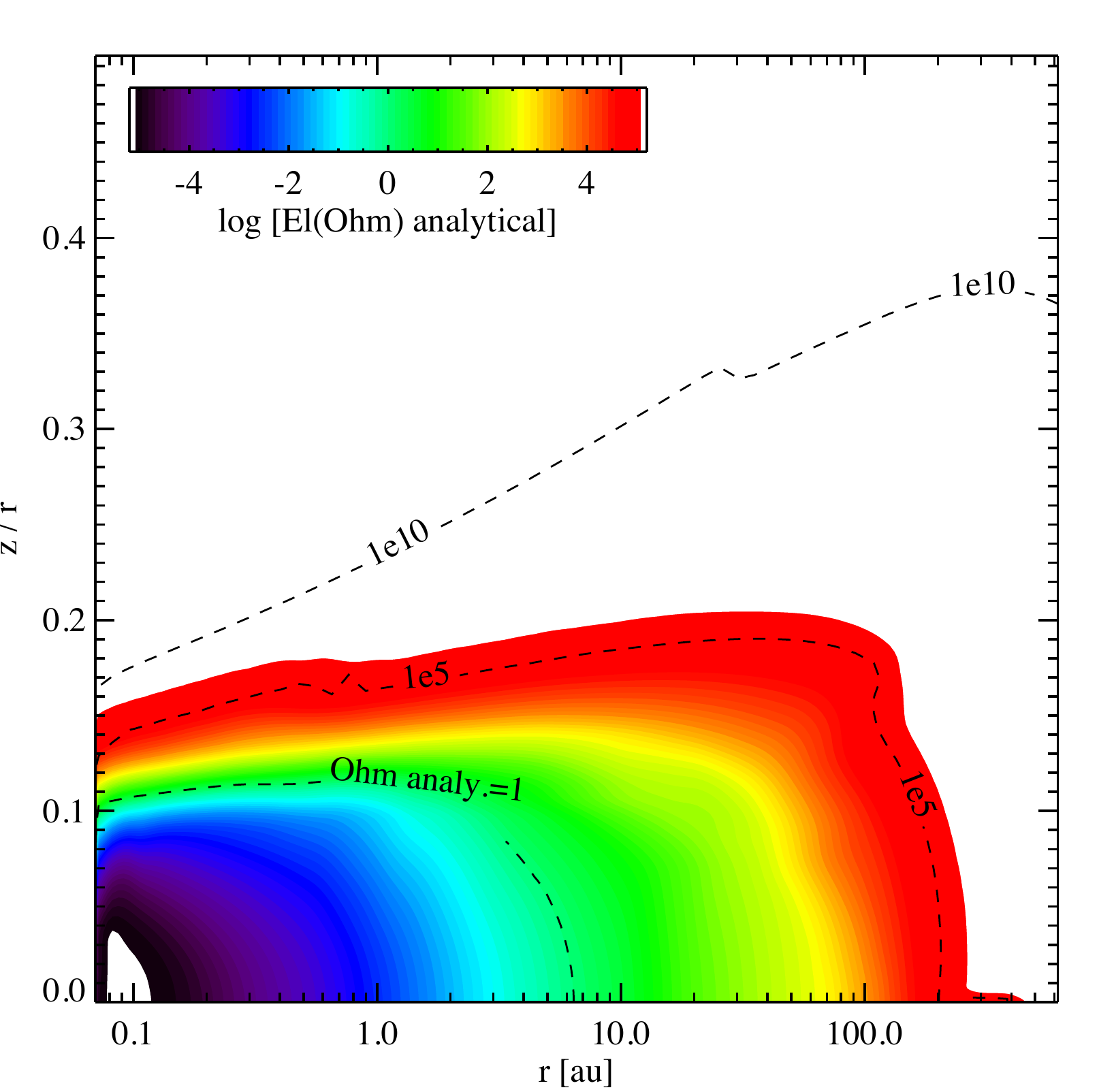} 
  \caption{Analytical approximation to the Ohm Elsasser number for a model with $\beta_{\mathrm{mid}}=10^4$.}
 \label{fig_Ohm_analytical}
\end{figure}

Fig.~\ref{fig_Ohm_analytical} shows the distribution of $\Lambda_{\rm Ohm} $ using an analytical approximation for $\beta_{\mathrm{mid}}=10^4$.

\end{appendix}

\end{document}